\documentclass[aps, superscriptaddress, showpacs,pra, twocolumn]{revtex4-1}
\usepackage{ORI_Group_style}
\usepackage{blindtext}
\usepackage[dvipsnames]{xcolor}
\usepackage{array}
\usepackage{subfigure}
\usepackage{float}
\usepackage{mathtools}

\newcommand{\cd}{c^\dagger}
\newcommand{\bd}{b^\dagger}

\hypersetup{colorlinks,linkcolor=red,citecolor=blue,urlcolor=blue}

\begin{document}

\title{Theory for Cavity Cooling of Levitated Nanoparticles via Coherent Scattering: Master Equation Approach}

 \author{C. Gonzalez-Ballestero}
 \affiliation{Institute for Quantum Optics and Quantum Information of the Austrian Academy of Sciences, A-6020 Innsbruck, Austria.}
 \email{carlos.gonzalez-ballestero@uibk.ac.at}
 \affiliation{Institute for Theoretical Physics, University of Innsbruck, A-6020 Innsbruck, Austria.}
 \author{P. Maurer}
 \affiliation{Institute for Quantum Optics and Quantum Information of the Austrian Academy of Sciences, A-6020 Innsbruck, Austria.}
 \affiliation{Institute for Theoretical Physics, University of Innsbruck, A-6020 Innsbruck, Austria.}
 \author{D.~\surname{Windey}}	
\affiliation{Photonics Laboratory, ETH Z{\"u}rich, 8093 Z{\"u}rich, Switzerland}
\author{L.~\surname{Novotny}}
    \affiliation{Photonics Laboratory, ETH Z{\"u}rich, 8093 Z{\"u}rich, Switzerland}
    \author{R.~\surname{Reimann}}
	\affiliation{Photonics Laboratory, ETH Z{\"u}rich, 8093 Z{\"u}rich, Switzerland}
 \author{O. Romero-Isart}
  \affiliation{Institute for Quantum Optics and Quantum Information of the Austrian Academy of Sciences, A-6020 Innsbruck, Austria.}
 \affiliation{Institute for Theoretical Physics, University of Innsbruck, A-6020 Innsbruck, Austria.}

\begin{abstract}
We develop a theory for cavity cooling of the center-of-mass motion of a levitated nanoparticle through coherent scattering into an optical cavity. We analytically determine the full coupled Hamiltonian for the nanoparticle, cavity, and free electromagnetic field. By tracing out the latter, we obtain a Master Equation for the cavity and the center of mass motion, where the decoherence rates ascribed to recoil heating, gas pressure, and trap displacement noise are calculated explicitly. Then, we benchmark our model by reproducing published experimental results for three-dimensional cooling. Finally, we use our model to demonstrate the possibility of ground-state cooling along each of the three motional axes. Our work illustrates the potential of cavity-assisted coherent scattering to reach the quantum regime of levitated nanomechanics.
\end{abstract}


\maketitle


\section{Introduction}

Initially conceived as a way to minimize clamping losses in mechanical resonators \cite{RomeroIsartNJP2010,ChangPNAS2010}, the study of levitated nanoparticles (NPs), or {\em levitodynamics}, has branched into a wide research field in the last years. On the one hand, this is due to the wide range of particles available for levitation, such as dielectrics \cite{LiNatPhys2011,GieselerPRL2012,KiesePNAS2013,MillenPRL2015,FonsecaPRL2016,JainPRL2016}, nanocrystals containing quantum emitters \cite{KuhlickeAPL2014,RahmanNatPhot2017,DelordPRL2018,ConanglaNanoLett2018,AldaAPL2018,FrimmerPRA2017}, 
nanomagnets~\cite{RusconiPRL2017} and superconducting spheres \cite{RomeroIsartPRL2012,CirioPRL2012,PinoQST2018}, as well as the large variety in NP shapes \cite{AsenbaumNatComm2013,HoangPRL2016,AhnPRL2018,KuhnOptica2017}. On the other hand, many recent experiments have demonstrated a very precise control of both the center of mass (COM) motion \cite{JainPRL2016,DiehlPRA2018,AhnPRL2018,TebbenjohannsArXiv2018} and the rotation
\cite{ReimannPRL2018,KuhnNatCom2017,KuhnNanoLett2015} of levitated NPs, as well as the integration of levitated NPs with optical emitters \cite{RahmanNatPhot2017,ConanglaNanoLett2018,DelordPRL2018}, and levitation in ultra high vacuum \cite{JainPRL2016}. Such achievements pave the way toward interesting new possibilities, such as using NPs as inertial or force sensors \cite{RanjitPRA2016,MonteiroPRA2017,HebestreitPRL2018},
studying microscopic thermodynamics of the COM motion of a NP \cite{GieselerNatNano2014,MillenNatNano2014,GieselerEntropy2018}, or the potential to use levitated NPs for optomechanics \cite{RomeroIsartNJP2010,ChangPNAS2010,RomeroIsartPRA2011} and for the preparation of large quantum superposition states \cite{RomeroIsartPRL2011,RomeroIsart2011b,BatemanNatComm2014}.

The promising prospects of levitodynamic-based quantum applications rely on the ability of cooling the COM motion down to the ground state. Although this milestone has not yet been attained, significant advances toward this goal have been achieved, such as parametric feedback cooling \cite{LiNatPhys2011,GieselerPRL2012,DiehlPRA2018} or cavity-assisted optomechanical cooling \cite{AsenbaumNatComm2013,KiesePNAS2013,MillenPRL2015}. Among these, a particularly interesting option, initially proposed for cold atoms and molecules \cite{VuleticPRA2001,DomokosPRA2002,RitschRMP2013}, is to cool the COM motion via coherent light scattering into an optical cavity, as recently demonstrated in two experiments \cite{WindeyArXiv2018,DelicArXiv2018}. In this configuration, a NP is optically trapped by an optical tweezer \cite{AshkinPRL1970,AshkinAPL1976}, whose photons are scattered into a blue-detuned cavity, reducing the mechanical energy of the COM in the process. This cooling method presents several advantages such as reduced technological complexity, a high level of control and the possibility of cooling the COM motion along the three motional axes. It therefore appears as a strong candidate for realizing the COM ground state in levitated nanomechanical systems.

\begin{figure}
	\centering
	\includegraphics[width=0.8\linewidth]{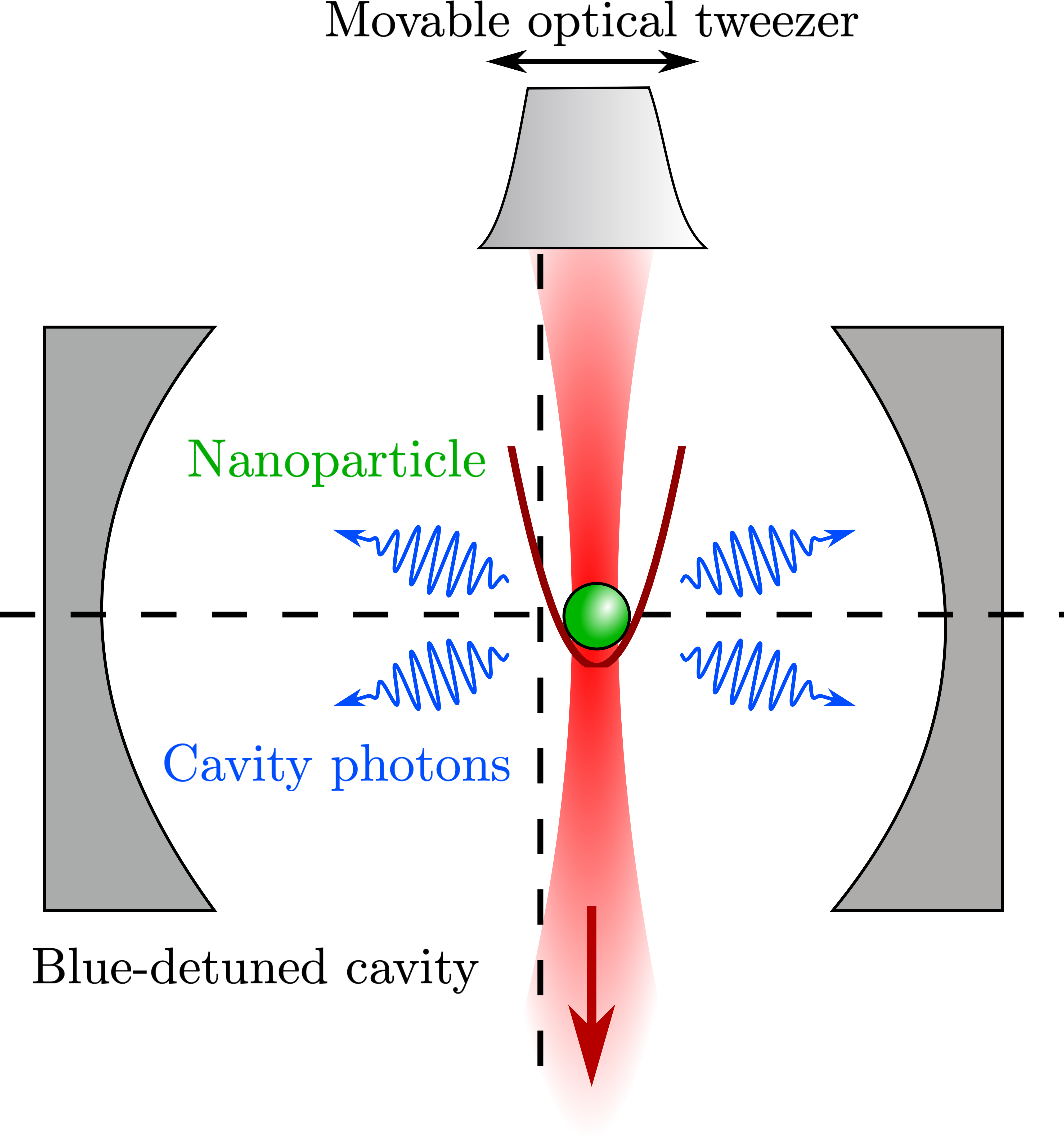}
	\caption{(Color online). Scheme of the system under study. A nanoparticle (NP) is levitated by a propagating optical tweezer (in red), and placed into an independent blue-detuned cavity (blue). The position of the particle along the cavity axis can be controlled by moving the tweezer focus.}\label{figsetup}
\end{figure}

Previous theoretical works have addressed the method of cavity-assisted COM cooling by coherent scattering: either semiclassically for cold atoms \cite{HechenblaiknerPRA1998,DomokosJPhysB2001,VuleticPRA2001}, or using quantum optomechanical theory to describe the cooling of the COM of atoms along one or two of the trapping axes \cite{LevPRA2008,VukicsJOptB2004}, and of optically levitated dielectric NPs along one trapping axis \cite{RomeroIsartPRA2011}. However, a full quantum theory of three-dimensional cooling of optically levitated dielectric NPs via coherent scattering, as well as a detailed study of the relevant heating and decoherence mechanisms, is lacking so far. On the one hand, developing such a theory will contribute to understand the limitations of present experiments \cite{WindeyArXiv2018,DelicArXiv2018} and the way to overcome them in order to achieve ground state cooling. On the other hand, once the ground state is reached, a full quantum theory including a detailed quantum description of the decoherence sources will be essential not only to describe this state, but to implement further quantum protocols and applications to prepare, for instance, non-Gaussian quantum states.

In this paper, we develop a quantum theory of three-dimensional COM cooling via cavity-assisted coherent scattering. First, in Sec. \ref{secHamiltonian} we derive the full optomechanical Hamiltonian in the long wavelength approximation. In Sec. \ref{secReducedDynamics}, we obtain an effective equation of motion for the reduced subsystem formed by cavity and COM motion, and introduce three relevant heating rates, namely the recoil heating, the heating related to the background gas pressure, and the trap displacement noise. In Sec. \ref{secResults}, we focus on a case study and characterize the system dynamics for a realistic experimental setup \cite{WindeyArXiv2018}. We continue in Sec. \ref{secGScooling} by analyzing the possibility of ground state cooling along each motional axes in state-of-the-art experiments. Finally, our conclusions are presented in Sec. \ref{secConclusion}.

\section{Hamiltonian of the total system}\label{secHamiltonian}

The system under study, schematically depicted in Fig.~\ref{figsetup}, consists of a single dielectric NP with radius $R$, and homogeneous and isotropic relative permittivity $\varepsilon$. The NP is trapped at the focus of a high-intensity optical tweezer propagating along the $z$~axis whose frequency is $\omega_0 = 2\pi c/\lambda_0$, where $c$ is the vacuum speed of light and $\lambda_0$ the tweezer wavelength. Additionally, the NP is placed inside an optical cavity and is coupled to a mode with frequency $\omega_c$. The cavity axis is orthogonal to the propagation direction of the tweezer. We include the two degenerate polarization states of the non-birefringent cavity with frequency $\omega_c$, although, as we will see below, only one will significantly contribute to COM cooling.
In this section we describe the system and the fundamental light-matter Hamiltonian governing its dynamics. We
then derive, for a NP confined close to the tweezer focus, a second quantization form for such Hamiltonian. 
Finally, we reduce this Hamiltonian to a quadratic form by transforming into a frame where non-quadratic terms become negligible.

The system detailed above, and depicted in Fig.~\ref{figsetup}, can be described by a Hamiltonian containing the kinetic energy of the free NP, the energy of the electromagnetic (EM) field, and the NP-field interaction, i.e.,
\begin{equation}\label{firstH}
    \hat{H} = \frac{\hat{\mathbf{P}}^2}{2m} + \hat{H}_F + \hat{H}_{\rm Int}.
\end{equation}
Here, $\hat{\mathbf{P}}$ denotes the COM momentum operator, and $m$ denotes the mass of the NP. The free Hamiltonian of the EM field is given by
\begin{equation}\label{Hfield0}
    \hat{H}_F = \frac{\varepsilon_0}{2}\int \text{d}^3\mathbf{r}  \left[\hat{\mathbf{E}}^2(\mathbf{r}) + c^2\hat{\mathbf{B}}^2(\mathbf{r})\right],
\end{equation}
with the transverse electric and magnetic field operators $\hat{\mathbf{E}}(\mathbf{r})$ and $\hat{\mathbf{B}}(\mathbf{r})$ at position $\mathbf{r}$.
The NP-EM field interaction term in Eq.~\ref{firstH} reads \cite{RomeroIsartPRA2011}
\begin{equation}\label{Hint1st}
    \hat{H}_{\rm Int} =-\frac{1}{2}\alpha \hat{\mathbf{E}}^2(\hat{\mathbf{R}}),
\end{equation}
where we have introduced the NP polarizability $\alpha = \varepsilon_0 \epsilon_c V$, with the NP volume $V$, the vacuum permittivity $\varepsilon_0$, and $\epsilon_c \equiv 3(\varepsilon-1)/(\varepsilon+2)$. 
The above expression is valid in the long wavelength approximation, i.e., when $R\ll \lambda$ for all relevant wavelengths $\lambda$ \cite{Novotnybook}. The motion-light interaction arises from evaluating the electric field operator at the COM position of the NP, $\hat{\mathbf{R}}$. Note that matter is treated classically in this work, as the optical response of the NP is described exclusively by its polarizability. 

In order to write the Hamiltonian in second quantization, we need to determine the structure of the EM field. Both in Eqs. \ref{Hfield0} and \ref{Hint1st}, the electric and magnetic field operators should be written in terms of the eigenmodes of the corresponding EM structure. Since solving Maxwell's equations in all space in the presence of a cavity is rather involved, it is common in photonics to 
approximate the EM field as independent free-space modes plus some extra modes representing the structure, i.e. the cavity. Here we use the same approximation, thus writing the electric field as
  \begin{equation}\label{Efield1st}
      \hat{\mathbf{E}}(\mathbf{r}) \approx \boldsymbol{\mathcal{E}}_{\rm tw}(\mathbf{r},t)+\hat{\mathbf{E}}_{\rm cav}(\mathbf{r})+\hat{\mathbf{E}}_{\rm free}(\mathbf{r}),
 \end{equation}
 whose three components correspond to the tweezer field, the cavity field, and the free EM field, respectively.  Since we assume the tweezer field to be in a highly populated coherent state, we approximate it as a classical quantity by its mean value in the rotating frame,
 \begin{equation}\label{Etweezertime}
     \boldsymbol{\mathcal{E}}_{\rm tw}(\mathbf{r},t) = \frac{1}{2}\left[\boldsymbol{\mathcal{E}}_{\rm tw}(\mathbf{r})e^{\im\omega_0 t} + \boldsymbol{\mathcal{E}}^*_{\rm tw}(\mathbf{r})e^{-\im\omega_0 t}\right].
 \end{equation}
On the other hand, the electric fields of the cavity and the free EM modes are described quantum mechanically, i.e.,
\begin{equation}\label{Ecavquantum}
        \hat{\mathbf{E}}_{\rm cav}(\mathbf{r}) = \sum_{\alpha}\sqrt{\frac{\hbar\omega_c}{2\varepsilon_0 V_c}}\left[\mathbf{E}_{c\alpha}(\mathbf{r})\hat{c}_\alpha + \text{H.c.}\right],
\end{equation}
 \begin{equation}\label{Efree}
     \hat{\mathbf{E}}_{\rm free}(\mathbf{r}) = \sum_{\mathbf{k}\varepsilon}\sqrt{\frac{\hbar\omega_\mathbf{k}}{2\varepsilon_0 \mathcal{V}}}\!\left[\boldsymbol{\varepsilon}_\mathbf{k}e^{\im\mathbf{k}\mathbf{r}}\hat{a}_{\varepsilon\mathbf{k}} + \text{H.c.}\right]\!.
 \end{equation}
 Here,  $V_c$, $\mathbf{E}_{c\alpha}(\mathbf{r})$ and $\hat{c}_\alpha$ represent the cavity mode volume, the normalized classical electric field of the cavity mode, and the annihilation operator for a cavity mode with polarization $\alpha$, respectively. Analogously, $\mathcal{V}$ and $\hat{a}_{\varepsilon\mathbf{k}}$ represent the quantization volume for the free EM field and the annihilation operator of a free EM mode with wavevector $\mathbf{k}$ and polarization $\boldsymbol{\varepsilon}_\mathbf{k}$. 
 Note that, in our approximation, the Hamiltonian of the EM field in Eq.~\ref{Hfield0} can be written as 
\begin{equation}\label{Hfield2Qv0}
    \hat{H}_F \approx  \hbar\omega_c \sum_{\alpha} \hat{c}_\alpha^\dagger \hat{c}_\alpha+\hbar\sum_{\mathbf{k}\varepsilon} \omega_\mathbf{k}\hat{a}_{\varepsilon\mathbf{k}}^\dagger\hat{a}_{\varepsilon\mathbf{k}}.
\end{equation}
However, in order to account for the cavity losses due to transmission through the mirrors, which play a relevant role in the cooling, it is convenient to refine the above Hamiltonian by explicitly including an additional set of EM modes to which the cavity is coupled \cite{RomeroIsartPRA2011,GardinerZollerQuantumNoise}, i.e., we take
\begin{equation}\label{Hfield2Q}
\begin{split}
    \hat{H}_F &=  \hbar\omega_c \sum_{\alpha} \hat{c}_\alpha^\dagger \hat{c}_\alpha+\hbar\sum_{\mathbf{k}\varepsilon} \omega_\mathbf{k}\hat{a}_{\varepsilon\mathbf{k}}^\dagger\hat{a}_{\varepsilon\mathbf{k}}
    \\
    &
    \!\!\!\!+\!\hbar\int_0^\infty \!\!\!\text{d}\omega\bigg[\omega\hat{a}_0^\dagger(\omega)\hat{a}_0(\omega) +\im\gamma(\omega)\!\left[\hat{a}_0(\omega)\hat{c}^\dagger \!-\! \text{H.c.}\right]\!\!\bigg]
   .
\end{split}
\end{equation}
Here, the continuum of modes described by the photonic operators $\hat{a}_0(\omega)$ would in principle represent a fraction of the free-space modes. Thus, by considering them a separate degree of freedom, we are double-counting some of the states. This overcounting, very common in quantum optics \cite{DutraCQEDbook}, has nevertheless a negligible effect on the correctness of the solution, since the ensemble of extra modes $\hat{a}_0(\omega)$ has zero measure \cite{RomeroIsartPRA2011}. The coupling constant $\gamma(\omega)$ is directly related to the cavity linewidth $\kappa$, here defined as the cavity field decay rate, and can be considered constant across a wide frequency range centered at $\omega_c$ with a value $\gamma(\omega) =\sqrt{\kappa/\pi}$ \cite{GardinerZollerQuantumNoise}. This equality can be readily certified by obtaining the Born-Markov master equation of the cavity in the presence of the environment composed by the modes $\hat{a}_0(\omega)$ (see e.g. Appendix \ref{appendixTRACEOUT}).

\begin{figure}
	\centering
	\includegraphics[width=\linewidth]{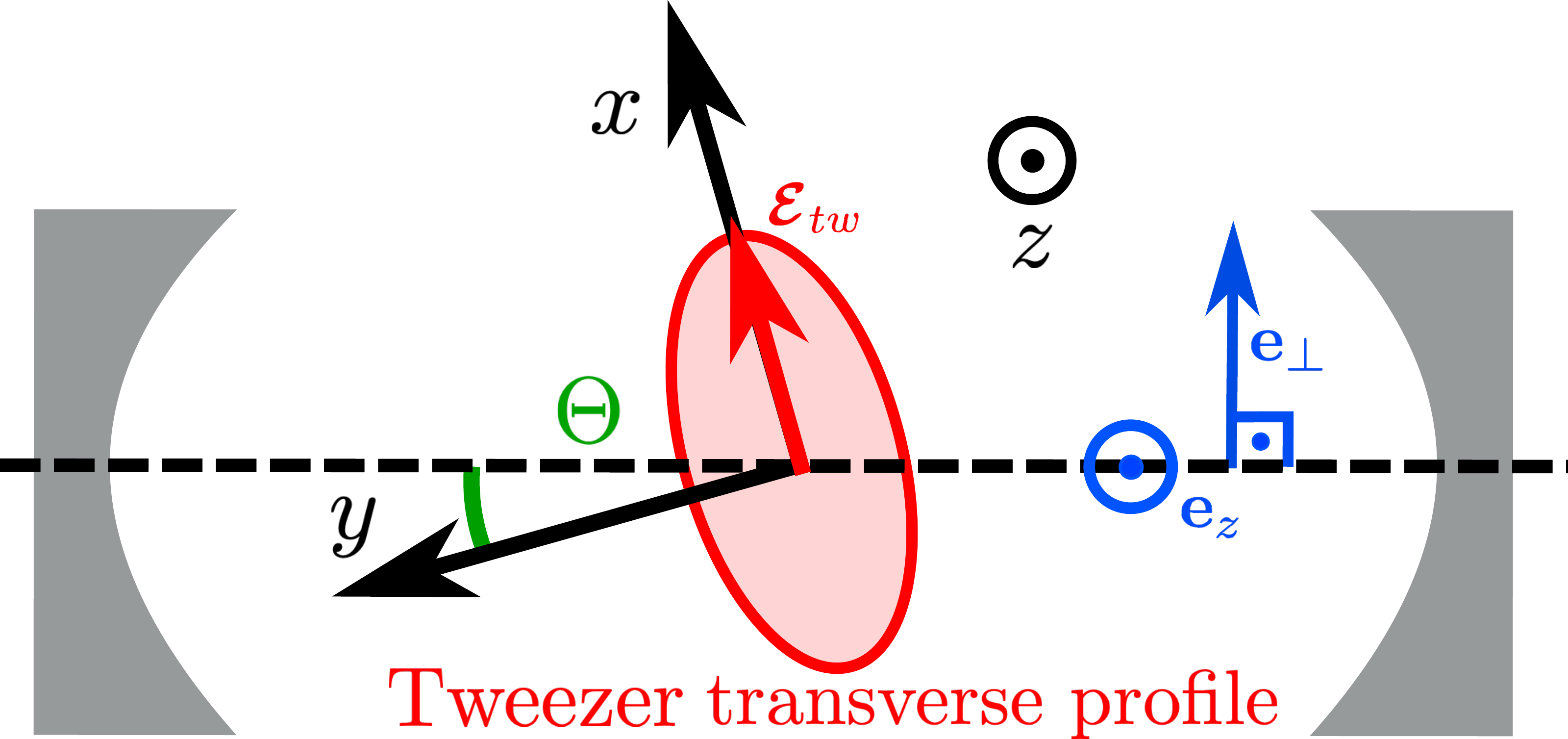}
	\caption{(Color online). Illustration of the chosen coordinate system. The tweezer propagates along the $z$ axis with a cylindrically asymmetric transverse profile modelled as an ellipse. We assume for simplicity linear polarization along one of the main axis, which we  choose as the $x$ axis. Then, both the cavity axis and the cavity polarizations appear rotated by an angle $\Theta$ in the $x-y$ plane. The origin of coordinates is taken at the intensity maximum of the tweezer.}\label{figaxis}
\end{figure}

In order to calculate the second-quantized interaction Hamiltonian $\hat{H}$ using the electric fields described above, it is convenient to determine the explicit form of the tweezer and the cavity field profiles. Let us start by the tweezer field, namely  
 $\boldsymbol{\mathcal{E}}_{\rm tw}(\mathbf{r})$ in Eq.~\ref{Etweezertime}. Our first step is to fix the $x$ and $y$ axis in our coordinate system. Those axes are aligned with the symmetry axes of the tweezer-formed trapping potential. As pointed out in studies with atomic clouds \cite{VuleticPRA2001}, the possibility of 3D cooling requires that none of these axes is parallel to the cavity axis. 
 In the present case, this is achieved for tweezer polarizations being 
 neither purely parallel nor purely orthogonal to the cavity axis, as schematically depicted in Fig.~\ref{figaxis}. Since the tweezer polarization axis determines the trapping directions of the levitated particle, it is convenient to choose a coordinate system centered on the tweezer intensity maximum, and aligned with the tweezer polarization. We therefore choose the tweezer to be polarized along the $x$ axis (see Fig.~\ref{figaxis}), i.e., $\boldsymbol{\mathcal{E}}_{\rm tw}(\mathbf{r}) = \mathbf{e}_x \mathcal{E}_{\rm tw}(\mathbf{r})$, and modelled in the paraxial approximation by a classical, zeroth order, propagating Hermite-Gaussian beam \cite{EriksonPRE94},
\begin{equation}\label{Etweezer}
        \mathcal{E}_{\rm tw}(\mathbf{r}) = \! E_0\frac{W_t}{W(z)}e^{-x^2/W_x^2(z)}e^{-y^2/W_y^2(z)}e^{\im k_0z}e^{\im \phi_t(\mathbf{r})}\!.
\end{equation}
In the above equation, $W(z) = W_t\sqrt{1+(z/z_R)^2}$, where $W_t$ is the tweezer waist at the focus and $z_R = \pi W_t^2/\lambda_0$ is the Rayleigh range. The asymmetry of the transverse beam profile, which becomes more relevant for strongly focused tweezers, is encoded in the different extensions of the beam along the transverse axis, $W_{x,y}(z) = A_{x,y}W(z)$ with $A_{x,y}$ adimensional. 
The phase factor $\phi_t(\mathbf{r})$ is given by
\begin{equation}
    \phi_t(\mathbf{r}) = \arctan\left(\frac{z}{z_R}\right) -\frac{k_0z}{2}\frac{x^2+y^2}{z^2+z_R^2} \approx 0,
\end{equation}
in the vicinity of the origin, i.e., for $x,y,z \ll z_R \approx 1\,\mu\rm m$ in typical experiments \cite{WindeyArXiv2018}.
Finally, the field amplitude $E_0$ can be related to the tweezer power $P_t$ by first obtaining the paraxial expression for the associated magnetic field \cite{EriksonPRE94},
 \begin{equation}\label{Htweezertime}
     \boldsymbol{\mathcal{H}}_{\rm tw}(\mathbf{r},t) = \frac{1}{2}\mathbf{e}_y\left[\mathcal{H}_{\rm tw}(\mathbf{r})e^{\im\omega_0 t} + \mathcal{H}^*_{\rm tw}(\mathbf{r})e^{-\im \omega_0 t}\right],
 \end{equation}
 where $\mathcal{H}_{\rm tw}(\mathbf{r}) = \sqrt{\varepsilon_0/\mu_0}\mathcal{E}_{\rm tw}(\mathbf{r})$. One then defines the tweezer power in terms of the Poynting vector as
 \begin{equation}
     P_t = \int \text{d}\mathbf{S} \left\langle \boldsymbol{\mathcal{E}}_{\rm tw}(\mathbf{r},t) \times \boldsymbol{\mathcal{H}}_{\rm tw}(\mathbf{r},t) \right\rangle_T,
 \end{equation}
 where $d\mathbf{S}$ is the surface element perpendicular to the Poynting vector (in this case, $d\mathbf{S} = dxdy\mathbf{e}_z$), and $\langle\rangle_T$ denotes the time average. Using this expression we find
 \begin{equation}
     E_0 = \sqrt{\frac{4P_t}{\pi\varepsilon_0 c W_t^2 A_x A_y}}.
 \end{equation}
In this way we can relate all the beam parameters to experimental input parameters \cite{WindeyArXiv2018}, such as power, wavelength and beam waist (or, equivalently, numerical aperture of the focusing lens).

Let us now shift our attention to the cavity field $\mathbf{E}_{c\alpha}(\mathbf{r})$ in Eq.~\ref{Ecavquantum}. We will also approximate its profile by a zeroth-order Hermite-Gauss beam, this time in a standing wave configuration. Because of our choice of coordinate system (Fig.~\ref{figaxis}), the axis of this beam will not be parallel to $x$ or $y$, but instead lie along along $\mathbf{e}_{\rm cav} = \sin\Theta\mathbf{e}_x + \cos\Theta\mathbf{e}_y$. The expression for the two degenerate polarization modes is thus given by
\begin{equation}\label{Ecav}
    \begin{split}
        \mathbf{E}_{c\alpha}(\mathbf{r}) &= \mathbf{e}_\alpha \frac{W_c}{W(y')}
        e^{-(x'^2+z^2)/W^2(y')}
        \\
        &\times \cos[k_c y'+\phi_c(\mathbf{r})-\phi].
    \end{split}
    \end{equation}
Here, $k_c = \omega_c/c$, and $W_c$ is the cavity waist which, together with the cavity length $L_c = \pi c/(2\kappa\mathcal{F})$ ($\mathcal{F}$ being the Finesse), determines the mode volume as $V_c = \pi W_c^2 L_c/4$. The above beam contains the rotated coordinates $x'$ and $y'$, which are given by
     \begin{equation}
        x' = x\cos\Theta  - y\sin\Theta .
    \end{equation}
    \begin{equation}
        y' = x\sin\Theta  + y\cos\Theta .
    \end{equation}
Importantly, the phase factor $\phi$ in Eq.~\ref{Ecav}
determines the field intensity at the position of the NP, i.e. at the origin of coordinates. For $\phi = 0(\pi/2)$ the particle is at a maximum(minimum) of the cavity intensity profile.
Finally, the functions $W(y')$ and $\phi_c(\mathbf{r})$ in Eq.~\ref{Ecav} are given by
    \begin{equation}
        W(y')= W_c\sqrt{1+\left(y'/y_R\right)^2},
    \end{equation}
    \begin{equation}
        \phi_c(\mathbf{r}) = \arctan\left(\frac{y'}{y_R}\right)-\frac{k_c y'}{2} \frac{x'^2+z^2}{y'^2+y_R^2},
    \end{equation}
    with $y_R = \pi W_c^2/\lambda_c$ being the Rayleigh range of the cavity beam.
    Since we will be interested in the cavity field close to the origin, i.e., for $x,y \ll y_R \sim\,$mm in typical experiments \cite{WindeyArXiv2018}, we will assume $\phi_c(\mathbf{r}) \approx 0$. Finally, the polarization vectors in Eq.~\ref{Ecavquantum} are the two unit vectors orthogonal to the cavity axis, and will be labeled as
    \begin{equation}\label{polvecsCAVITY}
        \mathbf{e}_\alpha = \left\{
        \begin{array}{l}
             \mathbf{e}_z  \\
             \mathbf{e}_\perp = \mathbf{e}_x\cos\Theta - \mathbf{e}_y\sin\Theta .
        \end{array}
        \right.
    \end{equation}
Note that, by taking $\Theta \to 0$, we recover the usual Hermite-Gauss standing wave along the $y$ axis.

We are now in a position to derive the second quantized form of the total Hamiltonian, assuming that the NP always remains close to the origin of coordinates, i.e., close to the tweezer intensity maximum. We will follow the treatment done in Ref. \cite{RomeroIsartPRA2011}, although in this paper our main focus will be a different optomechanical coupling, namely that arising from the coherent scattering from the tweezer into the cavity.
We start by introducing the expression for the total electric field operator, Eq.~\ref{Efield1st}, into the interaction Hamiltonian, Eq.~\ref{Hint1st}, to obtain
\begin{equation}\label{Hintexpression}
    \hat{H}_{\rm Int}\! = -\frac{\varepsilon_0\epsilon_c V}{2}\!\left[\boldsymbol{\mathcal{E}}_{\rm tw}(\hat{\mathbf{R}},t)+\hat{\mathbf{E}}_{\rm cav}(\hat{\mathbf{R}})+\hat{\mathbf{E}}_{\rm free}(\hat{\mathbf{R}})\!\right]^2\!\!\!.
\end{equation}
The expression above results in six different terms, which give rise to the different physical interactions in this system. Let us analyze them separately.

First, the tweezer-tweezer term $\propto \boldsymbol{\mathcal{E}}^2_{\rm tw}(\hat{\mathbf{R}},t)$ gives rise to the trapping of the NP. The interaction energy arising from this term is 
\begin{equation}\label{Htt}
\begin{split}
        \hat{H}_{\rm t-t} &= - \frac{\varepsilon_0\epsilon_c V}{2}\boldsymbol{\mathcal{E}}^2_{\rm tw}(\hat{\mathbf{R}},t)  .
\end{split}
\end{equation}
Since the NP will be confined close to the origin of coordinates, we approximate the electric field of the tweezer by its expansion close to $\hat{\mathbf{R}} = 0$. Using Eqs. \ref{Etweezertime} and \ref{Etweezer}, this expansion is carried out to yield
\begin{equation}\label{etw2nonRWA}
\begin{split}
    \boldsymbol{\mathcal{E}}^2_{\rm tw}(\hat{\mathbf{R}},t)& \approx E_0^2\cos^2(\omega_0 t)
    \times
    \\
    &\times\left[1-\left(\frac{2\hat{X}^2}{A_x^2W_t^2}+\frac{2\hat{Y}^2}{A_y^2W_t^2}+\frac{\hat{Z}^2}{z_R^2}\right)
    \right]
    \\
    &
    +
    E_0^2(k_0\hat{Z})\left[k_0\hat{Z}\cos(2\omega_0 t)-\sin(2\omega_0t)\right].
\end{split}
\end{equation}
As the tweezer frequency $\omega_0$ is very large compared to the typical coupling rates between the different degrees of freedom, we can perform a Rotating Wave Approximation by neglecting all the rapidly oscillating terms $\exp(\pm 2i\omega_0 t)$. Within this approximation, we introduce the above electric field
 into Eq.~\ref{Htt} and, after neglecting constant energy contributions, we obtain the usual harmonic potential,
\begin{equation}
    \hat{H}_{\rm t-t} = \sum_{j=x,y,z}\frac{1}{2}m\Omega_j^2\hat{R}_j^2.
\end{equation}
This potential represents the trapping of the NP by the tweezer, with trapping frequencies
\begin{equation}\label{COMfreqsdefinition}
    \left[
    \begin{array}{c}
         \Omega_x  \\
         \Omega_y \\
         \Omega_z
    \end{array}
    \right]
    =
    \sqrt{\frac{\varepsilon_0\epsilon_cE_0^2}{2 \rho W_t^2}}\left[
    \begin{array}{c}
          \sqrt{2}/A_x \\
         \sqrt{2}/A_y \\
         \lambda_0/(\pi W_t)
    \end{array}
    \right],
\end{equation}
where $\rho = m/V$ is the mass density of the NP.
Note that the presence of this trapping potential allows for the definition of the COM phonons. Specifically, we can combine the above interaction term with the kinetic energy of the COM, namely the first term in Eq.~\ref{firstH}, to obtain
\begin{equation}\label{HCOM}
    \frac{\hat{\mathbf{P}}^2}{2m} + \hat{H}_{\rm t-t} = \hbar \sum_{j=x,y,z}\Omega_j\hat{b}^\dagger_j\hat{b}_j \equiv \hat{H}_{\rm COM}.
\end{equation}
Here, the bosonic operators $\hat{b}_j$ are defined in terms of the position and momentum operators through
\begin{equation}
    \hat{R}_j = r_{j0}\left(\hat{b}^\dagger_j+\hat{b}_j\right)  \hspace{0.3cm} ; \hspace{0.3cm} \hat{P}_j = \im m\Omega_jr_{j0}\left(\hat{b}^\dagger_j-\hat{b}_j\right).
\end{equation}
where we have defined the zero-point motion $r_{j0} = (\hbar/2m\Omega_j)^{1/2}$. 

The second contribution we consider is given by the square of the cavity field,
\begin{equation}
\begin{split}
    \hat{H}_{\rm c-c} &= - \frac{\varepsilon_0\epsilon_c V}{2}\hat{\mathbf{E}}^2_{\rm cav}(\hat{\mathbf{R}}). 
\end{split}    
\end{equation}
Introducing Eq.~\ref{Ecavquantum} together with the field $\mathbf{E}_{c\alpha}(\mathbf{r}) = \mathbf{e}_\alpha E_c(\mathbf{r})$ from Eq.~\ref{Ecav}, and keeping only the non-rotating terms \footnote{This Rotating Wave Approximation (RWA) is analogous to the one taken for the tweezer field, and equally valid since $\omega_c \approx \omega_0$. Note that we take it at this point for simplicity, but one could keep the counter-rotating terms until Eq.~\ref{Hlong1}, where they would vanish under the general RWA undertaken immediately afterwards.}, we find
\begin{equation}
\begin{split}
    \hat{H}_{\rm c-c} 
    = - \hbar\frac{\omega_c\epsilon_c V}{2 V_c} \vert E_{c}(\hat{\mathbf{R}})\vert^2  \left( \sum_\alpha\hat{c}^\dagger_\alpha\hat{c}_\alpha +\frac{1}{2}\right).
\end{split}    
\end{equation}
We now expand the square modulus in powers of the COM position, $\hat{\mathbf{R}}$, up to first order, obtaining
\begin{equation}
     \begin{split} 
     \hat{H}_{\rm c-c} &= - \hbar\frac{\omega_c\epsilon_c V}{2 V_c} \cos^2(\phi) \sum_\alpha\hat{c}^\dagger_\alpha\hat{c}_\alpha
     \\
     &
     -\hbar\frac{\omega_c\epsilon_c V}{2 V_c} k_c\sin(2\phi)
     \hat{Y}'\left( \sum_\alpha\hat{c}^\dagger_\alpha\hat{c}_\alpha +\frac{1}{2}\right).
     \end{split}
\end{equation}
where we have defined for convenience the rotated $y$~operator
\begin{equation}
    \hat{Y}' \equiv  \sin\Theta\hat{X}+ \cos\Theta\hat{Y}.
\end{equation}
The interaction Hamiltonian $\hat{H}_{\rm c-c}$ contains three physically different effects: first, a shift of the cavity frequency that depends on the NP position inside the cavity,
\begin{equation}
    \omega_c \to \omega_c - \Delta_c = \omega_c\left(1-\frac{\epsilon_c V}{2 V_c}\cos^2\phi\right).
\end{equation}
Second, an optomechanical coupling between COM and cavity modes,
\begin{equation}\label{OMcouplingcavity}
    -\hbar \sum_\alpha\hat{c}^\dagger_\alpha\hat{c}_\alpha \left(g_{\rm cx} \hat{b}^\dagger_x + g_{\rm cy} \hat{b}^\dagger_y + {\rm H.c.}\right),
\end{equation}
with
\begin{equation}\label{gcx}
    g_{\rm cx} = \frac{\omega_c\epsilon_c V}{2 V_c} k_c x_0\sin(2\phi)\sin\Theta,
\end{equation}
\begin{equation}\label{gcy}
    g_{\rm cy} = \frac{\omega_c\epsilon_c V}{2 V_c} k_c y_0\sin(2\phi)\cos\Theta.
\end{equation}
Lastly, a force on the COM along the $X-Y$ plane,
\begin{equation}
    -(\hbar/2)  \left(g_{\rm cx} \hat{b}^\dagger_x + g_{\rm cy} \hat{b}^\dagger_y + {\rm H.c.}\right).
\end{equation}
For common experimental values \cite{WindeyArXiv2018}, we find that $g_{\rm cj} \approx 2\pi \times 1$Hz. This allows us to neglect this third term, as it corresponds to a negligible shift in the equilibrium position of the COM motion ($\sim g_{\rm cj}/\Omega_j \sim 10^{-5}$ times the zero point motion). Note that the optomechanical coupling in Eq.~\ref{OMcouplingcavity} could still be relevant if the cavity occupation is large, so this term must be retained. We thus write the contribution of the term  $\hat{\mathbf{E}}^2_{\rm cav}(\hat{\mathbf{R}})$ as
\begin{equation}\label{Hcc}
\begin{split}
    \hat{H}_{\rm c-c} &\approx -\hbar \sum_\alpha\hat{c}^\dagger_\alpha \hat{c}_\alpha
    \Big[\Delta_c + \sum_{j=x,y}g_{\rm cj}(\hat{b}^\dagger_j+ \hat{b}_j)\Big].
\end{split}
\end{equation}

The third contribution to Eq.~\ref{Hintexpression} stems from the square of the free EM field, and is given by
\begin{equation}
\begin{split}
    \hat{H}_{\rm f-f} &= -\frac{\varepsilon_0\epsilon_c V}{2} \hat{\mathbf{E}}^2_{\rm free}(\hat{\mathbf{R}}) =
    \\
    &
    = -\hbar\frac{\epsilon_c V}{4\mathcal{V}}\sum_{\mathbf{k}\varepsilon}\sum_{\mathbf{k}'\varepsilon'}\sqrt{\omega_\mathbf{k}\omega_{\mathbf{k}'}}
        \boldsymbol{\varepsilon}_\mathbf{k} \boldsymbol{\varepsilon}_{\mathbf{k}'} 
        \\
        &
        \times \left(e^{\im\mathbf{k}\hat{\mathbf{R}}}\hat{a}_{\varepsilon\mathbf{k}} + \text{H.c.}\right)\left(e^{\im\mathbf{k}'\hat{\mathbf{R}}}\hat{a}_{\varepsilon'\mathbf{k}'} + \text{H.c.}\right),
\end{split}
\end{equation}
where we have used Eq.~\ref{Efree}. This term has been proven to become negligible for particles smaller than the relevant wavelengths \cite{PflanzerPRA2012}, and thus will be ignored hereafter.

Finally, we address the contributions arising from the three cross terms, namely 
\begin{equation}\label{Htcdefinition}
    \hat{H}_{\rm t-c} = -\varepsilon_0\epsilon_c V\boldsymbol{\mathcal{E}}_{\rm tw}(\hat{\mathbf{R}},t)\cdot\hat{\mathbf{E}}_{\rm cav}(\hat{\mathbf{R}}),
\end{equation}
\begin{equation}
    \hat{H}_{\rm t-f} = -\varepsilon_0\epsilon_c V\boldsymbol{\mathcal{E}}_{\rm tw}(\hat{\mathbf{R}},t)\cdot\hat{\mathbf{E}}_{\rm free}(\hat{\mathbf{R}}),
\end{equation}
and
\begin{equation}
    \hat{H}_{\rm c-f} = -\varepsilon_0\epsilon_c V\hat{\mathbf{E}}_{\rm free}(\hat{\mathbf{R}})\cdot\hat{\mathbf{E}}_{\rm cav}(\hat{\mathbf{R}}),
\end{equation}
which result in interactions between the three system components. They are all constructed in the same way, namely by expanding the corresponding electric fields close to the origin,
\begin{equation}
    \boldsymbol{\mathcal{E}}_{\rm tw}(\hat{\mathbf{R}},t)\approx \mathbf{e}_x E_0(\cos\omega_0t -k_0\hat{Z}\sin\omega_0 t),
\end{equation}
\begin{equation}\label{Ecavapprox}
    \hat{\mathbf{E}}_{c\alpha}(\hat{\mathbf{R}})\! \approx \!\! \sum_{\alpha}\!\sqrt{\!\frac{\hbar\omega_c}{2\varepsilon_0 V_c}}\mathbf{e}_\alpha \!\!\left[\cos\phi + k_c\hat{Y}'\sin\phi \right]\!\hat{c}_\alpha \! +  \text{H.c.},
\end{equation}
\begin{equation}
\begin{split}
    \hat{\mathbf{E}}_{\rm free}(\hat{\mathbf{R}}) &\approx \sum_{\mathbf{k}\varepsilon}\sqrt{\frac{\hbar\omega_\mathbf{k}}{2\varepsilon_0 \mathcal{V}}}\boldsymbol{\varepsilon}_\mathbf{k}\!\left[\!(1\!+\!\im\mathbf{k}\cdot\hat{\mathbf{R}})\hat{a}_{\varepsilon\mathbf{k}}\! + \!\text{H.c.}\right]\!,
\end{split}
\end{equation}
and keeping terms of up to first order in the COM position. 
Among the three cross terms, the most important is the tweezer-cavity contribution Eq.~\ref{Htcdefinition}, which will responsible for the cooling via coherent scattering. It is given by
\begin{equation}\label{Htc}
\begin{split}
    \hat{H}_{\rm t-c} &= -\hbar\sum_\alpha G_\alpha \left(\hat{c}^\dagger_\alpha + \hat{c}_\alpha\right)\left[k_c\hat{Y}'\sin(\phi)\cos(\omega_0t) \right.
    \\
    &
    \left.+\cos(\phi)\left(\cos(\omega_0t)-k_0\hat{Z}\sin(\omega_0t)\right)\right],
\end{split}
\end{equation}
where
\begin{equation}\label{Galpha}
    G_\alpha = \varepsilon_0\epsilon_c V E_0\left(\mathbf{e}_x\cdot\mathbf{e}_\alpha\right)\sqrt{\frac{\omega_c}{2\hbar\varepsilon_0 V_c}}.
\end{equation} 
Note that $\hat{H}_{\rm t-c}$ contains both a 
displacement of the cavity modes induced by the trapping field, and an interaction between cavity and COM degrees of freedom, the latter of which will ultimately be responsible for the cooling. Note that it follows from Eq.~\ref{polvecsCAVITY} that the coupling $G_z$ vanishes, i.e. the $z-$polarized cavity mode plays a negligible role in the dynamics since it will not be populated by the $x-$polarized tweezer. Because of this argument, we will neglect this mode and reduce the cavity to a single mode polarized along $\mathbf{e}_\perp$.

The second cross term, namely the one coming from the tweezer-free field product, reads
\begin{equation}\label{Htf}
\begin{split}
    \hat{H}_{\rm t-f} &= -\hbar\sum_{\mathbf{k}\varepsilon}G_0(\mathbf{k})
    \bigg\{\!\!-k_0\hat{Z}\sin\omega_0 t \left(\hat{a}^\dagger_{\varepsilon\mathbf{k}} + \hat{a}_{\varepsilon\mathbf{k}}\right)
    \\
    &
    \!\!\!+
    \!\cos\omega_0 t\!\left[\!\left(\hat{a}^\dagger_{\varepsilon\mathbf{k}} + \hat{a}_{\varepsilon\mathbf{k}}\right)\!+\!i\mathbf{k}\!\cdot\!\hat{\mathbf{R}}\!\left(-\hat{a}^\dagger_{\varepsilon\mathbf{k}} + \hat{a}_{\varepsilon\mathbf{k}}\right)\right] \!\!\bigg\},
\end{split}
\end{equation}
with the coupling rate
\begin{equation}
    G_0(\mathbf{k}) = \varepsilon_0\epsilon_c V E_0 \sqrt{\frac{\omega_\mathbf{k}}{2\hbar\varepsilon_0\mathcal{V}}}\left(\mathbf{e}_x\cdot\boldsymbol{\varepsilon}_\mathbf{k}\right).
\end{equation}
The term $\hat{H}_{\rm t-f}$ contains a displacement of the free space EM modes, and a tweezer-mediated interaction between free space modes and COM motion. 
As we will see below, the latter interaction will result in the recoil heating of the NP \cite{RomeroIsartPRA2011,JainPRL2016}. 

Finally, the product between cavity and free EM field adds a contribution
\begin{align}\label{Hcf}
\begin{split}
    \hat{H}_{\rm c-f} &= -\hbar\sum_\mathbf{ k\varepsilon,\alpha}\!G_\alpha(\mathbf{k})\!\left(\hat{c}^\dagger_\alpha + \hat{c}_\alpha\right)
    \!\left[\im\mathbf{k}\hat{\mathbf{R}}\!\left(\!-\hat{a}^\dagger_{\varepsilon\mathbf{k}}\! + \!\hat{a}_{\varepsilon\mathbf{k}}\right)\right.
    \\
    &
    +
    \left.\left(\hat{a}^\dagger_{\varepsilon\mathbf{k}} + \hat{a}_{\varepsilon\mathbf{k}}\right)\left(
    \cos\phi+ k_c\hat{Y}'\sin\phi
    \right)\right],
\end{split}
\end{align} 
with a coupling factor
\begin{equation}
    G_\alpha(\mathbf{k}) = \frac{\epsilon_c V}{2}\sqrt{\frac{\omega_c\omega_\mathbf{k}}{V_c\mathcal{V}}}\left(\mathbf{e}_\alpha\cdot\boldsymbol{\varepsilon}_\mathbf{k}\right).
\end{equation}
This interaction term is more involved than the previous two. On the one hand, it contains a quadratic interaction between cavity modes and free field, responsible for extra losses of the cavity into free space. Since these losses are mediated by the NP, they depend on its position through $\cos\phi$. Note that this term is small for sub-wavelength particles, as otherwise the cavity linewidth would be modified by the presence of the NP, an effect not observed in experiments \cite{WindeyArXiv2018}.
On the other hand, $\hat{H}_{\rm c-f}$ contains a non-quadratic, three body interaction between free space photons, cavity photons, and COM phonons. This contribution, however, can be safely neglected since it is of order $\sim kr_{j0}, k_cr_{j0} \sim 10^{-6},10^{-5}$ times smaller than the quadratic contribution at optical frequencies. This argument might fail if the cavity is largely populated by the tweezer, where this three-body interaction results in an effective free field-COM coupling which contributes to the recoil heating. However, this extra, cavity-induced recoil will be negligible compared to the recoil induced by the much highly occupied tweezer (Eq.~\ref{Htf}), which is known to dominate \cite{RomeroIsartPRA2011,JainPRL2016}. Therefore, we can safely approximate
\begin{equation}\label{Hcf2}
\begin{split}
    \hat{H}_{\rm c-f} &\!\approx\! -\hbar\sum_{\mathbf{k}\varepsilon,\alpha}\!G_\alpha(\mathbf{k})\!\!
    \left(\hat{c}^\dagger_\alpha + \hat{c}_\alpha\right)
    \!\!(\hat{a}^\dagger_{\varepsilon\mathbf{k}} + \hat{a}_{\varepsilon\mathbf{k}})\!
    \cos\phi.
\end{split}
\end{equation} 

The full Hamiltonian of the system is obtained by adding up all the above contributions, namely Eqs. \ref{Hfield2Q}, \ref{HCOM}, \ref{Hcc}, \ref{Htc}, \ref{Htf}, and \ref{Hcf}. As discussed above, from now on we will ignore the $z-$polarized cavity mode and refer to the remaining creation operators simply as $\hat{c}_\perp \equiv \hat{c}$.
In terms of the COM creation and annihilation operators, the full Hamiltonian reads
\begin{widetext}
\begin{equation}\label{Hlong1}
\begin{split}
    \hat{H}/\hbar &\approx \tilde{\omega}_c \hat{c}^\dagger \hat{c}+\sum_{\mathbf{k}\varepsilon} \omega_\mathbf{k}\hat{a}_{\varepsilon\mathbf{k}}^\dagger\hat{a}_{\varepsilon\mathbf{k}}+\sum_{j}\Omega_j\hat{b}^\dagger_j\hat{b}_j +\int \text{d}\omega \omega\hat{a}_0^\dagger(\omega)\hat{a}_0(\omega)
    -\cos(\omega_0 t)\bigg[G\cos(\phi)\hat{c}^\dagger + \sum_{\mathbf{k}\varepsilon}G_0(\mathbf{k})\hat{a}^\dagger_{\varepsilon\mathbf{k}}+\text{H.c.}\bigg]
     \\
    &
     -G \left(\hat{c}^\dagger + \hat{c}\right)\left[k_c\left(\sin(\Theta) x_0\hat{b}^\dagger_x+ \cos(\Theta) y_0\hat{b}^\dagger_y + \text{H.c.}\right)\sin(\phi)\cos(\omega_0t)- k_0z_0(\hat{b}^\dagger_z + \hat{b}_z)\cos(\phi)\sin(\omega_0t)\right]
    \\
    &
    +\sum_{\mathbf{k}\varepsilon}G_0(\mathbf{k})\left[k_0z_0(\hat{b}^\dagger_z + \hat{b}_z)\sin(\omega_0 t) (\hat{a}^\dagger_{\varepsilon\mathbf{k}} + \hat{a}_{\varepsilon\mathbf{k}}) -\im\sum_jk_j r_{0j}(\hat{b}^\dagger_j + \hat{b}_j)
     \cos(\omega_0 t) (-\hat{a}^\dagger_{\varepsilon\mathbf{k}} + \hat{a}_{\varepsilon\mathbf{k}})\right] 
    \\
    &
    -\sum_{\mathbf{k}\varepsilon}G(\mathbf{k})\left(\hat{c}^\dagger + \hat{c}\right)
    \left(\hat{a}^\dagger_{\varepsilon\mathbf{k}} + \hat{a}_{\varepsilon\mathbf{k}}\right)\cos(\phi) -\hat{c}^\dagger\hat{c}\left[g_{\rm cx}(\hat{b}^\dagger_x+ \hat{b}_x)+g_{\rm cy}(\hat{b}^\dagger_y+ \hat{b}_y)\right]+\im\int \text{d}\omega\gamma(\omega)\left[\hat{a}_0(\omega)\hat{c}^\dagger - \text{H.c.}\right],
\end{split}
\end{equation}
\end{widetext}
where $G \equiv G_\perp$, $G(\mathbf{k}) \equiv G_\perp (\mathbf{k})$, and $\tilde{\omega}_c = \omega_c - \Delta_c$.

The above Hamiltonian, though simpler than the original, remains very challenging to solve, but it can be further simplified. First, we elliminate the time dependence by virtue of the same Rotating Wave Approximation undertaken after Eq.~\ref{etw2nonRWA}. In order to do this, we first
perform a unitary transformation into a frame rotating with the tweezer frequency, i.e.,
\begin{equation}
    \hat{\mathcal{U}} = \exp\left(\im\omega_0 t\hat{A}\right),
\end{equation}
with
\begin{equation}
    \hat{A} =  \hat{c}^\dagger\hat{c} + \sum_{\mathbf{k}\varepsilon}\hat{a}^\dagger_{\varepsilon\mathbf{k}}\hat{a}_{\varepsilon\mathbf{k}}+\int_0^\infty \text{d}\omega\hat{a}^\dagger_0(\omega)\hat{a}_0(\omega).
\end{equation}
After applying this transformation, the Hamiltonian will contain two different kinds of terms, namely the non-rotating terms which do not depend on time, and contributions rotating at frequencies $\pm 2\omega_0$. We then take the Rotating Wave Approximation by neglecting the latter. After such approximation, the Hamiltonian is reduced to
\begin{widetext}
\begin{equation}\label{Hrwa}
\begin{split}
    \hat{H}/\hbar &\approx  \hat{c}^\dagger \hat{c}\left[\tilde{\delta}-g_{\rm cx}(\hat{b}^\dagger_x+ \hat{b}_x)-g_{\rm cy}(\hat{b}^\dagger_y+ \hat{b}_y)\right]+\sum_{\mathbf{k}\varepsilon} \Delta_\mathbf{k}\hat{a}_{\varepsilon\mathbf{k}}^\dagger\hat{a}_{\varepsilon\mathbf{k}}+\sum_{j}\Omega_j\hat{b}^\dagger_j\hat{b}_j +\int \text{d}\omega \Delta_0(\omega)\hat{a}_0^\dagger(\omega)\hat{a}_0(\omega)
    \\
    &
    - \frac{G}{2} \left(\hat{c}^\dagger + \hat{c}\right)
    \cos\phi -\sum_{\mathbf{k}\varepsilon}\frac{G_0(\mathbf{k})}{2}\left(\hat{a}^\dagger_{\varepsilon\mathbf{k}} + \hat{a}_{\varepsilon\mathbf{k}}\right)-\sum_{\mathbf{k}\varepsilon}G(\mathbf{k})\left(\hat{c}\hat{a}^\dagger_{\varepsilon\mathbf{k}} + \hat{c}^\dagger\hat{a}_{\varepsilon\mathbf{k}}\right)\cos\phi
     \\
    &
     -\frac{G}{2} \left(\hat{c}^\dagger + \hat{c}\right)k_c\left(\sin(\Theta) x_0\hat{b}^\dagger_x+ \cos(\Theta) y_0\hat{b}^\dagger_y + {\rm H.c.}\right)\sin\phi+  \im \frac{G}{2} \left(\hat{c}^\dagger - \hat{c}\right)k_0z_0(\hat{b}^\dagger_z + \hat{b}_z)\cos\phi
    \\
    &
    -\sum_{\mathbf{k}\varepsilon}\frac{G_0(\mathbf{k})}{2}\im\left(\hat{a}_{\varepsilon\mathbf{k}} - \hat{a}^\dagger_{\varepsilon\mathbf{k}}\right)\left[-k_0z_0(\hat{b}^\dagger_z + \hat{b}_z)  +\sum_jk_j r_{0j}(\hat{b}^\dagger_j + \hat{b}_j)
     \right]+\im\int d\omega\gamma(\omega)\left[\hat{a}_0(\omega)\hat{c}^\dagger \!-\! \text{H.c.}\right] ,
\end{split}
\end{equation}
\end{widetext}
where we have defined the detunings
\begin{equation}\label{deltatildedefinition}
    \tilde{\delta} = \tilde{\omega}_c - \omega_0,
\end{equation}
\begin{equation}
    \Delta_\mathbf{k} = \omega_\mathbf{k} - \omega_0,
\end{equation}
\begin{equation}
    \Delta_0(\omega) = \omega-\omega_0.
\end{equation}
Note that
the Hamiltonian contains a displacement of the cavity and the free EM modes (second line of Eq.~\ref{Hrwa}). As usual in quantum optics, it is convenient to remove such displacements by means of a second unitary transformation, which displaces all the system modes at once:
\begin{equation}\label{adisplacement}
    \hat{a}_{\varepsilon\mathbf{k}} \to \hat{a}_{\varepsilon\mathbf{k}} + \alpha_\mathbf{k},
\end{equation}
\begin{equation}\label{bdisplacement}
    \hat{b}_j \to \hat{b}_j + \beta_j,
\end{equation}
\begin{equation}\label{cdisplacement}
    \hat{c} \to \hat{c} + \alpha_c.
\end{equation}
\begin{equation}\label{a0displacement}
    \hat{a}_0(\omega) \to \hat{a}_0(\omega) + \alpha_0(\omega).
\end{equation}
After substitution of the above operators into Eq.~\ref{Hrwa}, we neglect the constant energy shifts and set the terms proportional to $\hat{a}_{\varepsilon\mathbf{k}}$, $\hat{b}_j$, $\hat{c}_\alpha$, and $\hat{a}_0(\omega)$ to zero, so that any displacement will vanish from the transformed Hamiltonian. In this way we obtain a system of equations relating the coefficients $\alpha_\mathbf{k}$, $\beta_j$, $\gamma_\alpha$, and $\alpha_0(\omega)$, whose solution is given in Appendix \ref{appendixDISPSS} together with the general form of the transformed Hamiltonian. 

After the above displacement, the operators appearing in the Hamiltonian represent the fluctuations of the corresponding degree of freedom above a classical value. Since these fluctuations are small, we can neglect the remaining non-quadratic terms in the transformed Hamiltonian. Then, in the transformed frame, the Hamiltonian can be written in a very compact form as
\begin{equation}\label{Hdisplin}
    \hat{H} = \hat{H}_S+ \hat{H}_{RA} +  \hat{H}_{RB} + \hat{V}_A + \hat{V}_{B1} + \hat{V}_{B2}.
\end{equation}
Here, we are describing the system in the usual notation in open quantum systems \cite{BreuerPetruccione}, where we divide our degrees of freedom into a system S, composed by the cavity and the COM modes, and two reservoirs, RA and RB, composed by the free EM modes and the output modes of the cavity. Such distinction facilitates the procedure outlined in the next section, namely the tracing out of the reservoir modes to get a reduced equation of motion only for the cavity and COM subsystem. In this system plus reservoir picture,  the Hamiltonian of the system S is defined as
\begin{equation}
    \hat{H}_S/\hbar = (\tilde{\delta}-2g_{\rm cx}\beta_x-2g_{\rm cy}\beta_y) \hat{c}^\dagger \hat{c}+\sum_{j}\Omega_j\hat{b}^\dagger_j\hat{b}_j + \hat{V}_0,
\end{equation}
where 
$\hat{V}_0= \sum_jg_j\hat{c}^\dagger(\hat{b}^\dagger_j+\hat{b}_j) + \text{H.c.}$ is the interaction between the system degrees of freedom, with coupling rates
\begin{equation}\label{OMcouplings}
    \left[
    \begin{array}{c}
         g_x  \\
         g_y  \\
         g_z
    \end{array}
    \right]=
     -\left[
    \begin{array}{c}
         (G/2)k_cx_0\sin\phi\sin\Theta +\alpha_cg_{\rm cx}  \\
         (G/2)k_cy_0\sin\phi\cos\Theta +\alpha_cg_{\rm cy}  \\
         -\im(G/2)k_0z_0\cos\phi
    \end{array}
    \right].
\end{equation}
On the other hand, the two reservoirs are governed by the Hamiltonians
\begin{equation}
    \hat{H}_{RA}/\hbar = \int \text{d}\omega \Delta_0(\omega)\hat{a}_0^\dagger(\omega)\hat{a}_0(\omega),
\end{equation}
\begin{equation}
    \hat{H}_{RB}/\hbar = \sum_{\mathbf{k}\varepsilon} \Delta_\mathbf{k}\hat{a}_{\varepsilon\mathbf{k}}^\dagger\hat{a}_{\varepsilon\mathbf{k}}.
\end{equation}
Finally, the interaction between system and reservoirs is given by three independent terms, namely
\begin{equation}
    \hat{V}_A/\hbar = \im\int \text{d}\omega\gamma(\omega)\left[\hat{a}_0(\omega)\hat{c}^\dagger \!-\! \text{H.c.}\right],
\end{equation}
\begin{equation}
    \hat{V}_{B1}/\hbar = \sum_{\mathbf{k}\varepsilon} \left(g_{\varepsilon \mathbf{k}}\hat{c}^\dagger\hat{a}_{\varepsilon\mathbf{k}} + \text{H.c.}\right),
\end{equation}
and
\begin{equation}
    \hat{V}_{B2}/\hbar =\sum_{ j,\mathbf{k}\varepsilon} \left[g_{j\varepsilon \mathbf{k}}\hat{a}^\dagger_{\varepsilon\mathbf{k}}\left(\hat{b}_j^\dagger+\hat{b}_j\right) + \text{H.c.}\right],
\end{equation}
where the coupling rates are given, respectively, by
\begin{equation}\label{gvaerpsilonk}
    g_{\varepsilon\mathbf{k}} = -G(\mathbf{k})\cos\phi,
\end{equation}
\begin{equation}
    g_{j\varepsilon\mathbf{k}} = \im\frac{G_0(\mathbf{k})}{2}\left(k_jr_{j0}-\delta_{jz}k_0z_0\right).
\end{equation}
By using the above definitions, the quadratic Hamiltonian describing both the system and the reservoirs, Eq.~\ref{Hdisplin}, takes a suitable form for tracing the latter out.

\section{Reduced dynamics of the cavity and the center-of-mass modes}\label{secReducedDynamics}

The Hamiltonian in Eq.~\ref{Hdisplin} describes an infinite system of coupled harmonic oscillators. Since we are only interested in the reduced dynamics of the system formed by COM and cavity, in this section we trace out the two sets of continuum EM modes to obtain an effective equation of evolution for such subsystem. 
The procedure for tracing out the reservoir degrees of freedom starts by transforming into the Interaction Picture with respect to the free evolution of all degrees of freedom, i.e. with respect to $\hat{H}_{RA}+\hat{H}_{RB}+\hat{H}_S-\hat{V}_0$. In this picture, the evolution of the total density matrix $\rho$ is given by the von Neumann Equation \cite{BreuerPetruccione},
\begin{equation}\label{vonNeumanneq1}
\begin{split}
    \dot{\rho} = -\frac{\im}{\hbar}\left[\hat{V}_0(t)+\hat{V}(t),\hat{\rho}(t)\right],
\end{split}
\end{equation}
where $\hat{V}_0(t)$ and $\hat{V}(t)=\hat{V}_A(t)+\hat{V}_{B1}(t)+\hat{V}_{B2}(t)$ represent the two interaction potentials in the Interaction Picture. In order to solve the above equation, we undertake the weak coupling or Born approximation, which consists on approximating the full density matrix as a product state, i.e., $\hat{\rho}(t) = \hat{\mu}(t)\otimes \hat{\rho}_R$. Here, $\hat{\mu}(t) = \text{Tr}_R\hat{\rho}(t)$ is the reduced density matrix of the system, $\text{Tr}_R$ denoting the partial trace over the reservoir modes. On the other hand, $\hat{\rho}_R$ is the reduced density matrix of the reservoir, which is assumed constant on the basis of the environment being composed by an infinitely large amount of modes, whose state is therefore only negligibly modified by the presence of the system. We will assume that the density matrix describing the free EM reservoirs is a thermal state at room temperature,
\begin{equation}\label{thermalstate}
    \hat{\rho}_R \propto e^{-\left(\hat{H}_{RA}+\hat{H}_{RB}\right)/k_B T}.
\end{equation}
Note that, in principle, the assumption of a thermal state for the reservoirs is only legitimate in the original frame, and such thermal state should be transformed accordingly into the displaced frame given by the transformation $\{\hat{a}_\mathbf{\varepsilon k},\hat{a}_0(\omega)\} \to \{\hat{a}_\mathbf{\varepsilon k}+\alpha_\mathbf{k},\hat{a}_0(\omega)+\alpha_0(\omega)\}$. However, this state remains a good approximation in the transformed frame since, as evidenced by our results in Appendix \ref{appendixDISPSS}, the contribution of the reservoir modes is only relevant at frequencies close to $\omega_0$. At these frequencies, the presence of a highly occupied classical tweezer field makes any other dynamics of the EM field negligible. Because of this reason, the thermal state in Eq.~\ref{thermalstate} remains a good approximation for reproducing experimental results, as we will see below.

Under the Born approximation and with the reservoirs in a thermal state, we can formally solve Eq.~\ref{vonNeumanneq1}, reinsert it into itself, and take the trace over the reservoir modes, obtaining
\begin{equation}\label{vonNeumanneq2}
\begin{split}
    \dot{\mu}(t) &= -\frac{\im}{\hbar}\left[\hat{V}_0(t),\hat{\mu} (0)-\frac{\im}{\hbar}\int_0^t \text{d}s\left[\hat{V}_0(s) ,\hat{\mu}(s)\right]\right]
    \\
    &
    -
    \frac{1}{\hbar^2}\text{Tr}_R\int_0^t \text{d}s\left[\hat{V}(t),\left[ \hat{V}(s),\hat{\mu}(s)\otimes\hat{\rho}_R\right]\right].
\end{split}
\end{equation}
The second argument inside the commutator in the first line above can be identified as $\hat{\mu}(t)$, as can be readily checked by taking the partial trace and formally solving for $\hat{\mu}(t)$ in Eq.~\ref{vonNeumanneq1}. Moreover, in the second line of Eq.~\ref{vonNeumanneq2}, it is customary to undertake the Markov approximation, which assumes the reservoir correlation functions decay much faster than the system-reservoir interaction rate. Formally, this is equivalent to approximating $\hat{\mu}(s) \approx \hat{\mu}(t)$ and extending the upper integration limit to infinity \cite{BreuerPetruccione}. The final Master equation reads
\begin{equation}\label{vonNeumanneq3}
\begin{split}
    \dot{\mu}(t) &= -\frac{\im}{\hbar}\left[\hat{V}_0(t),\hat{\mu}(t)\right]
    \\
    &
    -
    \frac{1}{\hbar^2}\text{Tr}_R\int_0^\infty \text{d}s\left[\hat{V}(t),\left[ \hat{V}(s),\hat{\mu}(t)\otimes\hat{\rho}_R\right]\right].
\end{split}
\end{equation}

The Master Equation \ref{vonNeumanneq3} now has the desired form, namely an equation of motion involving only the system degrees of freedom, i.e. the cavity and the COM modes. This equation contains the free evolution of the system plus some extra terms, describing the effect of the reservoirs (second line of Eq.~\ref{vonNeumanneq3}). The explicit calculation of such terms, 
 carried out in Appendix \ref{appendixTRACEOUT}, leads to the following Master Equation in the Schr\"odinger Picture,
\begin{equation}\label{MasterEqmu}
\begin{split}
    \dot{\mu}(t)  &= -\frac{\im}{\hbar}\left[\hat{H}_S',\hat{\mu}(t)\right]+\mathcal{D}[\hat{\mu}].
\end{split}
\end{equation}
Here, the Hamiltonian $\hat{H}_S'$ has the same form as the Hamiltonian $H_S$, but with all the involved frequencies re-normalized (shifted) by the effect of the reservoirs,
\begin{equation}\label{Hsprime}
\begin{split}
    \hat{H}_S'/\hbar &= \delta'\hat{c}^\dagger\hat{c} + \sum_j\Omega_j'\hat{b}_j^\dagger\hat{b}_j
    +\!\sum_j\left(g_j'\hat{c}^\dagger\hat{q}_j + \text{H.c.}\right),
\end{split}
\end{equation}
where we define $\hat{q}_j \equiv \hat{b}_j^\dagger + \hat{b}_j$, and the expressions for $\delta'$, $\Omega_j'$, and $g_j'$ are given in Appendix \ref{appendixTRACEOUT}. Note that the shifts in the frequencies and coupling rates represent the conservative part of the reservoir-induced system dynamics.
On the other hand, the term $\mathcal{D}[\hat{\mu}]$ contains the incoherent (or non-conservative) dissipators, which include cavity losses at a rate $\kappa'$, recoil heating of the COM modes at a rate $\Gamma_j^{(r)}$, and incoherent cavity-COM interaction at a rate $\Upsilon$:
\begin{equation}\label{dissipator0}
\begin{split}
    \mathcal{D}[\hat{\mu}] &= 2\kappa' \!\left[\hat{c}\hat{\mu}\hat{c}^\dagger-\frac{1}{2}\left\{ \hat{c}^\dagger\hat{c},\hat{\mu}\right\}\!\right]
     \!-\!\sum_j\Gamma_j^{(r)}\left[\hat{q}_j,\left[\hat{q}_j,\hat{\mu}\right]\right]
   \\
   &
   +\left[\Upsilon\left(2\hat{q}_z\hat{\mu}\hat{c}^\dagger-\left\{\hat{c}^\dagger\hat{q}_z,\hat{\mu}\right\}\right)+\text{H.c.}\right],
\end{split}
\end{equation}
where the curly brackets denote the anti-commutator. Here, the cavity linewidth $\kappa'$ contains in principle also a small correction due to the presence of the NP. The explicit expressions for $\kappa'$, $\Gamma_j^{(r)}$, and $\Upsilon$ are given in Appendix \ref{appendixTRACEOUT}.

\subsection{Other noise sources}

In a typical levitodynamics experiment, the noise sources associated to thermal free photons are not the only ones. Indeed, some noise sources, stemming from different degrees of freedom not accounted for in our original Hamiltonian (Eq.~\ref{firstH}), can play a relevant role in the system dynamics. Therefore, we must include such sources in our equation of motion for the cavity and COM degrees of freedom. In this work we focus on two particular decoherence channels, namely displacement noise in the trap and residual gas pressure.

We first focus our attention on the displacement noise associated with the ``shaking'' of the center of the trap. This mechanism has been discussed in detail in the literature \cite{PinoQST2018,JoosBookDecoherence,BreuerPetruccione,HenkelAPB1999}  and results in an extra dissipator in the Master Equation of the position localization or Brownian motion form,
\begin{equation}\label{DissipatorVIB}
\begin{split}
    \mathcal{D}_{\rm d}[\hat{\mu}] &= \!-\!\sum_j\!\Lambda_j\! \left[\!\hat{R}_j,\!\left[\hat{R}_j,\hat{\mu}\right]\!\right]\!=
    \\
    &
    =\!-\!\sum_j\!\Lambda_jr_{j0}^2 \left[\hat{q}_j,\left[\hat{q}_j,\hat{\mu}\right]\right]\!,
\end{split}
\end{equation}
where $\hat{R}_j = r_{j0} \hat{q}_j$ is the $j-$component of the COM position operator. The magnitude $\Lambda_j r_{j0}^2$ is the corresponding dissipation rate, and can be related to observables in the following way: let us describe the trap displacements along each direction $j$ by means of independent classical fluctuating variables $\xi_j(t)$, such that $\hat{R}_j \to \hat{R}_j + \xi_j(t)$. We assume such variables to have zero mean and nonzero fluctuations, i.e. $\langle \xi_j(t)\rangle_T = 0$ and $\langle\xi_j(t)\xi_j(t')\rangle_T \ne 0$ \cite{PinoQST2018}. In terms of these variables, we can define the following two-sided power spectral density (PSD)
\begin{equation}\label{noisePSD}
    S_{jj}^{(d)}(\omega) = \frac{1}{2\pi}\int_{-\infty}^\infty \text{d}\tau\langle \xi_j(t+\tau)\xi_j(t)\rangle_T e^{\im\omega \tau},
\end{equation}
which has units of m${}^2/$Hz. For the particular case $S_{jj}^{(d)}(\Omega_j) = S_{jj}^{(d)}(-\Omega_j)$ one can demonstrate, by averaging over the stochastic force generated by $\xi_j$, that the displacements result in the dissipator Eq.~\ref{DissipatorVIB}, with
\begin{equation}
    \Lambda_j =\frac{m^2\Omega_j^4}{\hbar^2}\pi S_{jj}^{(d)}(\Omega_j).
\end{equation}
The dissipation rates associated to the trap displacement, $\Gamma_j^{(d)}\equiv\Lambda_j r_{0j}^2$, can therefore be written as
\begin{equation}\label{GammaVdefinition}
    \Gamma_j^{(d)} = \pi\frac{\Omega_j}{4}\left(\frac{S_{jj}^{(d)}(\Omega_j)}{\Omega_j^{-1}r_{0j}^2}\right)\equiv \pi\frac{\Omega_j}{4} \sigma_j^2,
\end{equation}
i.e., the ratio $\Gamma_j^{(d)}/\Omega_j$ is proportional to the PSD in units of $r_{0j}^2/\Omega_j$, whose square root we define as $\sigma_j$. Note that the above rate scales linearly with the NP mass, therefore taking much higher values for a NP than for a trapped atom. Indeed, as we will see below, in recent levitodynamic cooling experiments \cite{WindeyArXiv2018}, this mechanism could represent a relevant source of heating.

Let us now focus on the effect of gas pressure which, by inducing an extra heating of the COM motion, limits the cooling power of the experiment.  
The pressure of the gas is modelled  through a combination of two dissipators \cite{JoosBookDecoherence,RodenburgOptica2016},
\begin{equation}
    \mathcal{D}_{\text{pressure}}[\hat{\mu}] = \mathcal{D}_R[\hat{\mu}] + \mathcal{D}_p[\hat{\mu}].
\end{equation}
Here, the first term also takes the form of a position localization dissipator,
\begin{equation}
\begin{split}
    \mathcal{D}_R[\hat{\mu}]&=  -\frac{m\gamma k_B T}{\hbar^2} \sum_j\left[\hat{R}_j,\left[\hat{R}_j,\hat{\mu}\right]\right]=
    \\
    &
    =
    -\frac{m\gamma k_B T}{\hbar^2}\sum_j r_{j0}^2 \left[\hat{q}_j,\left[\hat{q}_j,\hat{\mu}\right]\right],
\end{split}
\end{equation}
with $T$ the temperature of the gas,
whereas the second term describes viscous friction, and is given by
\begin{equation}
\begin{split}
    \mathcal{D}_p[\hat{\mu}] &= -\im\frac{\gamma}{2\hbar}\sum_j\left[\hat{R}_j,\left\{\hat{P}_j,\hat{\mu}\right\}\right]=
    \\
    & 
    = \frac{\gamma}{4}\sum_j\left[\hat{q}_j,\left\{\hat{p}_j,\hat{\mu}\right\}\right],
\end{split}
\end{equation}
where we define the conjugate variable $\hat{p}_j \equiv \hat{b}^\dagger_j - \hat{b}_j$. The rate $\gamma$ can be obtained from the kinetic theory of gases \cite{BeresnevJFM1990,LiNatPhys2011}, and reads
\begin{equation}\label{gammadefinition}
    \gamma = 0.619\frac{6\pi R^2}{m\bar{l}}\eta_g = 0.619\frac{6\pi R^2}{m}P\sqrt{\frac{2m_0}{\pi k_BT}},
\end{equation}
where $\eta_g=\bar{l}P(2m_0/\pi K_B T)^{1/2}$ is the viscosity of the gas, $\bar{l}$ the mean free path of the air molecules, $P$ the pressure and $m_0$ the molecular mass of the gas.

\begin{table}[t]
	\centering
	\begin{tabular}{ p{5cm} | p{3cm} }
		\hline
		\hline
		Parameter & Value \\
		\hline
		Tweezer power & $P_t = 0.5$ W \\
		Tweezer wavelength & $\lambda_0 = 1.55\,\mu$m \\
		Tweezer waist & $W_t \sim 1\,\mu$m \\
		\hline
		Cavity length & $L_c = 6.46\,{\rm mm}$ \\
		Cavity waist & $W_c = 48\,\mu$m \\
		Cavity linewidth & $2\kappa = 2\pi \times 1.06\,{\rm MHz}$ \\
		\hline
		Radius of NP & $R\sim (70 \pm 20)\, {\rm nm}$  \\
		$\text{SiO}_2$ permittivity at $1.55\,\mu$m  \hspace{0.1cm}\cite{MalitsonJOSA1965} &  $\varepsilon = 2.07$ \\
		Density of $\text{SiO}_2$ \hspace{0.1cm} \cite{HaynesBookChemPhys} & $\rho = 2200\, \text{kg/m}^3$ \\
		\hline
		\hline
	\end{tabular}
	\caption{Input parameters for our model. All the values except for the last two are taken from Ref. \cite{WindeyArXiv2018}. Although the tweezer waist is not directly measured, an estimation of its order of magnitude can be drawn (see main text). Note that, because of the chosen convention, the rate $\kappa$ in our theory corresponds to half the value of $\kappa$ reported in Ref. \cite{WindeyArXiv2018}.} \label{tab:1}
\end{table}

Both the first dissipator associated with the gas pressure, $\mathcal{D}_R[\mu]$, and the displacements of the trap center, $\mathcal{D}_\text{d}[\mu]$, have the same form as the recoil heating term in Eq.~\ref{dissipator0}, and can thus be grouped together. The final Master Equation then reads
\begin{equation}\label{MasterEqmu2}
\begin{split}
    \dot{\mu}  &= -\frac{\im}{\hbar}\left[\hat{H}_S',\hat{\mu}(t)\right]+\mathcal{D}'[\hat{\mu}],
\end{split}
\end{equation}
where $\hat{H}_S'$ is given by Eq.~\ref{Hsprime}, and the final dissipator reads
\begin{equation}\label{dissipatorfinal}
\begin{split}
    \mathcal{D}'[\hat{\mu}] &= 2\kappa' \!\left[\hat{c}\hat{\mu}\hat{c}^\dagger-\frac{1}{2}\left\{ \hat{c}^\dagger\hat{c},\hat{\mu}\right\}\!\right]
     \!-\!\sum_j\Gamma_j\left[\hat{q}_j,\left[\hat{q}_j,\hat{\mu}\right]\right]
   \\
   &
   +\left[\Upsilon\left(2\hat{q}_z\hat{\mu}\hat{c}^\dagger-\left\{\hat{c}^\dagger\hat{q}_z,\hat{\mu}\right\}\right)+\text{H.c.}\right]+\mathcal{D}_p[\hat{\mu}],
\end{split}
\end{equation}
with 
\begin{equation}\label{Gammaj}
    \Gamma_j = \Gamma_j^{(r)} + \Gamma_j^{(d)} + \Gamma_j^{(p)}.
\end{equation}
and
\begin{equation}\label{GammaPdefinition}
    \Gamma_j^{(p)} = \frac{m k_B T}{\hbar^2}r_{j0}^2 \gamma.
\end{equation}
Equation \ref{MasterEqmu2} is the final equation of evolution for the system formed by cavity and COM motion. As a reading guide, a compilation of the most relevant parameters governing the system evolution is shown in Table \ref{TableparamsAppendix} (Appendix \ref{appendixTABLE}).
Note that, although we are neglecting any parametric noise induced by fluctuations in the trapping frequencies, such noise could be incorporated in our model by means of a Brownian motion dissipator for the operators $\hat{R}_j^2$ \cite{SchneiderPRA1999}. This, however, lies beyond the scope of the present work since, as will be shown below, the three heating sources introduced here, namely gas pressure, recoil heating, and displacement noise, are both necessary and sufficient for our model to be compatible with experimental observations. 

\section{Results: 3D cavity cooling in recent experimental setup}\label{secResults}

In this section we study the behavior of the system for realistic experimental parameters. 
In order to illustrate our model, we take as a case study the recent experiment with a SiO${}_2$ NP by Windey et al. \cite{WindeyArXiv2018}. The values of the parameters measured in this experiment are shown in Table \ref{tab:1}, together with the permittivity and density of Silica extracted from the literature. 
Note that not all the parameters in our model are measured directly, leaving some of them free to fit the experimental results. For instance, the radius of the NP is only known within a relatively wide range, and in the following we will take $R = 50\,\rm nm$ for the sake of definiteness.

 \begin{figure}
	\centering
	\includegraphics[width=\linewidth]{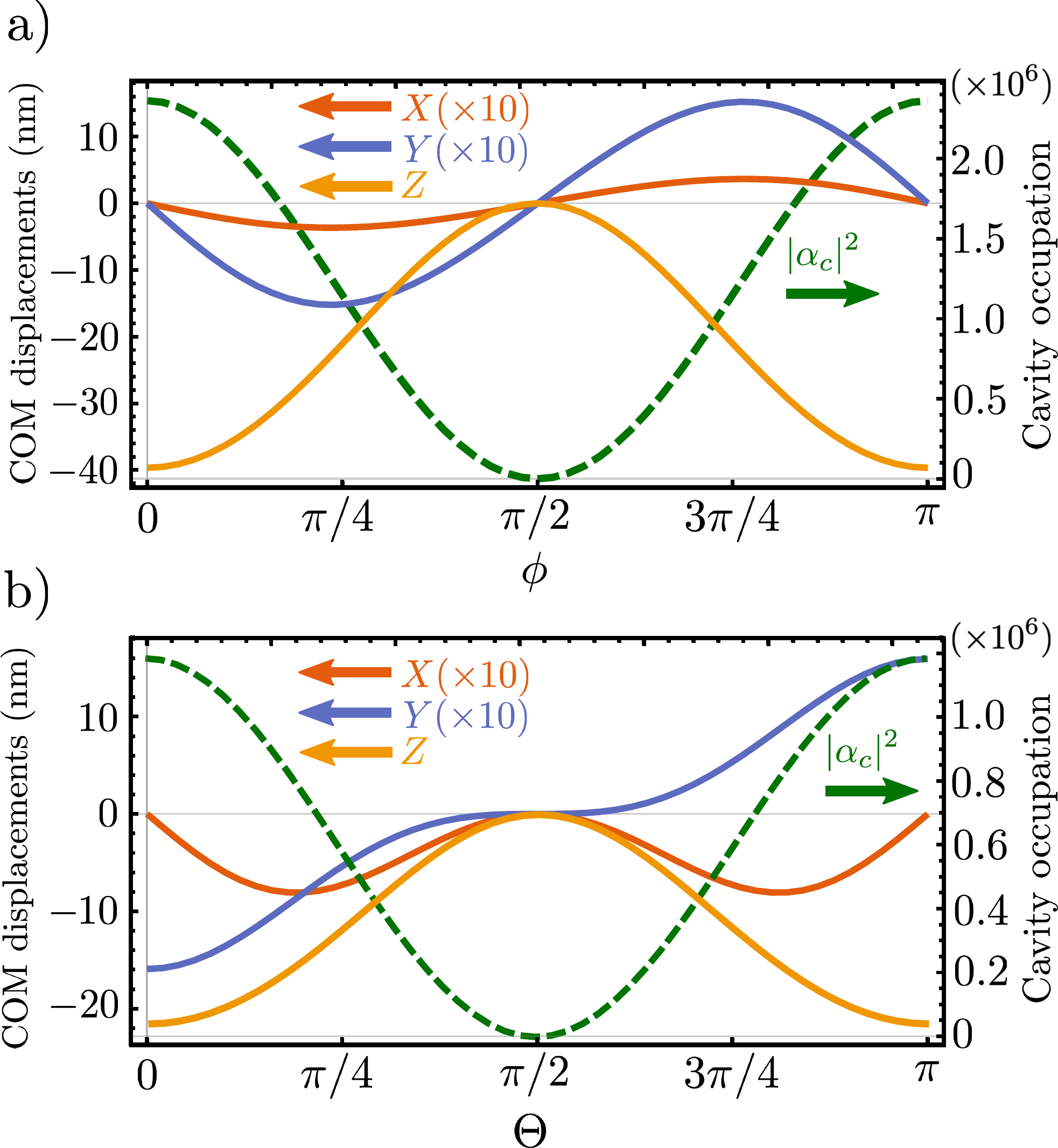}
	\caption{(Color online). Displacements of the COM equilibrium positions and of the cavity mode for the parameters of Table \ref{tab:1}. a) Displacements as a function of position inside the cavity, for $\Theta = 10^\circ$. b) Displacements as a function of tweezer polarization angle, for $\phi = \pi/4$. Note that the definition of the $X-Y$ axes is different for each value of $\Theta$ (see Fig.\ref{figaxis}). In both panels, the displacements along the X and Y directions are multiplied by a factor $10$ for better visualization.}\label{figdisplacements}
\end{figure}

Our first task is to evaluate all the parameters appearing in our effective equation of motion, Eq.~\ref{MasterEqmu2}, starting by the COM trapping frequencies $\Omega_j'$. To obtain them, we need to fix the values of the tweezer waist, $W_t$, and the asymmetry factors $A_x$ and $A_y$. Reasonable values for the former lie on the order of $W_t \sim \lambda_0/ (\pi \text{NA}) \sim 1\,\mu$m, where $\text{NA}\approx 0.8$ is the numerical aperture of the lens \cite{Novotnybook}. On the other hand, since the tweezer cross section is not expected to deviate significantly from a cylindrically symmetric spot, both $A_x$ and $A_y$ should be of the order of $1$. Within these bounds, we choose $W_t = 1.08 \,\mu$m, $A_x = 1.03$, and $A_y = 0.89$, which result in the following mechanical frequencies,
 \begin{equation}\label{mechanicalfrequencies}
     \left[
     \begin{array}{c}
          \Omega_x'  \\
          \Omega_y' \\
          \Omega_z'
     \end{array}
     \right] 
     =
     2\pi \times
     \left[
     \begin{array}{c}
          0.12  \\
          0.14 \\
          0.04
     \end{array}
     \right]
     \text{MHz}.
 \end{equation}
 These values agree with the measurements in Ref. \cite{WindeyArXiv2018}. 
 
 \begin{figure}
	\centering
	\includegraphics[width=\linewidth]{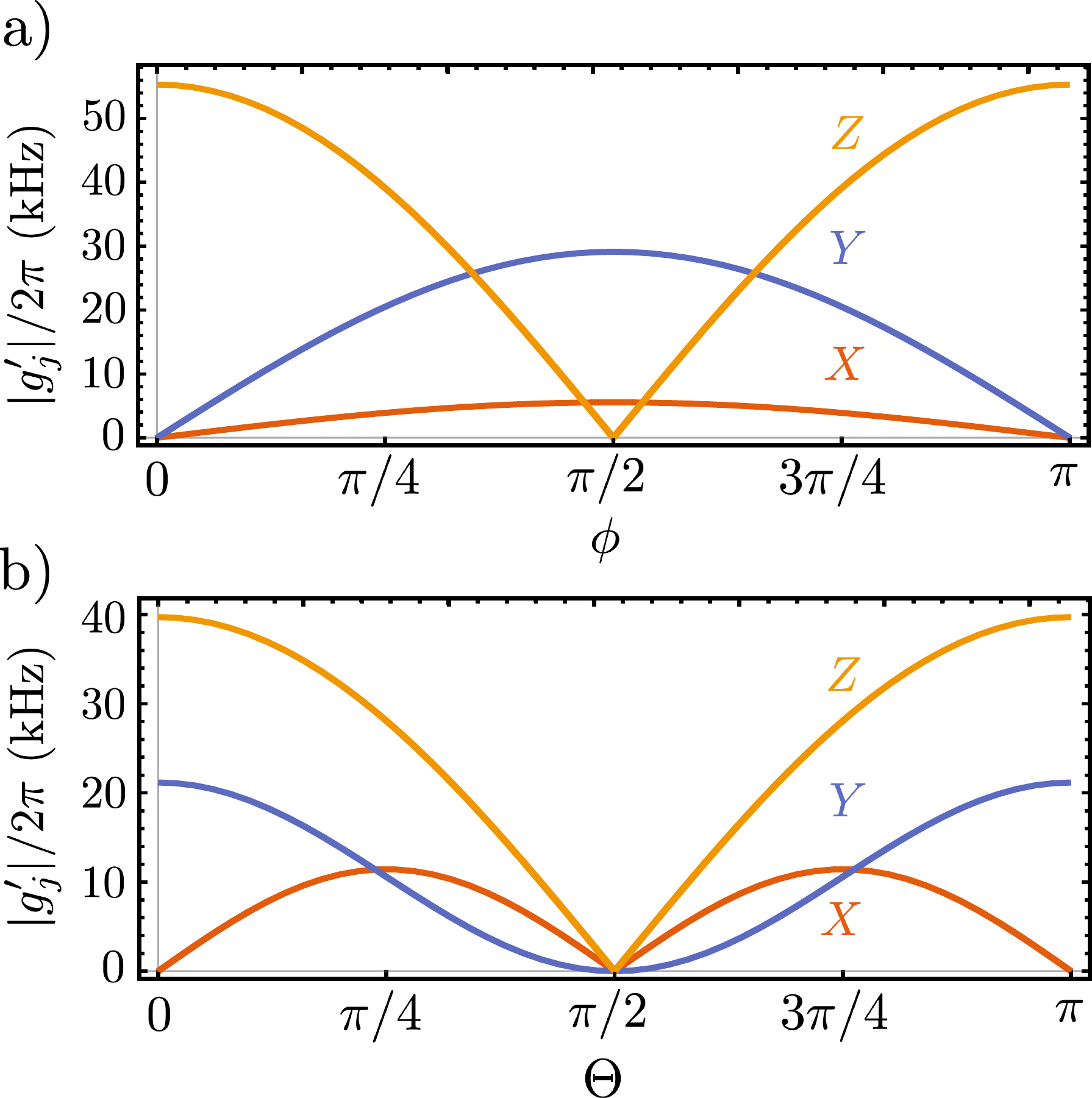}
	\caption{(Color online). Total optomechanical coupling rates in absolute value, for the parameters of Table \ref{tab:1} and $\delta' = 2\pi \times 400\,\rm kHz$. a) Couplings versus NP position along the cavity axis for $\Theta = 10^\circ$. b) Couplings versus tweezer polarization angle, for $\phi = \pi/4$.}\label{figcouplings}
\end{figure}
 
 We now focus on the cavity parameters in Eq.~\ref{MasterEqmu2}, namely the re-normalized detunings $\delta'$ and linewidth $\kappa'$. 
 Note that, due to the large cavity occupation induced by the high tweezer power, the cavity frequency can be significantly modified by the presence of the NP, thus allowing for the definition of two different detunings. Indeed, the detuning measured without the NP is defined as $\delta_{\rm bare} = \omega_c + \Delta_A - \omega_0$, whereas the detuning measured with the NP inside the cavity is given by $\delta' = \delta_{\rm bare} -\Delta_c + \Delta_{B1} -2g_{\rm cx}\beta_x-2g_{\rm cy}\beta_y$, where the expressions for all undefined shifts are given in Appendix \ref{appendixTRACEOUT}. 
 For the parameters in Table \ref{tab:1}, the difference between these two quantities is dominated by $\Delta_c$, and can reach a maximum value of $\vert \delta'-\delta_{\rm bare}\vert \approx 2\pi\times 3.4$ kHz when the NP is placed at an intensity maximum. 
 Since this value is $\sim10\%$ the mechanical frequency $\Omega_z$, a proper identification of which detuning is measured experimentally could be relevant for resolved sideband cooling \cite{AspelmeyerRMP2014}. Following the measurements in Ref. \cite{WindeyArXiv2018}, where the detuning is measured with the NP inside the cavity, in the following
  we will refer to $\delta'$ as the cavity detuning and set it to $\delta' = 2\pi \times 400$kHz unless stated otherwise \footnote{For the chosen parameters, a smaller detuning would result in the system becoming dynamically unstable, and the NP would abandon the trap. This phenomenon has also been experimentally observed in Ref. \cite{WindeyArXiv2018}.}. Finally, regarding the cavity linewidth, we find that $\kappa' = \kappa+\kappa_{B1}\approx \kappa$, since the correction to the bare cavity linewidth is negligible, $\kappa_{B1} = 2\pi\times (5\cos^2\phi)$Hz.
  
  \begin{table}[t]
	\centering
	\begin{tabular}{ p{2.4cm} | p{1.6cm} | p{1.6cm} | p{1.6cm} }
		\hline
		\hline
		Axis & X & Y & Z \\
		\hline
		$\Gamma_j^{(p)}/2\pi \text{ (GHz)}$ &  $28.6 P_{\rm mbar}$  &  $24.5 P_{\rm mbar}$  &  $85.7 P_{\rm mbar}$ \\
		$\Gamma_j^{(r)}/2\pi \text{ (kHz)}$ &  $0.09$  &  $0.15$  &  $1.89$ \\
		$\Gamma_j^{(d)}/2\pi \text{ (kHz)}$ &  $94\sigma_x^2$  &  $110\sigma_y^2$  &  $31\sigma_z^2$ \\
		\hline
		\hline
	\end{tabular}
	\caption{Heating rates contributing to the position localization rate $\Gamma_j$ in Eq.~\ref{Gammaj}, for the parameters in Table \ref{tab:1}. The second, third, and fourth rows show, respectively, the heating rate due to pressure ($P_{\rm mbar} =$ pressure in mbar), the photon recoil heating rate, and the trap displacement heating rate.} \label{tab:2}
\end{table}

The determination of the mechanical frequencies and the cavity parameters allows us to compute the displacements
of the cavity, $\alpha_c$, and of the COM motion, $\beta_j$, which arise after introducing the trapped NP into the optical cavity. The former can be expressed in terms of the tweezer-induced photonic occupation of the cavity mode, $\vert\alpha_c\vert^2$, whereas the latter quantifies the cavity-induced displacement of the COM equilibrium positions, $2\beta_j r_{0j}$. These displacements are shown in Fig.~\ref{figdisplacements} versus the NP position inside the cavity (panel a) and the tweezer polarization angle (panel b). 
All the displacements vanish when the NP sits at a cavity node ($\phi = \pi/2$), since the cavity field is zero at such position, and when the tweezer polarization is parallel to the cavity axis ($\Theta = \pi/2$), as no photons are scattered into the cavity direction by the dipolar NP. In general, the displacements of the COM equilibrium positions are orders of magnitude above the zero-point amplitudes $r_{0j} \sim 10^{-12}\,\rm m$. 
Moreover, the shift in the COM position along the tweezer axis ($Z$) has a larger magnitude and negative sign. Such traits can be ascribed, respectively, to the larger mechanical frequencies and to the reduction in the tweezer scattering force along $Z$, caused by the coherent scattering of tweezer photons into the cavity.
Finally, note that the cavity might be largely occupied, containing up to $\sim 10^6$ photons. This might result in a significant modification of the optomechanical coupling (see Eq.~\ref{OMcouplings}).

Using the displacements calculated above, we determine the remaining parameters in the equation of motion. First, we show in Fig.~\ref{figcouplings} the coherent optomechanical coupling rates between the cavity mode and the COM motion, namely $g_j'$ in the Hamiltonian Eq.~\ref{Hsprime}. As shown in panel \ref{figcouplings}a, for the NP sitting at the node of the cavity ($\phi=\pi/2$), the coupling vanishes for the $Z-$COM coordinate and reaches its maximum value for the transverse ($X-Y$) coordinates, while the opposite trend is observed at a cavity anti-node ($\phi=0$). Such behavior evidences the different origin of the couplings for the $Z$ and for the $X-Y$ directions, see e.g. Eq.~\ref{Htc}. Indeed, the interaction along the $Z-$ direction is proportional to the electric field intensity of the cavity, whereas the interaction along the $X-Y$ direction is proportional to its derivative. In Fig.~\ref{figcouplings}b, we show the dependence of the optomechanical couplings with the polarization angle of the tweezer. As expected, when the tweezer is polarized along the cavity axis ($\Theta=\pi/2$), no tweezer photons can be scattered into  the cavity, and the interaction rates along all three directions vanish. Moreover, as mentioned above, a tweezer polarization purely perpendicular to the cavity axis ($\Theta = 0,\pi$) results in the decoupling of the $X-$COM motion, and thus prevents the possibility of 3D cooling. 
Note that the coupling rates can reach values comparable or even higher than the mechanical frequencies, which may result in the system reaching dynamically unstable regimes \cite{KusturaArXiv2018}.

\begin{figure}
	\centering
	\includegraphics[width=\linewidth]{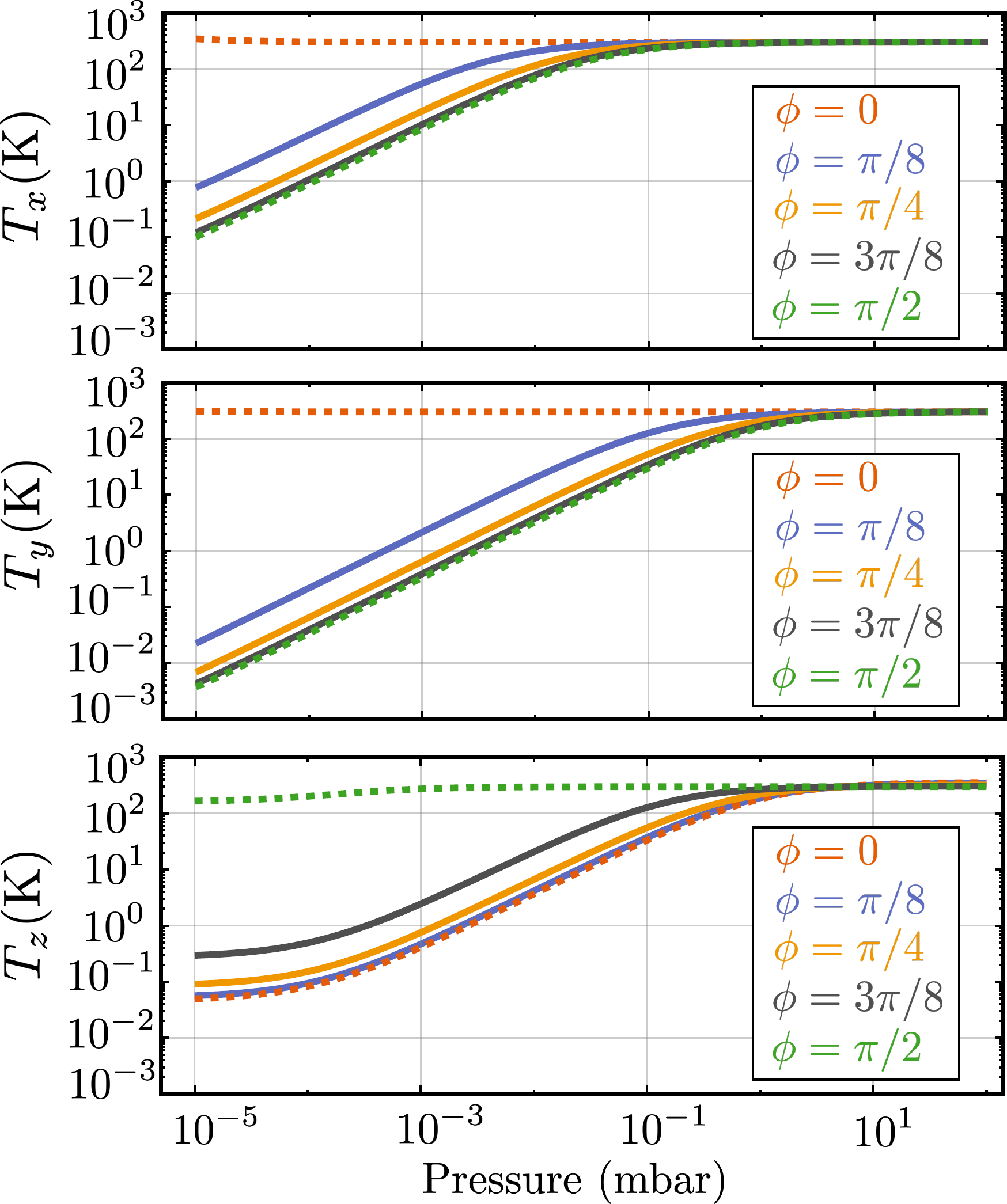}
	\caption{(Color online). Steady-state temperature of the COM motion along the three motional axes as a function of gas pressure, for different positions inside the cavity, $\phi$, detuning $\delta' = 2\pi \times 400\,\rm kHz$, and polarization angle $\Theta = 10^\circ$. The dashed lines represent the node ($\phi = \pi/2$) and anti-node ($\phi = 0$) of the cavity field.}\label{figTvsP}
\end{figure}

Finally, we calculate the remaining rates appearing in the dissipator Eq.~\ref{dissipatorfinal}, for the parameters in Table \ref{tab:1}. First, the incoherent coupling rate between the $Z-$COM mode and the cavity is given by $\Upsilon = 2\pi \times 82\cos\phi\cos\Theta \text{ Hz}$, and depends on the tweezer power as $P_t^{1/4}$. This rate can safely be neglected since it is much smaller than the corresponding coherent coupling rate, i.e., $\Upsilon \ll g_{z}'$. Regarding the dissipation associated to the viscous friction term, $\mathcal{D}_p$, we find that
\begin{equation}\label{gammaPV}
    \gamma = 2\pi\times (1.1 P_{\rm mbar})\,\text{kHz},
\end{equation}
where $P_{\rm mbar}$ is the gas pressure in mbar. From here, we can immediately determine the rates of position localization $\Gamma_j$ in Eq.~\ref{Gammaj}, which are independent on the  angles $\phi$ or $\Theta$. The three contributions to these rates are shown in Table \ref{tab:2}. Note that the displacement noise depends critically on the adimensional PSD, $\sigma_j$. In order to reproduce the experimental observations, these parameters are chosen as $\lbrace\sigma_x,\sigma_y,\sigma_z\rbrace =\lbrace0.67,0.26,18.6\rbrace$, 
which correspond to displacement PSD values of $\{\big[S_{xx}^{(d)}\big]^{1/2},\big[S_{yy}^{(d)}\big]^{1/2},\big[S_{zz}^{(d)}\big]^{1/2}\} \approx \{6,2,500\}\times 10^{-15} \,\text{m}\,\text{Hz}^{-1/2}$. These values are of the same order as those estimated in Ref. \cite{WindeyArXiv2018}. 

\begin{figure}
	\centering
	\includegraphics[width=\linewidth]{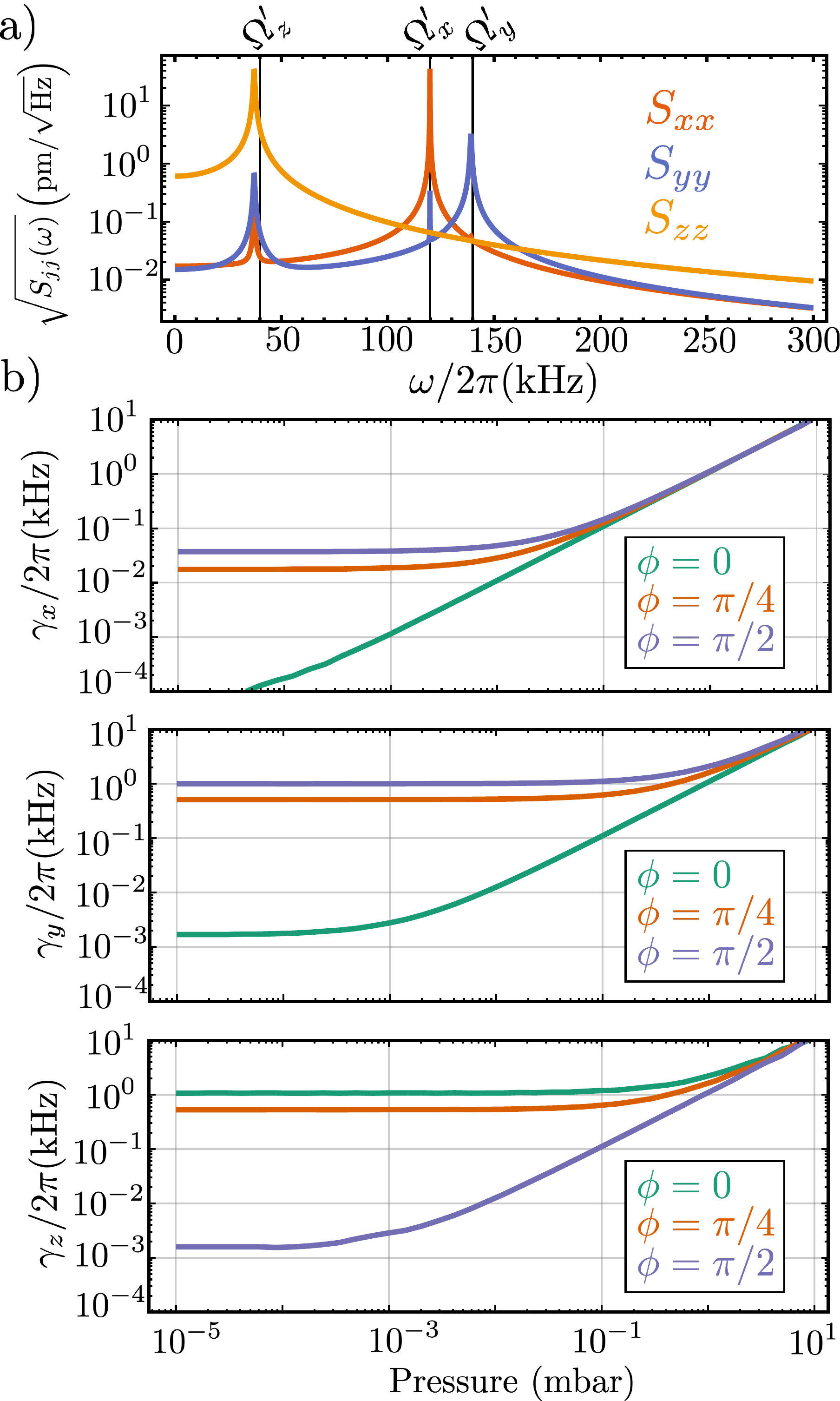}
	\caption{(Color online). a) Motional PSD along the three axes, for $P=3\cdot 10^{-3}$mbar and $\phi=\pi/4$. The vertical lines indicate the mechanical frequencies, Eq. \ref{mechanicalfrequencies}. b) FWHM of the main peaks in the PSD as a function of gas pressure. For all panels, $\Theta=10^\circ$.}\label{figPSDlinewidths}
\end{figure}

Once the parameters of the equation of motion Eq.~\ref{MasterEqmu2} have been calculated, we are in a position to study the system dynamics. 
The expected value of any system observable $\hat{O}$ can be obtained from the density matrix, $\hat{\mu}$, through the usual trace expression, $\langle \hat{O}\rangle(t) = \text{Tr}[\hat{O}\hat{\mu}(t)]$ \cite{BreuerPetruccione}. 
In the present case, there is no need to compute the whole density matrix, since our Master Equation is quadratic in the system creation and annihilation operators. Thus, any information about the state of the system is encoded in the first- and second-order momenta, namely the expected value of single system operators ($\langle\hat{c}\rangle,\langle\hat{c}^\dagger\rangle,\langle\hat{b}_x\rangle, \langle\hat{b}_x^\dagger\rangle, ...$) and the expected value of all their quadratic combinations ($\langle\hat{c}^2\rangle,\langle\hat{c}^\dagger\hat{c}\rangle,\langle\hat{c}^\dagger{}^2\rangle,\langle\hat{c}\hat{b}_x\rangle,\langle\hat{c}\hat{b}_x^\dagger\rangle...$).
As detailed in Appendix \ref{appendixEOMS}, from Eq.~\ref{MasterEqmu2} we derive the equations of motion for all the above expected values, which form a closed system of $44$ differential equations. After numerically solving such system of equations we can compute any observable as a function of their solutions, for instance the position operator $\langle \hat{R}_j\rangle = r_{0j}\big(\langle\hat{b}_j\rangle + \langle\hat{b}^\dagger_j\rangle\big)$. Specifically, the temperature of the motion along a given axis is defined as $T_j \equiv \hbar\Omega_j' \langle \hat{b}^\dagger_j\hat{b}_j\rangle/k_B$.

Our results for the steady-state temperature along each of the trapping axes as a function of the pressure of the surrounding gas are shown in Fig.~\ref{figTvsP}. As expected, the dependence of the temperatures on the position inside the cavity, $\phi$, mimics that of the optomechanical couplings in Fig.~\ref{figcouplings}. Specifically, the cooling along the $Z-$axis is maximally efficient at a cavity anti-node and vanishes at the cavity node, whereas the cooling along the $X-$ and $Y-$ directions obeys an opposite trend. For the parameters chosen above, the temperatures reached at $P = 10^{-5}$mbar and optimum cooling conditions are $T_x \approx 100\,{\rm mK}$, $T_y \approx 3\,{\rm mK}$, and $T_z \approx 60\,{\rm mK}$, which are consistent with the experimental observations \cite{WindeyArXiv2018}. According to our model, the saturation of $T_z$ at low pressures is limited by the trap displacement heating $\Gamma_z^{(d)}$, whereas at large pressures the gas heating rates $\Gamma_j^{(p)}$ dominate for all axes and prevent the motional cooling.
Interestingly, we have confirmed that for the chosen parameters the system is very close to a dynamical instability \cite{KusturaArXiv2018} due to the high tweezer powers, which can result in strong sensitivity to the system parameters. For instance, a slight increase in the size of the NP results in large heating rates at low pressures due to the critical ratio between the coupling rates and the optomechanical frequencies. Regarding an experimental implementation, such dynamical instabilities must be taken into account as they limit the parameter values for which efficient cooling is attainable.

It is common in optomechanics to define the cooling power by means of a \emph{cooling rate}, which determines both the minimal achievable temperature and the time required to reach it. This quantity can be defined by calculating the power spectral density (PSD) of the COM motion, defined as
\begin{equation}\label{motionalPSD}
\begin{split}
    S_{jj}(\omega) = \frac{r_{0j}^2}{2\pi}\int_{-\infty}^\infty d\tau\langle \hat{q}_j(t+\tau)\hat{q}_j(t)\rangle_{\rm ss} e^{i\omega \tau},
\end{split}
\end{equation}
where the sub-index \textit{ss} denotes the steady state.
 As detailed in Appendix \ref{appendixEOMS}, the motional PSD is computed by using the Quantum Regression Theorem \cite{CarmichaelBook} which, in its simplest form, relates the two-time correlation function $\langle\hat{q}_j(t+\tau)\hat{q}_j(t)\rangle_{\rm ss}$ to its value at zero delay, namely $\langle\hat{q}_j^2\rangle_{\rm ss}$. The calculation of the PSD is straightforward from here, since the zero-delay expected value has been already determined by numerically solving the system equations of motion. 

The motional PSD, Eq. \ref{motionalPSD}, is shown in Fig.~\ref{figPSDlinewidths}a for $\phi=\pi/4$. As evidenced here, the PSD along a given axis, $j$, shows a series of Lorentzian peaks, the most relevant of which is centered in the vicinity of $\Omega_j'$. The full width at half maximum (FWHM) of these peaks, shown in Fig.~\ref{figPSDlinewidths}b, determines the total rate at which the temperature evolves in time \cite{WindeyArXiv2018}. Note that, as demonstrated in Appendix \ref{appendixEOMS}, such FWHM is independent of the position localization rates $\Gamma_j^{(d)},\Gamma_j^{(p)}$, and $\Gamma_j^{(r)}$, which only affect the steady state temperature. Furthermore, the remaining source of heating provided by the rate $\gamma$ (Eq.~\ref{gammaPV}) becomes negligible at low pressures. Therefore, we can identify the cooling rates as the low-pressure limits of the curves in Fig.~\ref{figPSDlinewidths}, that is, $\gamma_x \sim 2\pi \times 30\,\rm Hz$ and $\gamma_{y,z} \sim 2\pi \times 1\,\rm kHz$ for optimal cooling angles $\phi$. These values are consistent with experimental observations \cite{WindeyArXiv2018} and, remarkably, they are of the same order as the rates predicted by adiabatic elimination of the cavity \cite{WilsonRaeNJP2008} ($\gamma_j \approx \vert g_j'\vert^2 \kappa /(\kappa^2+(\delta'-\Omega_j')^2)$), despite the fact that the assumptions for such elimination, namely $\kappa \gg \vert  g_j'\vert$, are not fulfilled in this case.

\begin{figure}
	\centering
	\includegraphics[width=\linewidth]{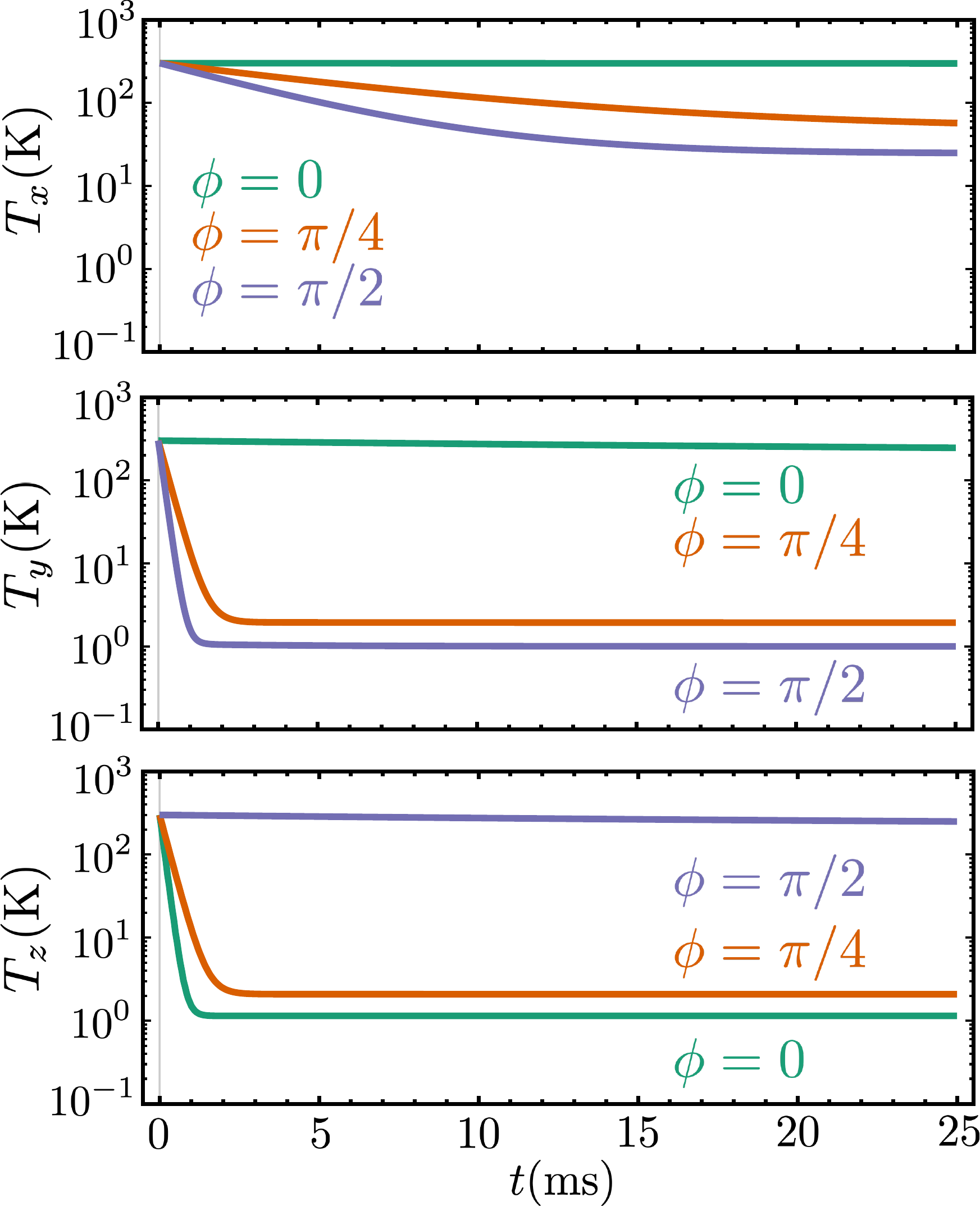}
	\caption{(Color online). Time evolution of the temperature of the COM motion along the three motional axes at pressure $P = 3 \cdot 10^{-3}\,\rm mbar$ and detuning $\delta' = 2\pi \times 400\, \rm kHz$, for different positions of the NP inside the cavity, $\phi$, and polarization angle $\Theta = 10^\circ$. The system is initialized in a thermal state at room temperature.}\label{figTvsTime}
\end{figure}

\begin{figure}
	\centering
	\includegraphics[width=\linewidth]{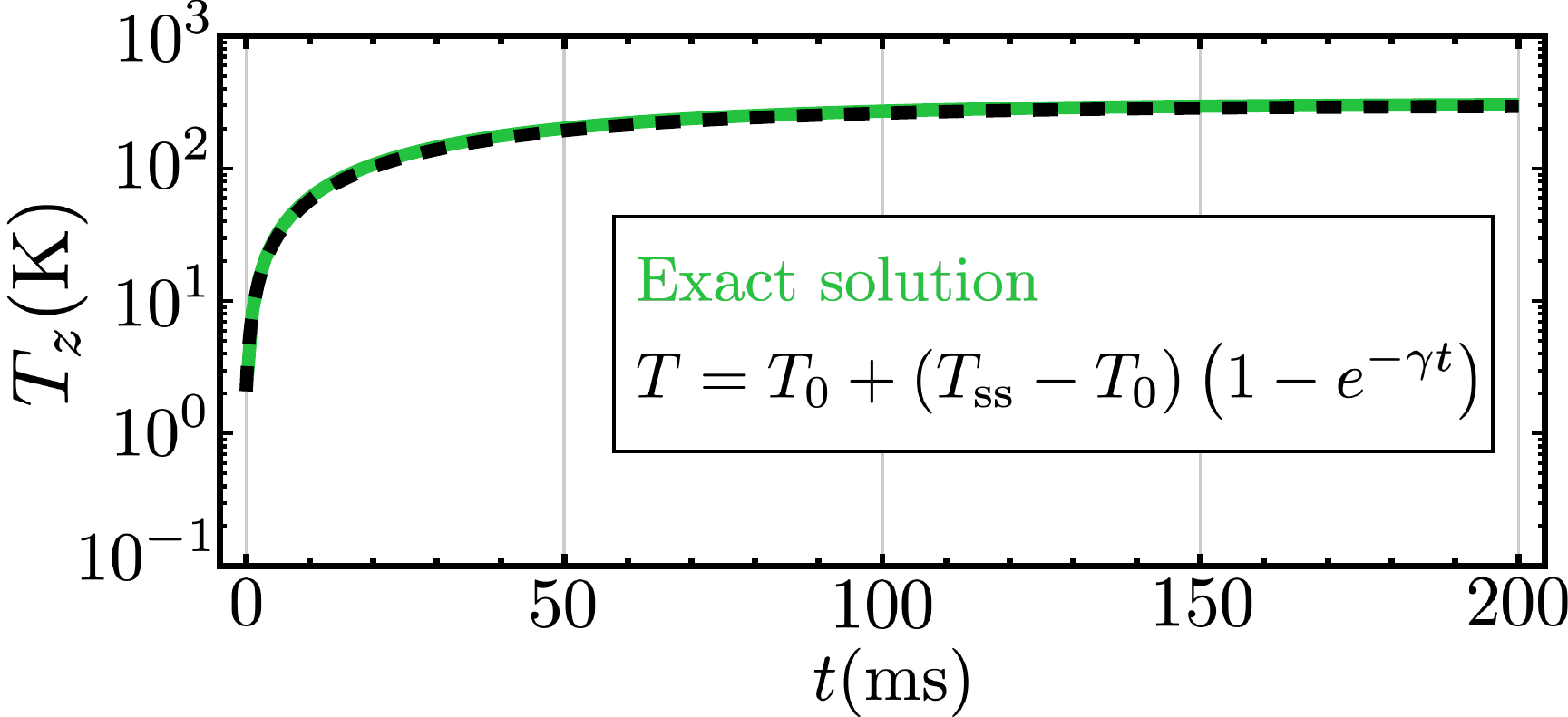}
	\caption{(Color online). Reheating curve: COM temperature along the $Z-$ axis, at pressure $P = 3 \cdot 10^{-3}\,\rm mbar$, detuning $\delta' = 2\pi \times 20\,\rm MHz$, and polarization angle $\Theta = 10^\circ$. The curves show similar behavior for every axis and every position $\phi$.}\label{figTvsTime_reheating}
\end{figure}

From the above argumentation and the curves in Fig.~\ref{figPSDlinewidths}, we can predict that, at pressures $P\lesssim 5\cdot10^{-3}\,\rm mbar$ and optimal cooling positions $\phi$, the temperatures will decay to $1/e$ times their initial value in around $\sim 1/\gamma_x \sim 5\,\rm ms$ along the $X-$ axis, and $\sim 1/\gamma_{y,z} \sim 0.2\,\rm ms$ along the $Y-$ and $Z-$ axes. This trend is indeed observed in the system dynamics, displayed in Fig.~\ref{figTvsTime}. Clearly, the dependence of the optomechanical couplings on the position of the NP inside the cavity not only influences the steady-state temperature, but the whole time evolution. Moreover, the efficiency of the cooling depends critically on $\phi$, with no cooling at all occurring in the less optimal configuration. In order to certify that the coupling to the cavity is responsible for the cooling, we display in Fig.~\ref{figTvsTime_reheating} the reheating of the COM motion along the $Z-$axis (similar curves are obtained for every axis). Here, we have initialized the system in its cooled steady state and effectively turned off the cavity cooling by setting the cavity detuning to $\delta' = 2\pi\times20\,\rm MHz$$\gg\kappa',\Omega_j'$. As expected, in this situation the dynamics is cavity-independent and governed only by gas pressure reheating, thus occurring on a timescale $\gamma^{-1} \sim 50\, \rm ms$, in agreement with the experiment \cite{WindeyArXiv2018}. 
Also in agreement with the above discussion, the reheating dynamics is independent on the position localization heating rates $\Gamma_j^{(d)},\Gamma_j^{(p)}$, and $\Gamma_j^{(r)}$. Finally, note that in this limit, namely $g_j' \to 0$, the equations of motion can be analytically solved in terms of $\gamma$, the steady-state temperature $T_{\rm ss}$, and the initial temperature $T_0$. The resulting analytical expression is represented by the dashed line in Fig.~\ref{figTvsTime_reheating}, and shows excellent agreement with the full numerical solution. Our results in Figs. \ref{figTvsTime} and \ref{figTvsTime_reheating} quantitatively agree with experimental measurements \cite{WindeyArXiv2018}.

\begin{figure}
	\centering
	\includegraphics[width=\linewidth]{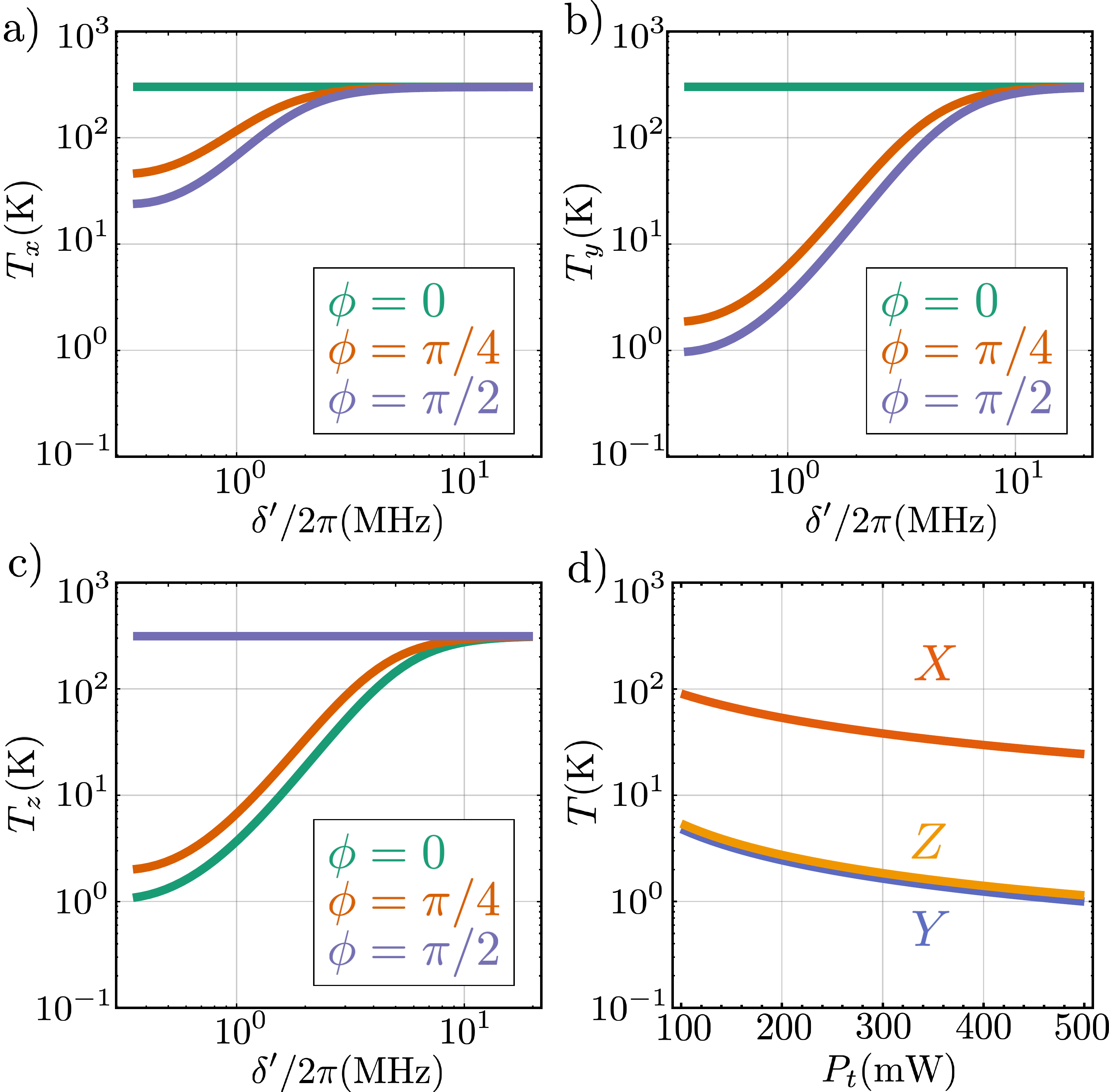}
	\caption{(Color online). a-c) Steady-state temperatures as a function of cavity detuning, for $P_t = 500\,\rm mW$. d) Steady-state temperature at the best cooling position (i.e., $\phi = \pi/2$ for $X-Y$, and $\phi = 0$ for $Z$) versus tweezer power, for $\delta' = 2\pi\times400\,\rm kHz$. In all panels, $P=3\cdot 10^{-3}\,\rm mbar$ and $\Theta=10^{\circ}$.}\label{figTvsdetuning}
\end{figure}

Because the above defined cooling rates are a function of the optomechanical couplings $g_j'$, they depend strongly on the cavity detuning $\delta'$, a dependence that stems from the tweezer-induced displacement $\alpha_c$ (see Eq.~\ref{solutionalphac}). In order to characterize this dependence, we show in Fig.~\ref{figTvsdetuning}(a-c) the steady-state temperature along the three motional axes as a function of $\delta'$. As expected, the cooling becomes significantly less efficient when the detuning becomes larger than the cavity linewidth, $\delta' \gtrsim \kappa' \approx 2\pi \times0.5$ MHz, and completely disappears at around $\delta' \approx 10\kappa'$. Moreover, the steady-state temperature saturates at small detunings due to the large optomechanical couplings $\vert g_j'\vert \approx \Omega_j'$. Indeed, as mentioned above, for such large couplings the system is near a dynamical instability regime, where further reduction of the detuning toward resonance with the COM modes pushes the system closer to the unstable regime. This results in additional heating (not shown in Fig.~\ref{figTvsdetuning}) and, eventually, in the loss of the NP from the trap as the system becomes unstable. For the current parameters, such instability might hinder the realization of ground state cooling, since reaching the ground state  usually requires the cavity detuning to be in resonance with the mechanical frequency. 

The dependence of the cooling rates on the optomechanical coupling rates $g_j'$ also allows for their tuning by means of the tweezer power. In Fig.~\ref{figTvsdetuning}d, we show the dependence of the steady-state temperature with the tweezer power, at $\delta' = 2\pi\times 400\,\rm kHz$ and at optimal cooling positions of the NP. Here, we observe an increase of the cooling efficiency with higher powers, as both the COM frequencies and the optomechanical couplings become more relevant with respect to the heating rates. As the power is increased, however, the decrease of the steady-state temperature becomes less pronounced, as the enhancement in the cooling rate starts to be compensated by the displacement heating rates $\Gamma_j^{(d)}\propto P_t^2$. Above a certain power ($P_t \approx 2\,\rm W$ for the parameters of Fig.~\ref{figTvsdetuning}), such heating rates become dominant, and the steady-state temperatures become a quadratically increasing function of the tweezer power. 

As shown by the results in this section, our model is able to reproduce all the experimental observations reported in Ref. \cite{WindeyArXiv2018}, both qualitatively and quantitatively. This shows the validity of such model to describe cavity-assisted cooling via coherent scattering.

\onecolumngrid
\begin{center}
\begin{figure}
 	\centering
 	\includegraphics[width=\linewidth]{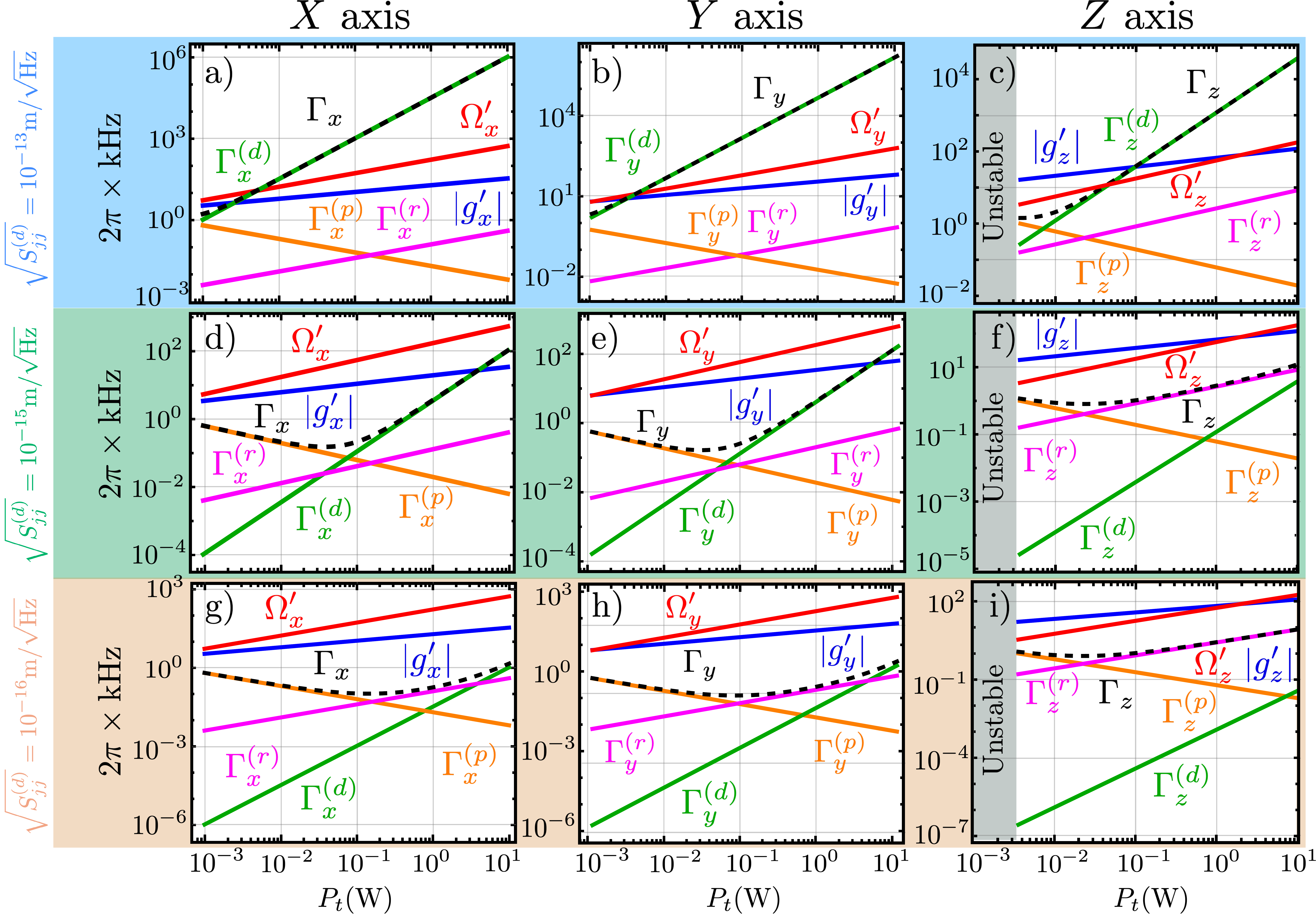}
 	\caption{(Color online). Comparison of the relevant system rates for cooling along the three motional axes, at $P=10^{-9}\,\rm mbar$. The rates are displayed as a function of tweezer power for three different values of the displacement PSD, $S_{jj}^{(d)}$, assumed equal along the three axes. For each axis we choose optimal cooling conditions, namely $\{\delta',\Theta,\phi\} = \{\Omega_x',\pi/4,\pi/2\}$ along $X$ (panels a,d,g), $\{\delta',\Theta,\phi\} = \{\Omega_y',0,\pi/2\}$ along $Y$ (panels b,e,h), and $\{\delta',\Theta,\phi\} = \{\Omega_z',0,0\}$ along $Z$ (panels c,f,i). All unspecified parameters have the same value as in Sec. \ref{secResults}.}\label{fig3x3panel}
\end{figure}
\end{center}
\twocolumngrid

\section{Requirements for ground state cooling}\label{secGScooling}

The parameter values chosen for the above section do not allow for ground state cooling, defined as the reduction of the COM phonon occupation to $\langle \hat{n}_j\rangle \equiv \langle \hat{b}^\dagger_j\hat{b}_j\rangle < 1$. The first reason is the high values of the gas pressure and the tweezer power, which result in large heating rates $\Gamma_j^{(d)}$ and $\Gamma_j^{(p)}$. Moreover, the optomechanical couplings $\vert g_j'\vert$, especially along the $Z-$axis, are comparable to the COM frequencies. As a result, a desirable condition for cooling, namely the cavity being in resonance with the COM modes ($\delta' = \Omega_j'$), cannot be reached without the system becoming dynamically unstable \cite{KusturaArXiv2018}. Another reason hindering ground state cooling is the unresolved sideband regime, i.e. the large linewidth of the cavity with respect to the mechanical frequencies.  In this section we use our model, which has been tested by reproducing experimental observations, to determine the conditions for reaching the mechanical ground state along any of the motional axes. 

Let us first focus on cooling the motion only along one of the axes, regardless of the dynamics along the remaining two. In order to simplify the physical interpretation of the results, for cooling along the $Y-$($Z-$) axis we take $\{\Theta,\phi\}=\{0,\pi/2\}$ ($\{\Theta,\phi\}=\{0,0\}$), such that the motion along the remaining two directions is uncoupled from the cavity. On the other hand, for cooling along the $X-$ axis, we can only uncouple the $Z-$ motion by choosing $\{\Theta,\phi\}=\{\pi/4,\pi/2\}$. In this section, we aim at modifying the system parameters to achieve ground state cooling in the resolved sideband  regime \cite{AspelmeyerRMP2014}, for which three requisites must be fulfilled: first, the total heating rates $\Gamma_j = \Gamma_j^{(r)}+ \Gamma_j^{(p)}+ \Gamma_j^{(d)}$ have to be minimized. Second, the cavity linewidth must be able to resolve the motional sidebands, i.e., $\kappa' \ll \Omega_j'$. Third, the optomechanical coupling must remain small, $\vert g_j'\vert \ll \Omega_j'$, so that the motional phonons and cavity photons do not hybridize appreciably. Let us focus on these three requisites independently.

We begin by the minimization of the heating rates. We will assume, on the one hand, a reduced gas pressure of  $P = 10^{-9}$mbar, a value within reach of current ultra-high vacuum technology. On the other hand, since the calculated COM temperatures in the previous section were limited by trap displacement heating, it is necessary to incorporate some isolation in our system, in order to reduce the corresponding noise PSDs, $S_{jj}^{(d)}$ (Eq.~\ref{noisePSD}). In Fig.~\ref{fig3x3panel}, we display the partial heating rates $\Gamma_j^{(r)}$, $\Gamma_j^{(p)}$, $\Gamma_j^{(d)}$, the total heating rate $\Gamma_j$, the frequencies $\Omega_j'$, and the optomechanical couplings $\vert g_j'\vert$ as a function of tweezer power and for three different values of the displacement PSD. 
Note that both the recoil heating rate and the displacement heating rate increase with the tweezer power, whereas $\Gamma_j^{(p)}$ obeys the opposite behavior, resulting in an optimal tweezer power for which the heating rates are minimized. At high values of the displacement PSD, the heating rate is completely dominated by the displacement heating rate, which become comparable or even larger than the mechanical frequencies. In this limit, ground state cooling becomes impossible. 
On the other hand, for $\big[S_{jj}^{(d)}\big]^{1/2}\sim 10^{-16}\,$m$/\text{Hz}^{-1/2}$, the trap displacement noise is effectively eliminated and the heating is dominated by the interplay between gas pressure and recoil heating. In this regime (lower row in Fig.~\ref{fig3x3panel}), the total heating rates reach minimum values of $\Gamma_j \sim 2\pi \times 100\,\rm Hz$, orders of magnitude smaller than the optomechanical couplings. Our results in Fig.~\ref{fig3x3panel} suggest that choosing an adequate tweezer power is crucial for ground state cooling.
We remark that all the results in Fig.~\ref{fig3x3panel} are practically independent on the cavity linewidth along a wide interval, at least $100 \,\text{Hz} \lesssim \kappa'/2\pi\lesssim 1\,\rm MHz$.

\begin{figure}
	\centering
	\includegraphics[width=\linewidth]{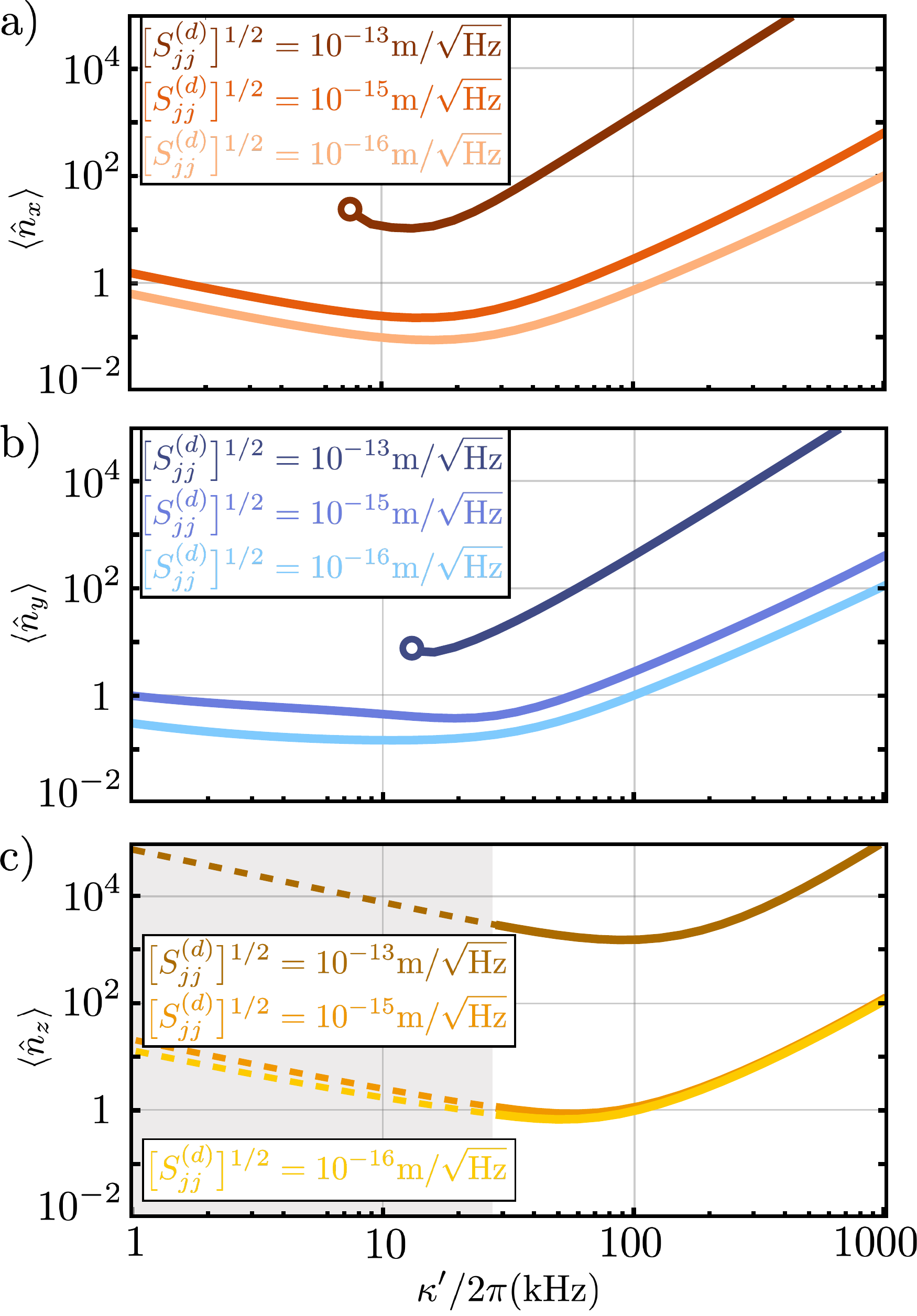}
	\caption{(Color online). Steady-state phonon occupation as a function of cavity linewidth. 
	For each curve in panels a-b, we take the parameters of Fig.~\ref{fig3x3panel} and the tweezer power that minimizes $\Gamma_j$. The open dots mark the linewidths below which the system becomes dynamically unstable. For the $Z-$axis (panel c) we set $P_t=3\,\rm W$ and $\phi=60^\circ$. The shaded region indicates the potentially unstable regime (see main text).}\label{figOptimumKappa}
\end{figure}

We now focus on the tuning of the cavity linewidth into the resolved sideband regime. Using our results in Fig.~\ref{fig3x3panel},
we calculate the steady state occupation along the $X$ and $Y$ axes (the $Z$ axis is more involved, and will be analyzed later) as a function of the cavity linewidth, choosing the values of the tweezer power that minimize the total heating rate. Such occupations are displayed in Fig.~\ref{figOptimumKappa}(a-b) for three different values of the displacement PSD. For high values of the PSD, the system becomes dynamically unstable at small $\kappa'$, as the dissipation rate cannot compensate for both the large heating rate and the high ratio $\vert g_j'\vert/\Omega_j' \approx 1$. On the other hand, for low values of the displacement PSD, ground state cooling is observed along both of the axes, with occupation numbers $\langle \hat{n}_x\rangle \approx 0.09$ and $\langle \hat{n}_y\rangle \approx 0.14$ respectively. The optimal cooling is not attained for arbitrarily small linewidths, but reaches a minimum at $\kappa' \sim \vert g_j'\vert$. As we will demonstrate below, this can be ascribed to the balance between the photon scattering rate into the cavity and the rate of photon loss through the mirrors. We emphasize that the minimal occupations observed in Fig.~\ref{figOptimumKappa}(a-b) are not necessarily the lowest occupations achievable in this system, as we will see below.

The last requirement for cooling is the reduction of the ratio $\vert g_j'\vert/\Omega_j'$. For the $X$ and $Y$ axes this becomes manifest in the fact that the tweezer powers chosen in Fig.~\ref{figOptimumKappa}(a-b) are not the optimal ones for cooling despite minimizing the heating rate. Indeed, one can check that for higher values of $P_t$, the advantageous reduction of $\vert g_j'\vert/\Omega_j' \sim P_t^{1/4}$ overcomes the detrimental increase in the heating rate, and gives rise to lower occupation numbers than the ones shown in Fig.~\ref{figOptimumKappa}(a-b). However, the critical importance of reducing the ratio $\vert g_j'\vert/\Omega_j'$ is most evidenced when cooling along the $Z-$axis, where this ratio takes very large values (See Fig.~\ref{fig3x3panel}). Such large values, combined with the recoil heating rate being $\sim 10\times$ larger along the $Z-$axis, make it impossible to cool the motion for tweezer powers below $P_t \sim 1-2$W, as the system becomes unstable for any value of $\kappa'$ within the resolved sideband regime. 
However, in order to efficiently cool the motion along the $Z$ axis, the ratio $\vert g_j'\vert/\Omega_j'$ can be reduced by either increasing the tweezer power or changing the position of the NP inside the cavity, $\phi$. A combination of both methods allows for ground state cooling along the $Z$ axis as well, as shown in Fig.~\ref{figOptimumKappa}c. Due to the larger recoil heating rates, the minimum occupations reached along $Z$, $\langle\hat{n}_z\rangle\approx0.66$, are roughly one order of magnitude larger than along the remaining two axes. Additionally, since recoil dominates the heating for displacement PSDs below $\sim 10^{-15}\,$m$/\sqrt{\text{Hz}}$ (see Fig.~\ref{fig3x3panel}), reducing such PSD below that value has little effect on the phonon occupation. Remarkably, for the parameters of Fig.~\ref{figOptimumKappa}c we find a regime of uncertain stability (shaded area), where some solutions of the polynomial equation for $\beta_y$ (see Appendix \ref{appendixDISPSS}) yield stable equations of motion and some do not. Physically, this is a regime where the dynamics is extremely sensitive to small perturbations, as evidenced by the fact that the dependence of the equations of motion on $\beta_y$ is weak. Discerning whether the system is stable or not in such regime would likely require to refine the treatment of the problem, including previously disregarded terms in the Hamiltonian. This, however, lies beyond the scope of the present work, as such uncertain stability regimes are seldom encountered in the parameter space.

Regardless of the cooling axis, all the steady-state occupation curves in Fig.~\ref{figOptimumKappa} follow a general trend, reaching a minimum at $\kappa' \approx \vert g_j' \vert$. 
This behavior can be explained from our equations of motion in the low-pressure limit $\gamma \ll \Omega_j',\kappa',\vert g_j'\vert$, where, assuming only one motional axis is coupled to the cavity, its steady-state occupation along the chosen direction can be analytically calculated as
\begin{equation}
    \langle \hat{b}^\dagger_j \hat{b}_j\rangle_{\rm ss} = \frac{A-B+C}{4\vert g \vert^2\delta\kappa\Omega \left[4\vert g \vert^2\delta - (\delta^2+\kappa^2)\Omega\right]},
\end{equation}
where
\begin{equation}
    A=2\vert g\vert^4\delta\left[\delta^2\kappa + \kappa^3 + 4\delta \Omega(\Gamma-\kappa)+2\kappa\Omega^2\right],
\end{equation}
\begin{equation}
    B = \Gamma \Omega (\delta^2+\kappa^2)\left[\kappa^2+(\delta-\Omega)^2\right]\left[\kappa^2+(\delta+\Omega)^2\right],
\end{equation}
\begin{equation}
\begin{split}
    C &=  \vert g \vert^2 \Big[-\kappa \Omega\left(\delta^2+\kappa^2\right)\left[\kappa^2+(\delta-\Omega)^2\right]+
    \\
    &
    +
    2\Gamma\delta\left[2\Omega^4+\Omega^2(3\kappa^2-5\delta^2)+(\delta^2+\kappa^2)^2\right]\Big],
\end{split}
\end{equation}
and we use the simplified notation
$g \equiv \vert g_j\vert$, $\delta \equiv \delta'$, $\Gamma \equiv \Gamma_j$, $\kappa\equiv\kappa'$, 
and $\Omega \equiv \Omega_j'$. Assuming the cavity is tuned at $\delta = \Omega$, we can find three simple limits for the above expression:
\begin{equation}
\langle \hat{n}_j\rangle_{\rm ss} \approx 
\left\{
\begin{array}{cc}
    \Gamma_j/\kappa' &  \text{for } \kappa' \ll \vert g_j' \vert \ll \Omega_j'\\
    2\Gamma_j/\kappa' &  \text{for } \kappa' \approx \vert g_j' \vert \ll \Omega_j'\\
     \kappa'\Gamma_j/\vert g_j' \vert^2& \text{for } \vert g_j' \vert \ll \kappa' \ll \Omega_j'.
\end{array}
\right.
\end{equation}
The above limits are easily understood as the ratio between the heating rate and a cooling rate, the latter of which depends on two factors: on the one hand, the ability of the NP to scatter photons inside the cavity, $\vert g_j'\vert$, and on the other hand, the ability of the cavity to dissipate such photons, $\kappa'$. For $\kappa' \ll \vert g_j'\vert$, the cooling power is limited by the small cavity loss, which is not enough to efficiently reduce the cavity occupation. This results in the increase in steady-state occupation at low $\kappa'$ observed in Fig.~\ref{figOptimumKappa}. In the opposite limit, $\kappa' \gg \vert g_j'\vert$, the cooling rate is given by $\vert g_j'\vert^2/\kappa'$, which coincides with the rate obtained through adiabatic elimination \cite{WilsonRaeNJP2008}. In this case, the cooling efficiency is limited by the low value of the optomechanical coupling and, for the system under study, ground state cooling becomes ultimately inefficient as $\kappa'$ increases. Finally, in the intermediate regime $\kappa' \approx \vert g_j'\vert$ the cooling is optimized, as the rate at which photons are scattered into the cavity is equal to the rate at which they are lost through the cavity mirrors. This explains the optimal values of $\kappa'$ found in Fig.~\ref{figOptimumKappa}, and gives an order-of-magnitude theoretical estimation for the lowest achievable occupations, namely $\langle \hat{n}_x,\hat{n}_y,\hat{n}_z\rangle \approx 2\Gamma_j/\vert g_j'\vert = \{0.02,0.02,0.2\}$.

\begin{figure}
	\centering
	\includegraphics[width=\linewidth]{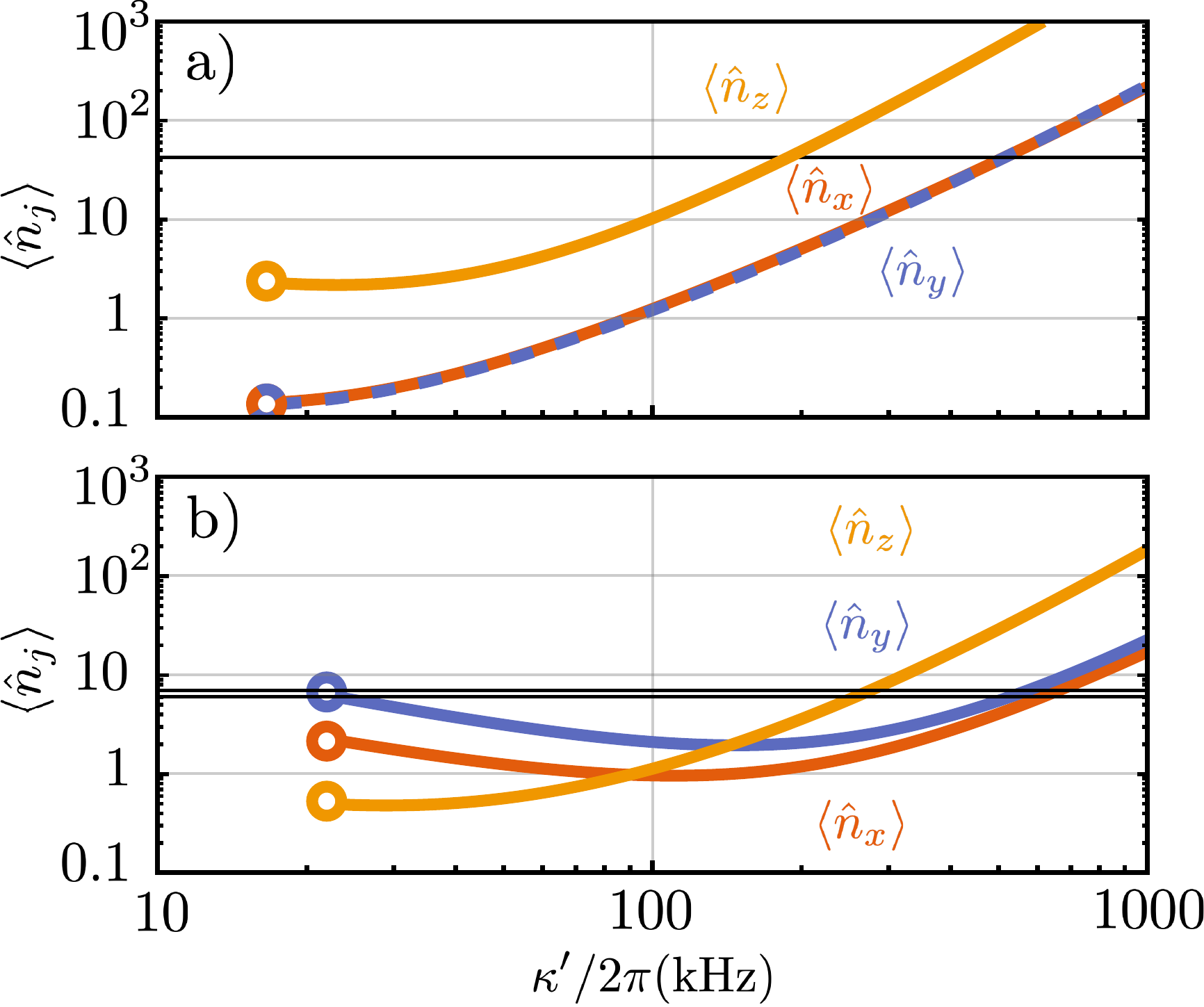}
	\caption{(Color online). Steady-state phonon occupation along the three motional axes as a function of cavity linewidth, for $\big[S_{jj}^{(d)}\big]^{1/2}=10^{-16}\,$m$/\text{Hz}^{-1/2}$ and $\Theta = \pi/4$. a) Ground state cooling along the $X-Y$ plane, at $\{P_t,\delta',\phi\}=\{75\,\text{mW},\Omega_y',65^\circ\}$. b) Ground state cooling along the $Z$ axis, at $\{P_t,\delta',\phi\}=\{3\,\text{W},\Omega_z',60^\circ\}$. Unspecified parameters are taken as in previous figures. As a guide, the black horizontal lines indicate the occupations corresponding to $T_j= 10\,\mu$K along the Z-axis (panel a) or the X-Y axes (panel b).}\label{fig3Dcooling}
\end{figure}

So far, we have demonstrated ground state cooling along each of the three motional axes, in order to have a deeper understanding of the underlying physics and relevant parameters. However, the results obtained above (e.g. in Fig.~\ref{figOptimumKappa}) are ill suited from an experimental point of view, as the 
motion along the remaining two axes is either in an unfavourable regime for cooling or even completely uncoupled from the cavity. As a result, the occupations along these two axes usually remain at their room temperature values, and in some cases drastically increase as the system is close to a dynamically unstable regime. From a practical perspective, it is convenient to find a configuration in which one of the motional axis is cooled to the ground state while the remaining two are cooled at least below $\sim 1$K, such that their motion does not probe the non-linearity of the trapping potential. This can be achieved by taking our results in Fig.~\ref{figOptimumKappa} as a starting point in parameter space, and modifying the angles $\{\Theta,\phi\}$ in order to couple the remaining axes to the cavity. As shown in Fig.~\ref{fig3Dcooling}, this method leads to regimes in which near-to-ground state cooling is attained along the three axes. Note that the $X-$ and $Y-$ motional directions can be simultaneously cooled to the ground state (Fig.~\ref{fig3Dcooling}a). In both panels of Fig.~\ref{fig3Dcooling}, the axes that do not reach the ground state are cooled below $10\,\mu$K. These results show that ground state cooling along each of the motional directions is experimentally feasible.

All the results in this section highlight the relevance of isolation from trap displacement noise for reaching the ground state, and provide strict bounds for the displacement noise level above which ground-state cooling becomes impossible. However, note that reducing the trap displacement noise might not be sufficient for reaching the ground state, as the system might be subject to some additional heating sources that, similar to the trap displacement, have no effect on the motional PSD. 
Thus, a proper identification and suppression of the relevant heating sources in a given experimental setup is crucial for ground state cooling.

\section{Conclusion}\label{secConclusion}

In this work, we have developed a theoretical model for the COM cooling of a levitated NP via coherent scattering into a cavity. We have done so by, first, obtaining a quadratic Hamiltonian for the NP and the electromagnetic field and, second, by tracing out the free field to obtain a Master Equation for the cavity and COM degrees of freedom. Heating due to gas pressure and trap displacement has been included. Our model has been shown to reproduce recent experimental observations, and is thus revealed as a useful tool for levitodynamics.

We have also used our model to study the possibility of ground state cooling of a levitated NP. The relevance of the heating due to trap displacement, which is often ignored, has been emphasized and evidences the necessity of isolation in order to reach the ground state. Our model demonstrates the potential of current experimental setups to reach the mechanical ground state of the COM motion of a NP. The requirements for ground state cooling given in this paper could bring about the quantum regime of levitodynamics, where the full quantum theory we have developed represents a necessary and powerful theoretical tool.

\begin{acknowledgments}
C.~G.~B. acknowledges funding from the EU Horizon 2020 program under the Marie Sk\l{}odowska-Curie grant agreement no.~796725.
This research was partly supported by the Swiss National Science Foundation (no.~200021L$\_$169319) and ERC-QMES (no.~338763). R.~R. acknowledges funding from the EU Horizon 2020 program under the Marie Skłodowska-Curie grant agreement no.~702172.  \\
We thank Mathieu L. Juan for insightful discussions.
\end{acknowledgments}

\appendix

\section{Calculation of the displacements}\label{appendixDISPSS}

In this Appendix we find the equations for the coefficients $\alpha_\mathbf{k}, \beta_j, \gamma_\alpha$ and solve them. We start by substituting the displaced operators \ref{adisplacement}, \ref{bdisplacement}, \ref{cdisplacement}, and \ref{a0displacement} into the Hamiltonian Eq.~\ref{Hrwa}. The transformed Hamiltonian reads $\hat{H}_0 + \Delta\hat{H}$, where $\hat{H}_0$ contains no displacements:
\begin{widetext}
\begin{equation}\label{Hrwa2}
\begin{split}
    \hat{H}_0/\hbar &\approx  \hat{c}^\dagger \hat{c}\left(\tilde{\delta}-2g_{\rm cx}\beta_x-2g_{\rm cy}\beta_y\right)+\sum_{\mathbf{k}\varepsilon} \Delta_\mathbf{k}\hat{a}_\mathbf{k}^\dagger\hat{a}_\mathbf{k}+\sum_{j}\Omega_j\hat{b}^\dagger_j\hat{b}_j -\hat{c}^\dagger\hat{c}\left[g_{\rm cx}(\hat{b}^\dagger_x+ \hat{b}_x)+g_{\rm cy}(\hat{b}^\dagger_y+ \hat{b}_y)\right]
     \\
    &
     -\frac{G}{2} \left(\hat{c}^\dagger + \hat{c}\right)k_c\left(\sin\Theta x_0\hat{b}^\dagger_x+ \cos\Theta y_0\hat{b}^\dagger_y +\text{ H.c.}\right)\sin\phi+  \im\frac{G}{2} \left(\hat{c}^\dagger - \hat{c}\right)k_0z_0(\hat{b}^\dagger_z + \hat{b}_z)\cos\phi
    \\
    &
    -\sum_{\mathbf{k}\varepsilon}\frac{G_0(\mathbf{k})}{2}\im\left(\hat{a}_\mathbf{k} - \hat{a}^\dagger_\mathbf{k}\right)\left[-k_0z_0(\hat{b}^\dagger_z + \hat{b}_z)  +\sum_jk_j r_{0j}(\hat{b}^\dagger_j + \hat{b}_j)
      \right] -\sum_{\mathbf{k}\varepsilon}G(\mathbf{k})\cos\phi\left(\hat{c}\hat{a}^\dagger_\mathbf{k} + \hat{c}^\dagger\hat{a}_\mathbf{k}\right)
      \\
      &
      -\left(\alpha_c\hat{c}^\dagger+\alpha_c^*\hat{c}\right)\left[g_{\rm cx}(\hat{b}^\dagger_x+ \hat{b}_x)+g_{\rm cy}(\hat{b}^\dagger_y+ \hat{b}_y)\right]+\int \text{d}\omega \Delta_0(\omega)\hat{a}_0^\dagger(\omega)\hat{a}_0(\omega)+\im\int \text{d}\omega\gamma(\omega)\left[\hat{a}_0(\omega)\hat{c}^\dagger \!-\! \text{H.c.}\right].
\end{split}
\end{equation}
The non-quadratic term in the first line of the above equation, $\sim \hat{c}^\dagger\hat{c}(\hat{b}_j+\text{H.c.})$, can be safely neglected, since in the displaced frame the cavity is in the vacuum state and thus the contribution of this term is of order $g_{cj} \sim 2\pi \times 1$Hz.
On the other hand, $\Delta\hat{H}$ contains all the single-operator terms:
\begin{equation}
    \begin{split}
    \Delta\hat{H}/\hbar &\approx \tilde{\delta} \hat{c}^\dagger \alpha_c+\sum_{\mathbf{k}\varepsilon} \Delta_\mathbf{k}\hat{a}_\mathbf{k}^\dagger\alpha_\mathbf{k}+\sum_{j}\Omega_j\hat{b}^\dagger_j\beta_j 
    -\sum_{\mathbf{k}\varepsilon}G(\mathbf{k})\cos\phi\left(\alpha_c\hat{a}^\dagger_\mathbf{k} + \hat{c}^\dagger\alpha_\mathbf{k}\right)
    - \frac{G}{2}\hat{c}^\dagger
    \cos\phi -\sum_{\mathbf{k}\varepsilon}\frac{G_0(\mathbf{k})}{2}\hat{a}^\dagger_\mathbf{k} 
     \\
    &
     -\frac{G}{2} \left(\alpha_c^* + \alpha_c\right)k_c\left(\sin\Theta x_0\hat{b}^\dagger_x+ \cos\Theta y_0\hat{b}^\dagger_y \right)\sin\phi+  \im\frac{G}{2} \left(\alpha_c^* - \alpha_c\right)k_0z_0\hat{b}^\dagger_z\cos\phi
     \\
    &
     -\frac{G}{2} \hat{c}^\dagger k_c\left[\sin\Theta x_0(\beta^*_x +\beta_x)+ \cos\Theta y_0(\beta^*_y +\beta_y)\right]\sin\phi+  \im\frac{G}{2} \hat{c}^\dagger k_0z_0(\beta^*_z +\beta_z)\cos\phi
    \\
    &
    +\sum_{\mathbf{k}\varepsilon}\frac{G_0(\mathbf{k})}{2}\im \hat{a}^\dagger_\mathbf{k}\bigg[-k_0z_0(\beta^*_z + \beta_z)  +\sum_jk_j r_{0j}(\beta^*_j + \beta_j)\bigg]
    -2\gamma \hat{c}^\dagger\left(g_{\rm cx}\beta_x+g_{\rm cy}\beta_y\right)
    -\vert\alpha_c\vert^2\left(g_{\rm cx}\hat{b}^\dagger_x+g_{\rm cy}\hat{b}^\dagger_y\right)
    \\
    &
    -\sum_{\mathbf{k}\varepsilon}\frac{G_0(\mathbf{k})}{2}\im\left(\alpha_\mathbf{k} - \alpha^*_\mathbf{k}\right)\!\bigg(\!-k_0z_0\hat{b}^\dagger_z   +\!\sum_jk_j r_{0j}\hat{b}^\dagger_j 
      \bigg)\!+\!\!
      \int\!\! \text{d}\omega\bigg[\hat{a}_0^\dagger(\omega)[\Delta_0(\omega)\alpha_0(\omega)-\im\alpha_c\gamma(\omega)] +\im\gamma(\omega)\alpha_0(\omega)\hat{c}^\dagger\bigg]
      \\
      &
     +\text{H.c.}.
\end{split}
\end{equation}
\end{widetext}
We can now make $\Delta\hat{H} = 0$ by setting the coefficients of each creation operator to zero, obtaining the following system of equations:
\begin{equation}
\begin{split}
    \Omega_x\beta_x &-\sin\phi\sin\Theta (k_cx_0)\frac{G}{2}(\alpha_c+\alpha_c^*) 
    \\
    &
    -\sum_\mathbf{k}\frac{G_0(\mathbf{k})}{2}\im(\alpha_\mathbf{k}-\alpha_\mathbf{k}^*)k_x x_0-g_{\rm cx}\vert\alpha_c\vert^2
    =0,
\end{split}
\end{equation}
\begin{equation}
\begin{split}
    \Omega_y\beta_y &-\sin\phi\cos\Theta (k_cy_0)\frac{G}{2}(\alpha_c+\alpha_c^*) 
    \\
    &
    -\sum_\mathbf{k}\frac{G_0(\mathbf{k})}{2}\im(\alpha_\mathbf{k}-\alpha_\mathbf{k}^*)k_y y_0 -g_{\rm cy}\vert\alpha_c\vert^2
    =0,
\end{split}
\end{equation}
\begin{equation}
\begin{split}
    \Omega_z\beta_z &+\cos\phi (k_0z_0)\frac{G}{2}i(\alpha_c^*-\alpha_c) 
    \\
    &
    -\sum_\mathbf{k}\frac{G_0(\mathbf{k})}{2}\im(\alpha_\mathbf{k}-\alpha_\mathbf{k}^*)(k_z-k_0) z_0
    =0,
\end{split}
\end{equation}
\begin{equation}\label{eqforgamma}
\begin{split}
    (\tilde{\delta}&-2g_{\rm cx}\beta_x-2g_{\rm cy}\beta_y)\alpha_c-\frac{G}{2}\cos\phi +\im\int d\omega\gamma(\omega)\alpha_0(\omega)
    \\
    &
    -G\sin\phi\sin\Theta(k_cx_0)\beta_x-G\sin\phi\cos\Theta(k_cy_0)\beta_y
    \\
    &
    +\im G(k_0z_0)\cos\phi\beta_z -\cos\phi\sum_\mathbf{k}\alpha_\mathbf{k} G(\mathbf{k})= 0,
\end{split}
\end{equation}
\begin{equation}\label{akdisplacements}
\begin{split}
    \sum_\mathbf{k}& \Big\{\im G_0(\mathbf{k})\left[k_xx_0\beta_x + k_yy_0\beta_y + (k_z-k_0) z_0 \beta_z\right]
    \\
    &
    + \Delta_\mathbf{k}\alpha_\mathbf{k} -\frac{G_0(\mathbf{k})}{2}-\cos\phi\alpha_c G(\mathbf{k})
    \Big\} = 0,
\end{split}
\end{equation}
\begin{equation}\label{eqforalpha0}
\begin{split}
    \int \text{d}\omega \left[\Delta_0(\omega)\alpha_0(\omega)-\im\alpha_c\gamma(\omega)\right] = 0,
\end{split}
\end{equation}
where we have used the fact that the coefficients $\beta_j$ can be chosen real without loss of generality.
From Eqs. \ref{akdisplacements} and \ref{eqforalpha0}, we immediately obtain \cite{RomeroIsartPRA2011}
\begin{equation}\label{solutionalpha0}
    \alpha_0(\omega) = -\pi\alpha_c\gamma(\omega_0)\delta(\Delta_0)-\im\alpha_c\gamma(\omega)\text{P.V.}\frac{1}{\Delta_0(\omega)},
\end{equation}
\begin{equation}\label{solutionalphak}
\begin{split}
    \!\!\alpha_\mathbf{k}& =\!\Big\{\!\!-\!iG_0(\mathbf{k})\big[k_xx_0\beta_x + k_yy_0\beta_y\! +\!(k_z\!-\!k_0) z_0 \beta_z\big]
    \\
    &+\frac{G_0(\mathbf{k})}{2}
    + \cos\phi G(\mathbf{k})\alpha_c
    \Big\} \text{P.V.}\frac{1}{\Delta_\mathbf{k}},
\end{split}
\end{equation}
where P.V. stands for the Cauchy principal value.
We now introduce the above results into Eq.~\ref{eqforgamma} to solve for $\alpha_c$. In the process, we must calculate two integrals. The first one is given by
\begin{equation}\label{gammaint}
\begin{split}
    \im\int \text{d}\omega\gamma(\omega)\alpha_0(\omega) &= -\im\pi\alpha_c\gamma^2(\omega_0)+
    \\
    &+\alpha_c\text{P.V.}\int \text{d}\omega\gamma^2(\omega)\frac{1}{\Delta_0(\omega)}. 
\end{split}
\end{equation}
The principal value integral above can be carried out by using the fact that $\gamma(\omega) \approx \sqrt{\kappa/\pi}$ on a broad range of frequencies. After changing variables from $\omega$ to $\omega-\omega_0$, we can approximate the second term of the above equation by
\begin{equation}\label{gammainttrick}
    \frac{\alpha_c \kappa}{\pi}\text{P.V.}\int_{-\omega_0}^\infty \text{d}\omega\frac{1}{\omega} \approx  \frac{\alpha_c \kappa}{\pi}\text{P.V.}\int_{-\infty}^\infty \text{d}\omega\frac{1}{\omega} = 0.
\end{equation}
The second integral to calculate is given by
\begin{equation}
    \sum_{\mathbf{k}\varepsilon} G(\mathbf{k})\alpha_\mathbf{k} = \frac{\mathcal{V}}{(2\pi)^3}\sum_\varepsilon\int \text{d}\Omega_\mathbf{k}\int \text{d}kk^2 \alpha(\mathbf{k}),
\end{equation}
where we have transformed the sum into an integral, $\sum_{\varepsilon\mathbf{k}} \to \sum_\varepsilon \mathcal{V}/(2\pi)^3 \int \text{d}\Omega_\mathbf{k}\int dk k^2$,  $\Omega_\mathbf{k}$ being the solid angle in wavevector space. Using the closure relation of the polarization vectors,
\begin{equation}
    \sum_\varepsilon (\boldsymbol{\varepsilon}_\mathbf{k}\cdot \mathbf{e}_1)(\boldsymbol{\varepsilon}_\mathbf{k}\cdot \mathbf{e}_2) = \mathbf{e}_1\cdot \mathbf{e}_2-\sum_{ij}
\frac{k_ik_j}{k^2}e_{1i}e_{2j},
\end{equation}
we can perform explicitly all the angular integrals above, and solve for the coefficient $\alpha_c$ as
\begin{equation}\label{solutionalphac}
\begin{split}
    \alpha_c &= \frac{1}{\tilde{\delta}-\im\kappa-2g_{\rm cx}\beta_x-2g_{\rm cy}\beta_y-\varpi\cos^2\phi}
    \\
    &\times\left[P_{xy} 
    +\im(k_0z_0)\cos\phi\beta_z\left(-G+\sqrt{2}\varpi\eta\right)\right],
    \end{split}
\end{equation}
where we have defined for compactness
\begin{equation}
\begin{split}
    P_{xy} &= \frac{G}{2}\cos\phi + \frac{\varpi\eta\cos\phi}{\sqrt{2}}+
    \\
    &+G\sin\phi\left(k_cx_0 \beta_x\sin\Theta + k_cy_0\beta_y\cos\Theta\right),
\end{split}
\end{equation}
\begin{equation}\label{defvarpi}
    \varpi \equiv\frac{\epsilon_c^2V^2}{12\pi^2}\frac{\omega_c}{V_c}\text{P.V.}\int_0^{K_c} \text{d}k k^3\frac{1}{k-k_0},
\end{equation}
\begin{equation}
    \eta = \cos\Theta\sqrt{\frac{\varepsilon_0V_cE_0^2}{\hbar\omega_c}}.
\end{equation}
In Eq.~\ref{defvarpi}, we have introduced a high-energy cutoff $K_c$ to prevent the integral from diverging. Physically, this cutoff restricts the integration domain to values of $k$ for which the long-wavelength approximation (and thus our model) remains valid. In our calculations we choose $K_c = 0.1 /R$. 

The last step is to introduce Eqs. \ref{solutionalpha0}, \ref{solutionalphak}, and \ref{solutionalphac} into the three equations for the coefficients $\beta_j$. In the same fashion as above, we first calculate the sums
\begin{equation}
\begin{split}
    \sum_{\mathbf{k}\varepsilon}&\frac{G_0(\mathbf{k})}{2}\im\left(\alpha_\mathbf{k}-\alpha_\mathbf{k}^*\right)\left[\begin{array}{c}
         k_xx_0  \\
         k_y y_0 \\
         (k_z-k_0) z_0
    \end{array}\right] =
    \\
    &
    =\!\varpi_2 \!\!\left[\!\!\begin{array}{c}
         (k_0x_0)^2\beta_x  \\
         2(k_0 y_0)^2\beta_y \\
         2(k_0 z_0)^2\beta_z
    \end{array}\!\!\right]\!\!+\!\text{Im}(\alpha_c)k_0z_0\cos\phi \!\!\left[\!\!\begin{array}{c}
         0  \\
         0 \\
         2\eta\varpi
    \end{array}\!\!\right]\!\!,
\end{split}
\end{equation}
where we define 
\begin{equation}
    \varpi_2 = \left(\frac{\epsilon_c V E_0}{\pi}\right)^2\frac{\varepsilon_0}{30\hbar k_0^2}\text{P.V.}\int_0^{K_c}\text{d}k\frac{k^5}{k-k_0}.
\end{equation}
The resulting system of equations for $\beta_j$ can be further simplified. Indeed, since the expression for $\alpha_c$ is linear in $\beta_z$, we can express this coefficient in terms of $\beta_x$ and $\beta_y$ as
\begin{equation}\label{eqbetaz}
\begin{split}
    \beta_z \!&=\! k_0z_0\!\cos\phi P_{xy}\kappa\!\bigg[\frac{\kappa^2+Q_{xy}^2}{\varpi\eta-G/2}\!\left[\Omega_z\!-\!2\varpi_2(k_0z_0)^2\right]
    \\
    &-Q_{xy} (\sqrt{2}\varpi\eta-G) (k_0z_0)^2\cos^2\phi\bigg]^{-1},
\end{split}
\end{equation}
where we define
\begin{equation}
\begin{split}
    Q_{xy} &= \tilde{\delta}-\varpi\cos^2\phi-2g_{\rm cx} \beta_x -2g_{\rm cy} \beta_y.
\end{split}
\end{equation}
Introducing all the above identities into the equations for $\beta_x$ and $\beta_y$ we obtain the following system of equations,
\begin{equation}\label{eqbetax}
\begin{split}
    (\kappa^2&+Q_{xy}^2)\left[\Omega_x-\varpi_2(k_0x_0)^2\right]\beta_x =\frac{G}{2}\sin\phi\sin\Theta
    \\
    &
   \times \!\!k_cx_0\!\left[\!P_{xy}Q_{xy}\!-\!\kappa(k_0z_0)\!\cos\phi(\sqrt{2}\eta\varpi-G)\beta_z\!\right]
    \\
    &
    +\!g_{\rm cx}\!\left[\!P_{xy}^2 + \beta_z(k_0z_0)^2\cos^2\phi(\sqrt{2}\eta\varpi-G)^2\right],
\end{split}
\end{equation}
\begin{equation}\label{eqbetay}
\begin{split}
    (\kappa^2&+Q_{xy}^2)\!\left[\Omega_y-2\varpi_2(k_0y_0)^2\right]\!\beta_y =\frac{G}{2}\sin\phi\cos\Theta
    \\
    &
   \times \!\!k_cy_0\!\left[\!P_{xy}Q_{xy}\!-\!\kappa(k_0z_0)\!\cos\phi(\sqrt{2}\eta\varpi-G)\beta_z\!\right]
    \\
    &
    +\!g_{\rm cy}\!\left[\!P_{xy}^2 + \beta_z(k_0z_0)^2\cos^2\phi(\sqrt{2}\eta\varpi-G)^2\right].
\end{split}
\end{equation}
Both the above equations are polynomial equations of degree $5$ in the variables $\beta_x$ and $\beta_y$. However, they can be reduced by introducing the explicit expressions for $g_{\rm cx}$ and $g_{\rm cy}$ (Eqs. \ref{gcx} and \ref{gcy}).  This allows us to take the ratio of both equations and show that, for $\Theta \ne \pi/2$, the two variables are easily related through
\begin{equation}\label{eqbetaxbetay}
    \beta_x = \sqrt{\frac{\Omega_y}{\Omega_x}}\tan\Theta\frac{\Omega_y-2\varpi_2(k_0y_0)^2}{\Omega_x-\varpi_2(k_0x_0)^2}\beta_y.
\end{equation}
After introducing this solution into either Eq.~\ref{eqbetax} or \ref{eqbetay}, we find a fifth degree polynomial equation for the remaining coefficient, $\beta_y$, which has to be solved numerically. Note that imaginary solutions for such polynomial equation are disregarded as $\beta_y \in \mathbb{R}$ by assumption. This univocally determines $\beta_y$ since in the stable regions the polynomial usually has a single real solution. After numerically obtaining $\beta_y$, the remaining coefficients can be easily computed using Eqs. \ref{solutionalpha0}, \ref{solutionalphak}, and \ref{solutionalphac}, \ref{eqbetaz}, and \ref{eqbetaxbetay}.

\section{Tracing out of the continuum reservoirs}\label{appendixTRACEOUT}

Our starting point in this section is Eq.~\ref{vonNeumanneq3} and, more specifically, the second term
\begin{equation}
    D\equiv -
    \frac{1}{\hbar^2}\text{Tr}_R\int_0^\infty \text{d}s\left[\hat{V}(t),\left[ \hat{V}(s),\hat{\mu}(t)\otimes\hat{\rho}_R\right]\right].
\end{equation}
As mentioned in the main text, the interaction potential contains three terms, resulting in $9$ independent contributions for the above commutator. However, these interaction terms correspond to two reservoirs considered independent, and therefore, in a thermal state, the cross terms between them will vanish after tracing,
\begin{equation}
    \text{Tr}_{R}\left(\hat{V}_A \hat{V}_{B1} \hat{\rho}_R\right)=\text{Tr}_{R}\left(\hat{V}_A \hat{V}_{B2} \hat{\rho}_R\right)=0.
\end{equation}
We thus reduce the problem to calculating four independent dissipators,
\begin{equation}\label{DA}
    D_A= -
    \frac{1}{\hbar^2}\text{Tr}_R\int_0^\infty \!\!\text{d}s\left[\hat{V}_A(t),\left[ \hat{V}_A(s),\hat{\mu}(t)\otimes\hat{\rho}_R\right]\right]\!,
\end{equation}
\begin{equation}
    D_{B1}\equiv \!-
    \frac{1}{\hbar^2}\text{Tr}_R\!\!\int_0^\infty \!\!\!\!\text{d}s\!\left[\hat{V}_{B1}(t),\!\left[ \hat{V}_{B1}(s),\hat{\mu}(t)\otimes\hat{\rho}_R\right]\right]\!\!,
\end{equation}
\begin{equation}
    D_{B2}\equiv \!-
    \frac{1}{\hbar^2}\text{Tr}_R\!\!\int_0^\infty \!\!\!\!\text{d}s\!\left[\hat{V}_{B2}(t),\!\left[ \hat{V}_{B2}(s),\hat{\mu}(t)\otimes\hat{\rho}_R\right]\right]\!\!,
\end{equation}
\begin{equation}
\begin{split}
    D_{B12}\equiv &\! -
    \!\frac{1}{\hbar^2}\text{Tr}_R\!\!\int_0^\infty \!\!\!\!\text{d}s\!\left[\hat{V}_{B1}(t),\!\left[ \hat{V}_{B2}(s),\hat{\mu}(t)\!\otimes\!\hat{\rho}_R\right]\right]
    \\
    &
    \!-
    \!\frac{1}{\hbar^2}\text{Tr}_R\!\!\int_0^\infty\!\!\!\! \text{d}s\!\left[\hat{V}_{B2}(t),\!\left[ \hat{V}_{B1}(s),\hat{\mu}(t)\!\otimes\!\hat{\rho}_R\right]\right]\!\!.
\end{split}
\end{equation}

Let us illustrate the calculation for the cavity dissipation given by $D_A$. The potential in the Interaction Picture is given by
\begin{equation}
    \hat{V}_A(t) = \im\int \text{d}\omega\gamma(\omega)\left[\hat{a}_0(\omega)\hat{c}^\dagger e^{\im(\delta_0-\Delta_0)t} \!-\! \text{H.c.}\right],
\end{equation}
where for simplicity we have defined
\begin{equation}\label{delta0definition}
    \delta_0 \equiv \tilde{\delta}-2g_{\rm cx}\beta_x-2g_{\rm cy}\beta_y.
\end{equation}
We now introduce this potential into Eq.~\ref{DA} and take the trace over the bath modes explicitly. In order to do so, we note that in a thermal state, 
\begin{equation}
\text{Tr}_B \left[\hat{a}_0(\omega)\hat{a}_0(\omega')\hat{\rho}_R\right]=\text{Tr}_B \left[\hat{a}_0^\dagger(\omega)\hat{a}_0^\dagger(\omega')\hat{\rho}_R\right]=0,
\end{equation}
\begin{equation}
\begin{split}
    \text{Tr}_B &\left[\hat{a}_0^\dagger(\omega)\hat{a}_0(\omega')\hat{\rho}_R\right]=
    \\
    &
    =\text{Tr}_B \left[\hat{a}_0(\omega)\hat{a}_0^\dagger(\omega')\hat{\rho}_R\right]-1= \delta(\omega-\omega')\bar{n}_\omega,
\end{split}
\end{equation}
where $\bar{n}_\omega$ is the mean occupation of the bath at frequency $\omega$, given by the Bose distribution. After tracing out we obtain
\begin{equation}
\begin{split}
    D_A &= -\int_0^\infty  \text{d}s\int_0^\infty \text{d}\omega \gamma^2(\omega)
    \\
    &\times\!\!\Big[\!(\bar{n}_\omega\!+\!1) \mathcal{A}_{\hat{c}^\dagger,\hat{c}}(\delta_0\!-\!\Delta_0)
    \!+\!\bar{n}_\omega \mathcal{A}_{\hat{c},\hat{c}^\dagger}(-\delta_0\!+\!\Delta_0)\!\Big]\!,
\end{split}   
\end{equation}
with
\begin{equation}
    \mathcal{A}_{\hat{a},\hat{b}}(\phi) = \hat{a}\hat{b}\mu e^{\im\phi} + \mu \hat{a}\hat{b}e^{-\im\phi}-2\hat{a}\mu\hat{b}\cos\phi.
\end{equation}
We now perform the integral in $s$ by using the identity
\begin{equation}
    \int_0^\infty \text{d}s e^{\pm \im s A} = \pi\delta(A) \pm \im\text{P.V.}\frac{1}{A},
\end{equation}
obtaining
\begin{equation}\label{DA2}
\begin{split}
    D_A &= 2\kappa\Big\{(\bar{n}_{\omega_0+\delta_0}+1)\mathcal{D}_{\hat{c}}[\mu]+
    \bar{n}_{\omega_0+\delta_0}\mathcal{D}_{\hat{c}^\dagger}[\mu]\Big\}
    \\
    &
    -2\im\left[\hat{c}^\dagger\hat{c},\mu\right]\text{P.V.}\!\!\int_0^\infty \!\! \text{d}\omega\gamma^2(\omega)\bar{n}_\omega \frac{1}{\delta_0+\omega_0-\omega},
\end{split}
\end{equation}
in terms of the Lindblad dissipator \cite{BreuerPetruccione,GardinerZollerQuantumNoise}
\begin{equation}
    \mathcal{D}_{\hat{a}}[\hat{\mu}] = \hat{a}\hat{\mu}\hat{a}^\dagger-\frac{1}{2}\left\{ \hat{a}^\dagger\hat{a},\hat{\mu}\right\},
\end{equation}
where the curly brackets denote the anti-commutator. To obtain the result Eq.~\ref{DA2}, we have used the same approximations used for Eqs. \ref{gammaint} and \ref{gammainttrick}. In order to estimate the Lamb shift of the cavity in Eq.~\ref{DA2}, we assume that, as most physical spectral densities, the coupling $\gamma$ satisfies $\gamma(\omega) \approx  0$ at low frequencies. Then, in the integration, the only relevant contribution will arise from frequencies close to $\omega = \delta_0+\omega_0 \approx\omega_0$. In this interval, we can approximate $\bar{n}_\omega \approx \exp(-\hbar\omega_0/k_BT)$ since $\exp(\hbar\omega_0/k_BT) \gg1$. Making also use of the identity $\gamma(\omega_0) =\sqrt{\kappa/\pi}$, we find after changing variables
\begin{equation}
\begin{split}
    2\text{P.V.}&\int_0^\infty \text{d}\omega\gamma^2(\omega)\bar{n}_\omega \frac{1}{\delta_0+\omega_0-\omega}\approx
    \\
    &
    \approx \frac{2\kappa}{\pi} e^{-u_0}\text{P.V.}\int_{-u_0}^\infty dye^{-y} \frac{1}{y}
\end{split}
\end{equation}
with $u_0 = \hbar(\omega_0+\delta_0)/k_B T\gg 1$. The integral above is named the Exponential Integral \cite{WongBookAsymptotic} $Ei(u_0)$, and its asymptotic expansion for $u_0 \gg 1$ is $E_i(u_0) \approx e^{u_0}[u_0^{-1}+O(u_0^{-2})]$. This identity allows us to approximate in this limit
\begin{equation}\label{DeltaAdefinition}
\begin{split}
    2\text{P.V.}&\int_0^\infty \text{d}\omega\gamma^2(\omega)\bar{n}_\omega \frac{1}{\delta_0+\omega_0-\omega}\approx
    \\
    &
    \approx \frac{2\kappa}{\pi} \frac{k_B T}{\hbar\omega_0} \equiv \Delta_A.
\end{split}
\end{equation}
where we have assumed $\omega_0 \gg \delta_0$, as is usual in cavity cooling experiments. Using this result, and noting again that, since the occupation of the environment at the frequency of the tweezer is negligible, i.e., $\exp(\hbar\omega_0/k_BT) \gg1$, we can finally write the dissipator $D_A$ in Eq.~\ref{DA2} as
\begin{equation}\label{DAfinal}
\begin{split}
    D_A[\hat{\mu}] &\approx 2\kappa\mathcal{D}_{\hat{c}}[\hat{\mu}]
    -\im\left[\Delta_A\hat{c}^\dagger\hat{c},\hat{\mu}\right].
\end{split}
\end{equation}
The above dissipator contains a frequency shift of the cavity and a Lindblad-type dissipation, which adds an exponential decay of the cavity photon occupation at rate $2\kappa$. Note that this frequency shift is not measurable independently or, in other words, a measurement of the frequency of the cavity in the absence of the NP would yield $\omega_c + \Delta_A$, this being the true frequency of the bare cavity.

The remaining contributions are calculated in the same fashion. First, the contribution $D_{B1}$ can be shown to have a similar form as Eq.~\ref{DAfinal}, as the operator $\hat{V}_{B1}$ also couples the cavity modes to the free EM field. This results again in both a frequency shift of the cavity and a Lindblad dissipation, this time induced by the presence of the NP:
\begin{equation}
\begin{split}
    D_{B1}[\hat{\mu}] &= -\im\left[\Delta_{B1}\hat{c}^\dagger\hat{c},\hat{\mu}\right] +
    \\
    &
    +2\kappa_{B1} \mathcal{D}_{\hat{c}}[\hat{\mu}]  +2\kappa_{B1}' \mathcal{D}_{\hat{c}^\dagger}[\hat{\mu}],
\end{split}
\end{equation}
where
\begin{equation}
    \Delta_{B1} =  \sum_{\mathbf{k}\varepsilon}\vert g_{\varepsilon\mathbf{k}}\vert^2(2\bar{n}_{\omega_\mathbf{k}}+1)\text{P.V.}\frac{1}{\delta_0-\Delta_\mathbf{k}},
\end{equation}
\begin{equation}
    \kappa_{B1} = \pi \sum_{\mathbf{k}\varepsilon}\vert g_{\varepsilon\mathbf{k}}\vert^2(\bar{n}_{\omega_\mathbf{k}}+1)\delta\left(\delta_0-\Delta_\mathbf{k}\right),
\end{equation}
\begin{equation}
    \kappa_{B1}' = \pi \sum_{\mathbf{k}\varepsilon}\vert g_{\varepsilon\mathbf{k}}\vert^2\bar{n}_{\omega_\mathbf{k}}\delta\left(\delta_0-\Delta_\mathbf{k}\right).
\end{equation}
By substituting above the expression for the couplings, Eq.~\ref{gvaerpsilonk}, and transforming the sum into an integral, we can determine the above coefficients as
\begin{equation}\label{DeltaB1definition}
\begin{split}
    \Delta_{B1} &= -\left(\frac{\epsilon_cV\cos\phi}{2\pi}\right)^2\frac{\omega_c}{3V_c c^3}
    \\
    &
    \times\text{P.V.}\int_0^{cK_c}\text{d}\omega(2\bar{n}_{\omega}+1)\frac{\omega^3}{\omega-(\omega_0+\delta_0)},
\end{split}
\end{equation}
\begin{equation}
\begin{split}
    \kappa_{B1} &= \!\left(\!\frac{\epsilon_cV\cos\phi}{2\pi}\!\right)^2\!\frac{\pi\omega_c}{3V_c }\!\left(\!\frac{\omega_0+\delta'}{c}\!\right)^3\!(\bar{n}_{\omega_0+\delta_0}+1),
\end{split}
\end{equation}
\begin{equation}
    \kappa_{B1}'=\kappa_{B1}\frac{\bar{n}_{\omega_0+\delta_0}}{\bar{n}_{\omega_0+\delta_0}+1}.
\end{equation}
For usual parameter values \cite{WindeyArXiv2018}, $\bar{n}_{\omega_0+\delta_0} \sim 10^{-13}$ and $\kappa_{B1}\approx 2\pi \times 100\cos^2\phi\text{Hz}$, which allows us to neglect $\kappa_{B1}'$ and express
\begin{equation}\label{kappaB1definition}
\begin{split}
    \kappa_{B1} &\approx \!\left(\!\frac{\epsilon_cV\cos\phi}{2\pi}\!\right)^2\!\frac{\pi\omega_c}{3V_c }\!\left(\!\frac{\omega_0+\delta_0}{c}\!\right)^3\!.
\end{split}
\end{equation}
The final contribution $D_{B1}$ thus reads
\begin{equation}
\begin{split}
    D_{B1}[\hat{\mu}] &= -\im\left[\Delta_{B1}\hat{c}^\dagger\hat{c},\hat{\mu}\right] 
    +2\kappa_{B1} \mathcal{D}_{\hat{c}}[\hat{\mu}] .
\end{split}
\end{equation}

Regarding the contribution $D_{B2}$, its calculation involves the approximation that both the coupling rates $g_{j\varepsilon\mathbf{k}}$ and the occupation numbers $\bar{n}_\omega$ are smooth enough functions to approximate their values at $\omega_0 \pm\Omega_j$ by their values at $\omega_0$. This approximation is extremely good in the regime $\Omega_j \ll \omega_0$, and has been checked numerically to be accurate up to $8$ digits in the final dissipation rates, for the parameters in Ref. \cite{WindeyArXiv2018}. By further neglecting the terms $\bar{n}_{\omega_0} \approx 0$ as done above, the resulting contribution reads
\begin{equation}\label{DB22}
\begin{split}
    D_{B2}[\hat{\mu}] &= -\im\sum_{jj'}\Delta_{jj'}\left[\hat{q}_j\hat{q}_{j'},\hat{\mu}\right]-\sum_j\Gamma_{jj}\left[\hat{q}_j,\left[\hat{q}_j,\hat{\mu}\right]\right]
    \\
    &-\sum_{j\ne j'}\Gamma_{jj'}\left(2\hat{q}_j\hat{\mu} \hat{q}_{j'}-\left\{\hat{q}_j\hat{q}_{j'},\hat{\mu}\right\}\right),
\end{split}
\end{equation}
with $\hat{q}_j \equiv \hat{b}_j^\dagger + \hat{b}_j$, dissipation rates
\begin{equation}
    \Gamma_{jj'} = \pi\sum_{\mathbf{k}\varepsilon}(\bar{n}_{\omega_\mathbf{k}}+1)\delta(\Delta_\mathbf{k})g_{j\varepsilon\mathbf{k}}^*g_{j'\varepsilon\mathbf{k}},
\end{equation}
and frequency shifts
\begin{equation}
    \Delta_{jj'} = -\sum_{\mathbf{k}\varepsilon}(\bar{n}_{\omega_\mathbf{k}}+1)\text{P.V.}\frac{1}{\Delta_\mathbf{k}}g_{j\varepsilon\mathbf{k}}^*g_{j'\varepsilon\mathbf{k}}.
\end{equation}
Although Eq.~\ref{DB22} has an apparent off-diagonal structure, by explicitly performing the angular integrals in the above rates we find that
\begin{equation}
    \Gamma_{jj'}, \Delta_{jj'} \propto \delta_{jj'}.
\end{equation}
Thus, we might write the contribution $D_{B2}$ in the more compact way
\begin{equation}\label{DB23}
\begin{split}
    D_{B2}[\hat{\mu}] &= -\im\sum_{j}\Delta_{j}\left[\hat{q}_j\hat{q}_{j'},\hat{\mu}\right]-\sum_j\Gamma_{j}^{(r)}\left[\hat{q}_j,\left[\hat{q}_j,\hat{\mu}\right]\right],
\end{split}
\end{equation}
which includes a frequency shift of the COM oscillations, and a dissipation of the Brownian Motion form \cite{GardinerZollerQuantumNoise} which we identify with the photon recoil heating and thus label with the superscript $(r)$. This kind of noise, different from the Lindblad dissipation appearing in $D_A$ and $D_{B1}$, induces heating without modifying the rate at which the COM occupations decay. The recoil heating rates are given by
\begin{equation}\label{Gammarecoildefinition}
    \left[\begin{array}{c}
         \Gamma_x^{(r)}  \\
         \Gamma_y^{(r)} \\
         \Gamma_z^{(r)}
    \end{array}\right] 
    = \frac{\pi\varepsilon_0}{30\hbar}\left(\frac{\epsilon_cVE_0}{2\pi}\right)^2
       k_0^5 \left[\begin{array}{c}
         x_0^2  \\
         2y_0^2 \\
         7z_0^2
    \end{array}\right], 
\end{equation}
and as expected are larger for the motion along the tweezer axis.
The frequency shifts, on the other hand, read
\begin{equation}\label{Deltajdefinition}
\begin{split}
    \left[\begin{array}{c}
         \Delta_x  \\
         \Delta_y \\
         \Delta_z
    \end{array}\right] 
    &= \frac{\varepsilon_0}{30\hbar c^5}\left(\frac{\epsilon_cVE_0}{2\pi}\right)^2
    \\
    &
    \times\!\text{P.V.}\!\int_0^{cK_c} \!\!d\omega \frac{\omega^3}{\omega_0-\omega} \!\left[\!\begin{array}{c}
         \omega^2x_0^2  \\
         2\omega^2y_0^2 \\
         (2\omega^2+5\omega_0^2)z_0^2
    \end{array}\!\right] \!.
\end{split}
\end{equation}

Finally, let us focus on the contribution $D_{B12}$. Under the same approximations undertaken above, this term can be shown to be
\begin{equation}
\begin{split}
    D_{B12}[\hat{\mu}] &= -\im\sum_j \left[\tilde{g}_j\hat{c}^\dagger\hat{q}_j+\text{H.c.},\hat{\mu}\right]
    \\
    &+\sum_j\bigg[\Upsilon_j\left(2\hat{q}_j\hat{\mu}\hat{c}^\dagger-\left\{\hat{c}^\dagger\hat{q}_j,\hat{\mu}\right\}\right)+\text{H.c.}\bigg],
\end{split}
\end{equation}
i.e., it contains a dissipative interaction between COM and cavity, with rate
\begin{equation}
\begin{split}
    \Upsilon_j &= \pi\sum_{\mathbf{k}\varepsilon}g_{\varepsilon\mathbf{k}}g_{j\varepsilon\mathbf{k}}\delta(\Delta_\mathbf{k})
    =\im\omega_c C_0 \cos\phi\cos\Theta \delta_{jz}
    ,
\end{split}
\end{equation}
and a coherent interaction at rate
\begin{equation}\label{gtildedefinition}
\begin{split}
    \tilde{g}_j &= -\sum_{\mathbf{k}\varepsilon}g_{\varepsilon\mathbf{k}}g_{j\varepsilon\mathbf{k}}\text{P.V.}\frac{1}{\Delta_\mathbf{k}}
    \\
    &
    =\im\omega_cC_0\cos\phi\cos\Theta\delta_{jz}\frac{1}{\pi k_0^3}\text{P.V.}\int_0^{K_c}\!\!\!\!dk\frac{k^3}{k_0-k},
\end{split}
\end{equation}
where we have defined the adimensional constant
\begin{equation}
    C_0 = \frac{\epsilon_c^2}{12\pi}(k_0z_0)(Vk_0^3)\sqrt{\frac{V}{V_c}\frac{\varepsilon_0V E_0^2}{2\hbar \omega_c}}.
\end{equation}
The contribution $D_{B12}$ is therefore simplified to
\begin{equation}
\begin{split}
    D_{B12}[\hat{\mu}] &= -\im\left[\tilde{g}_z\hat{c}^\dagger\hat{q}_z+\text{H.c.},\hat{\mu}\right]
    \\
    &+\bigg[\Upsilon_z\left(2\hat{q}_z\hat{\mu}\hat{c}^\dagger-\left\{\hat{c}^\dagger\hat{q}_z,\hat{\mu}\right\}\right)+\text{H.c.}\bigg].
\end{split}
\end{equation}
For typical experimental values \cite{WindeyArXiv2018}, $C_0\approx 10^{-13}$ so $\Upsilon \sim 2\pi \times 100$Hz.

After combining all the terms of this section into the von Neumann equation Eq.~\ref{vonNeumanneq3} and transforming back into the Schr\"odinger Picture, we find that we can regroup all the terms into the equation
\begin{equation}\label{generalMeq}
    \dot{\mu} = -\im\left[\hat{H}_S+\Delta\hat{H},\hat{\mu}\right] + \mathcal{D}[\hat{\mu}].
\end{equation}
The coherent part $\Delta\hat{H}$ contains all the coherent interactions and frequency re-normalizations,
\begin{equation}
\begin{split}
    \Delta\hat{H} &= \left(\Delta_A + \Delta_{B1}\right)\hat{c}^\dagger\hat{c} +
    \\
    &+\tilde{g}_{z}\hat{c}^\dagger\left(\hat{b}_z+\hat{b}_z^\dagger\right) +
    \sum_j\Delta_j (\hat{b}_j+\hat{b}_j^\dagger)^2.
\end{split}
\end{equation}
The first two terms in this equation can be reabsorbed into the frequencies and the couplings of $\hat{H}_S$, i.e., we define 
\begin{equation}\label{deltaprimedefinition}
    \delta' \equiv \delta_0 + \Delta_A + \Delta_{B1},
\end{equation}
\begin{equation}\label{gtjdefinition}
    g_{tj} = g_j + \tilde{g}_z\delta_{jz}.
\end{equation}
For the third term, we can do the same by redefining the phonon operators, i.e., we extract from $\hat{H}_S$ the Hamiltonian of the oscillators and add it up to this term, writing
\begin{equation}\label{eqprov1}
 \begin{split}
      \hbar\sum_j&\left[\Omega_j\hat{b}_j^\dagger\hat{b}_j + \Delta_j (\hat{b}_j+\hat{b}_j^\dagger)^2\right]=
      \\
      &
      =\sum_j\left[\frac{\hat{P}_j^2}{2m} +\frac{1}{2}m\Omega_j^2\left(1+\frac{4\Delta_j}{\Omega_j}\right)\hat{R}_j^2\right]=
      \\
      &
      =\sum_j\Omega_j'\hat{\tilde{b}}_j^\dagger\hat{\tilde{b}}_j.
 \end{split}
\end{equation}
From this equation, we can see that this term results in a re-normalization of the COM oscillator frequencies, which now read
\begin{equation}\label{renormalizedCOMfreqs}
    \Omega_j' = \Omega_j\sqrt{1+\frac{4\Delta_j}{\Omega_j}}.
\end{equation}
We have also defined new bosonic operators $\hat{\tilde{b}}_j$ relative to this frequency, so that the physical meaning of the position and momentum operators remains unchanged. In terms of these operators, every time a term $\hat{b}^\dagger_j + \hat{b}_j$ appears in the Hamiltonian, an extra multiplying factor  
\begin{equation}\label{chijdefinition}
    \chi_j = \sqrt{\frac{\Omega_j}{\Omega_j'}} = \left(1+\frac{4\Delta_j}{\Omega_j}\right)^{-1/4}
\end{equation}
must be added. This substitution allows us to write the full coherent part as
\begin{equation}\label{Hsprimedefinition}
\begin{split}
    \hat{H}_s + \Delta\hat{H} &= \delta'\hat{c}^\dagger\hat{c} + \sum_j\Omega_j'\hat{\tilde{b}}_j^\dagger\hat{\tilde{b}}_j+
    \\
    &
    +\sum_jg_j'\hat{c}^\dagger(\hat{\tilde{b}}^\dagger_j+\hat{\tilde{b}}_j) + \text{H.c.},
\end{split}
\end{equation}
where $g_j' \equiv g_{tj}\chi_j$, and we have explicitly written the expression for $\hat{V}_0$. In the main text, we remove the tildes from the operators in order to simplify the notation.

Regarding the dissipators in Eq.~\ref{generalMeq}, we can write them as
\begin{equation}\label{auxDISSIPATOR}
\begin{split}
    \mathcal{D}[\hat{\mu}] &= 2\kappa' \mathcal{D}_{\hat{c}}[\hat{\mu}]
   \\
   &
   +\left[\Upsilon\left(2\hat{q}_z\hat{\mu}\hat{c}^\dagger-\left\{\hat{c}^\dagger\hat{q}_z,\hat{\mu}\right\}\right)+\text{H.c.}\right]
    \\
    &
    -\sum_j\Gamma_{j}^{(r)}\left[\hat{\tilde{b}}^\dagger_j+\hat{\tilde{b}}_j,\left[\hat{\tilde{b}}^\dagger_j+\hat{\tilde{b}}_j,\hat{\mu}\right]\right],
\end{split}
\end{equation}
where we have defined $\kappa' \equiv \kappa+\kappa_{B1}$, $\Upsilon = \Upsilon_z\chi_z$, and the recoil heating $\Gamma_{j}^{(r)}\equiv \Gamma_j \chi_j^2$. Again, in the main text we remove the tildes from the operators to obtain the Master Equation \ref{MasterEqmu}.

\section{Equations of motion and motional PSD}\label{appendixEOMS}

In this appendix we detail how to solve the equation of motion \ref{MasterEqmu2} for the compound cavity+COM system,
\begin{equation}\label{eom1}
\begin{split}
    \dot{\mu}  &= -\frac{\im}{\hbar}\left[\hat{H}_S',\hat{\mu}(t)\right]+\mathcal{D}'[\hat{\mu}],
\end{split}
\end{equation}
where $\hat{H}_S'$ is given by Eq.~\ref{Hsprime}, and the dissipator reads
\begin{equation}\label{dissipator1}
\begin{split}
    \mathcal{D}'[\hat{\mu}] &= 2\kappa' \!\left[\hat{c}\hat{\mu}\hat{c}^\dagger-\frac{1}{2}\left\{ \hat{c}^\dagger\hat{c},\hat{\mu}\right\}\!\right]
     \!-\!\sum_j\Gamma_j\left[\hat{q}_j,\left[\hat{q}_j,\hat{\mu}\right]\right]
   \\
   &
   +\frac{\gamma}{4}\sum_j\left[\hat{q}_j,\left\{\hat{p}_j,\hat{\mu}\right\}\right].
\end{split}
\end{equation}
Here, we have neglected the incoherent interaction $\propto \Upsilon$ since, in the cases we consider in the main text, it is negligible. Otherwise, including this term is straightforward. In principle, one could aim at solving the above equation for the full density matrix $\mu$, but this is not necessary in the present case. Indeed, we note that the above Master Equation is quadratic, i.e. it only contains quadratic combinations of creation and annihilation operators. This property implies that the Gaussian character of any initial state is preserved, i.e., if the initial state is Gaussian, any system observable will be determined exclusively by the first- and second- order momenta ($\langle \hat{c}\rangle, \langle \hat{c}^\dagger\rangle,...$ and $\langle \hat{c}^\dagger\hat{c}\rangle, \langle \hat{c}^2\rangle,...$, respectively) \emph{at all times} \cite{GardinerZollerQuantumNoise}. We thus can derive a \emph{closed} system of differential equations containing only the expected values of single creation/annihilation operators, and their quadratic combinations.

In order to obtain the desired system of equations, we start by determining the time evolution of the expected value for a general operator $\hat{O}$ in the Schr\"odinger Picture, given by
\begin{equation}
\begin{split}
    \frac{\text{d}}{\text{dt}}\langle \hat{O} \rangle = \frac{\text{d}}{\text{dt}}\text{Tr}[\hat{O}\mu] =  \text{Tr}[\hat{O}\dot{\mu}].
\end{split}
\end{equation}
Introducing Eq.~\ref{eom1} and using the cyclic invariance of the trace, we can write
\begin{equation}
\begin{split}
    \frac{\text{d}}{\text{dt}}\langle \hat{O} \rangle &= 
    \im\langle[\hat{H}_S',\hat{O}]\rangle+2\kappa' \langle [ \cd \hat{O} c- \lbrace \cd c,\hat{O}/2\rbrace]\rangle 
    \\
    &- \!\sum_j \Gamma_j \langle  [\hat{q}_j,[\hat{q}_j,\hat{O}]]\rangle-\frac{\gamma}{4}\!\sum_j  \langle \lbrace \hat{p}_j,[\hat{q}_j,\hat{O}] \rbrace\rangle.
\end{split}
\end{equation}
This relation allows us to substitute any operator $\hat{O}$ and get its equation of motion, by using the fundamental commutation relations
 $[c,\cd]=1$ and $[b_i,\bd_j]=\delta_{ij}$. In this way, we obtain two independent systems of equations, namely one for the single-operator expected values, and a second for the quadratic combinations. The first one is given by the equations
 \begin{equation}\label{singleop1}
     \frac{\text{d}}{\text{dt}}\langle  c   \rangle=(-\im \delta' -\kappa')\langle c\rangle- \im \sum_j g_j'\langle \bd_j + b_j\rangle,
 \end{equation}
 \begin{equation}\label{singleop2}
 \begin{split}
     \frac{\text{d}}{\text{dt}}\langle  b_k   \rangle&=(- \im \Omega_k' -\gamma/2) \langle b_k \rangle 
     \\
     &- \im g_k' \langle \cd \rangle - \im  g^*_k{}' \langle c \rangle+(\gamma/2)\langle \bd_k\rangle,
\end{split}
 \end{equation}
 and their Hermitian conjugates, which together form an $8\times 8$ system of equations. Note that all the expected values decay with time and, since there is no independent term to these equations, the steady-state value of the single-operator expected values will be zero.
 On the other hand, the second system of equations is given by
 \begin{equation}
     \frac{\text{d}}{\text{dt}}\langle  \cd c   \rangle=-2\kappa' \langle \cd c \rangle+\bigg(\im \sum_jg_j^*{}'\langle(\bd_j+b_j)c\rangle + \text{C.c}\bigg),
\end{equation}
 \begin{equation}
     \frac{\text{d}}{\text{dt}}\langle  cc  \rangle=(-2\im\delta'-2\kappa') \langle c c \rangle-2 \im \sum_j g_j'\langle (\bd_j + b_j)c\rangle, 
 \end{equation}
 \begin{equation}
\begin{split}
        \frac{\text{d}}{\text{dt}}\langle  \bd_k&b_l  \rangle=- \im (\Omega_l'-\Omega_k') \langle \bd_k b_l \rangle +2\Gamma_k\delta_{kl} -\im  g_l' \langle \bd _k\cd\rangle
        \\
        &
        -\im  g_l^*{}'\langle \bd _kc\rangle +\im  g_k'\langle b _l \cd\rangle+\im g^*_k{}'\langle b _l c\rangle\\
        &+(\gamma/2)(p_l\bd_k-p_kb_l),
\end{split}
\end{equation}
\begin{equation}
\begin{split}
    \frac{\text{d}}{\text{dt}}\langle  &b_kb_l  \rangle=-\im(\Omega_l' +\Omega_k')\langle b_kb_l\rangle -2\Gamma_k\delta_{kl}
    \\
    &
    -\im g_l' \langle  b_k \cd\rangle- \im g_l^*{}' \langle  b_k c \rangle -\im g_k' \langle  b_l \cd\rangle- \im g_k^*{}' \langle  b_l c \rangle \\
&+(\gamma/2) (p_lb_k+p_kb_l+ \delta_{kl}),
\end{split}
\end{equation} 
\begin{equation}
\begin{split}
    \frac{\text{d}}{\text{dt}}\langle  b_k\cd  \rangle&=[-\im(\Omega_k'-\delta')-\kappa'-\gamma/2]\langle b_k \cd \rangle+(\gamma/2)\langle \bd_k\cd\rangle
    \\
    &
    -\im g_k' \langle \cd \cd \rangle-\im g_k^*{}' \langle  \cd c\rangle +\im \sum_j g_j^*{}'\langle (\bd_j+b_j)b_k\rangle,
\end{split}
\end{equation}
\begin{equation}
\begin{split}
    \frac{\text{d}}{\text{dt}}\langle  b_kc  \rangle&=[-\im(\Omega_k'+\delta')-\kappa'-\gamma/2]\langle b_k c\rangle+\gamma/2\langle \bd_kc\rangle
    \\
    &
    -\im g^*_k{}' \langle cc\rangle-\im g_k' \langle c\cd \rangle-\im \sum_jg_j' \langle (\bd_j+b_j) b_k\rangle,
\end{split}
\end{equation}
and their Hermitian conjugates (here, $C.c.$ stands for complex conjugate).
This second system contains $36$ equations, and has to be solved numerically.

The above equations of motion allow us to calculate the power spectral density of the COM motion, defined as
\begin{equation}
\begin{split}
    S_{jj}(\omega) &= \frac{1}{2\pi}\int_{-\infty}^\infty \text{d}\tau\langle \hat{x}_j(t+\tau)\hat{x}_j(t)\rangle_{\rm ss} e^{\im\omega \tau}
    \\
    &= \frac{r_{0j}^2}{2\pi}\int_{-\infty}^\infty \text{d}\tau\langle \hat{q}_j(t+\tau)\hat{q}_j(t)\rangle_{\rm ss} e^{\im\omega \tau},
\end{split}
\end{equation}
where ``ss'' refers to the steady state. In order to calculate the above PSD, we first split the integral into two parts, and use the time-translation invariance of the steady state, i.e., $\langle \hat{A}(t)\hat{B}(t+\tau)\rangle_{\rm ss} = \langle \hat{A}(t-\tau)\hat{B}(t)\rangle_{\rm ss}$, to write in a more convenient notation
\begin{equation}
    S_{jj}(\omega) = S_{jj}^+(\omega) + S_{jj}^-(\omega), 
\end{equation}
with $S_{jj}^-(\omega) = [S_{jj}^+(\omega)]^*$, and 
\begin{equation}\label{PSDoneside}
\begin{split}
    S_{jj}^+(\omega) &= \frac{r_{0j}^2}{2\pi}\int_{0}^\infty d\tau e^{\im\omega \tau}
    \\
    &
    \times\Big[\langle\hat{b}_j(t)\hat{b}_j(t+\tau)\rangle_{\rm ss}+\langle\hat{b}_j(t)\hat{b}_j^\dagger(t+\tau)\rangle_{\rm ss}
    \\
    &
    +\langle\hat{b}_j^\dagger(t)\hat{b}_j(t+\tau)\rangle_{\rm ss}+\langle\hat{b}_j^\dagger(t)\hat{b}_j^\dagger(t+\tau)\rangle_{\rm ss}\Big].
\end{split}
\end{equation}
Expressed in this form, the PSD can be calculated directly by applying the Quantum Regression Theorem. We start by defining the vector containing the single operators of our equation of motion,
\begin{equation}
    \hat{\mathbf{v}} \equiv \left( \hat{c}, \hat{c}^\dagger , \hat{b}_x, \hat{b}_x^\dagger, \hat{b}_y, \hat{b}_y^\dagger, \hat{b}_z, \hat{b}_z^\dagger\right),
\end{equation}
whose expected value is governed by the matrix equation
\begin{equation}
    \frac{d}{dt}\langle\hat{\mathbf{v}}(t)\rangle = M_0 \langle\hat{\mathbf{v}}(t)\rangle,
\end{equation}
where the matrix $M_0$ can be directly extracted from Eqs. \ref{singleop1} and \ref{singleop2}. According to the Quantum Regression Theorem \cite{CarmichaelBook}, any two-time correlation function involving a general operator $\hat{A}$ obeys the equation
\begin{equation}
    \frac{d}{d\tau}\langle \hat{A}(t)\hat{\mathbf{v}}(t+\tau)\rangle = M_0 \langle \hat{A}(t)\hat{\mathbf{v}}(t+\tau)\rangle.
\end{equation}
which can be easily solved. For the steady-state values appearing in Eq.~\ref{PSDoneside}, we find
\begin{equation}\label{solutionEDO}
    \langle \hat{A}(t)\hat{\mathbf{v}}(t+\tau)\rangle_{\rm ss} = \sum_l c_{0l}(\hat{A})\boldsymbol{\lambda}_l e^{\lambda_l \tau},
\end{equation}
where we have defined the eigenvalues and eigenvectors of the matrix $M_0$,
\begin{equation}
    M_0 \boldsymbol{\lambda}_l = \lambda_l \boldsymbol{\lambda}_l,
\end{equation}
and the vector of coefficients $\mathbf{c}_0$ obeys the equation
\begin{equation}
    \Lambda \mathbf{c}_0(\hat{A}) = \langle \hat{A}\hat{\mathbf{v}}\rangle_{\rm ss}.
\end{equation}
Here, $\Lambda$ is the matrix whose columns are the eigenvectors $\boldsymbol{\lambda}_j$. Although the PSD has to be calculated numerically, two important properties can be noted right away. First, from Eq.~\ref{solutionEDO} we see that any two-time correlation function can be written as a sum of exponential functions which, after integration in $\tau$, yield a sum of Lorentzian profiles. Indeed, if we define the position of the operators $\hat{b}_j$ and $\hat{b}^\dagger_j$ inside the vector $\hat{\mathbf{v}}$ as $p_j$ and $\bar{p}_j$ respectively, we can easily obtain
\begin{equation}
    S_{jj}^+(\omega)=-\frac{r_{0j}^2}{2\pi}\sum_l c_{0l}(\hat{q}_j)\frac{\boldsymbol{\lambda}_l\cdot\left(\mathbf{e}_{p_j} + \mathbf{e}_{\bar{p}_j}\right)}{\lambda_l+i\omega},
\end{equation}
where the vector $\mathbf{e}_l$ represents the length $8$ vector with components $\mathbf{e}_l^{(k)} = \delta_{lk}$.
From the above equation, we can conclude that, in general, the motional PSD will be composed by a set of Lorentzian peaks centered at $\omega \propto \pm\text{Im}\lambda_l$ and with a linewidth $\Delta\omega \propto \text{Re}\lambda_l$. Furthermore, note that the only system parameters determining the positions and widths of the peaks are the entries of the matrix $M_0$ or, equivalently, the coefficients of Eqs. \ref{singleop1} and \ref{singleop2}. In these equations, the position diffusion coefficient $\Gamma_j$ of the dissipator does not appear and, henceforth, these two parameters do not depend on $\Gamma_j$. However, the steady-state values of the two-operator products do depend on $\Gamma_j$, which thus influences the relative weight of the PSD peaks.

\section{Summary of system parameters}\label{appendixTABLE}

\begin{table*}[h!]
\centering
 \begin{tabular}{ m{2.5cm} | m{8cm} | m{5cm} } 
   Symbol & 
   Physical meaning
   & 
   \vspace{0.4cm}
   Definition and related parameters
   \tabularnewline [2.5ex] 
 \hline\hline
   $\Omega_j$ $\left(\Omega_j'\right)$
   & 
   Bare (renormalized) COM motional frequencies
   & 
   Eq.~\ref{COMfreqsdefinition} (Eqs.~\ref{renormalizedCOMfreqs}, \ref{Deltajdefinition})
  \tabularnewline  [1ex] 
 \hline
   $\kappa$ $\left(\kappa'\right)$
   & 
   Bare (renormalized) COM motional frequencies
   & 
   Sec. \ref{secHamiltonian}, main text 
   \newline
   (Eqs. \ref{auxDISSIPATOR} and \ref{kappaB1definition})
 \tabularnewline  [1ex] 
 \hline
   $\delta'$ 
   & 
   Renormalized detuning between cavity and tweezer
   & 
   Eq. \ref{deltaprimedefinition} 
   \newline
   (see also Eqs. \ref{Hlong1}, \ref{deltatildedefinition}, \ref{delta0definition}, \ref{DeltaAdefinition}, \ref{DeltaB1definition})
 \tabularnewline  [1ex] 
 \hline
   $\beta_j$ $\left(\alpha_c\right)$
   & 
   Displacements of the COM operators (of the cavity operators)
   & 
   Eq. \ref{bdisplacement} (Eq. \ref{cdisplacement})\newline
   Appendix \ref{appendixDISPSS}
 \tabularnewline  [1ex] 
 \hline
   $\alpha_cg_{cj}$
   & 
   Bare OM coupling induced by cavity occupation
   & 
   Eqs. \ref{gcx} and \ref{gcy}
 \tabularnewline  [1ex] 
 \hline
   $G$
   & 
   Bare OM coupling induced by photon scattering from the tweezer into the cavity
   & 
  Eq. \ref{Galpha}
 \tabularnewline  [1ex] 
 \hline
   $g_j$ $\left(g_j'\right)$
   & 
   Total bare (renormalized) optomechanical coupling
   & 
   Eq. \ref{OMcouplings} (Eq. \ref{Hsprimedefinition}, see also Eqs. \ref{gtjdefinition}, \ref{chijdefinition}, and \ref{gtildedefinition})
 \tabularnewline  [1ex] 
 \hline
   $\gamma$
   & 
   Gas pressure damping coefficient
   & 
  Eq. \ref{gammadefinition}
 \tabularnewline  [1ex] 
 \hline
   $\Gamma_j^{(r)}$ / $\Gamma_j^{(p)}$ / $\Gamma_j^{(d)}$
   & 
   Heating rates associated to photon recoil / gas pressure / trap displacement noise
   & 
   Eq. \ref{Gammarecoildefinition} / Eq. \ref{GammaPdefinition} / Eq. \ref{GammaVdefinition}
 \tabularnewline  [1ex] 
 \hline
   $\Gamma_j$
   & 
   Total heating rate
   & 
  Eq. \ref{Gammaj}
 \tabularnewline  [1ex] 
 \hline
 \end{tabular}
 \caption{Summary of the most relevant parameters defined in the main text. The corresponding primed symbols denote the same parameters after the re-normalization induced by tracing out the free EM modes.}\label{TableparamsAppendix}
\end{table*}

\medskip
 

%

\end{document}